\pdfoutput=1
\RequirePackage{ifpdf}
\ifpdf 
\documentclass[pdftex]{sigma}
\else
\documentclass{sigma}
\fi

\numberwithin{equation}{section}

\newtheorem{Theorem}{Theorem}[section]
\newtheorem*{Theorem*}{Theorem}
\newtheorem{Corollary}[Theorem]{Corollary}
\newtheorem{Lemma}[Theorem]{Lemma}
\newtheorem{Proposition}[Theorem]{Proposition}
 { \theoremstyle{definition}
\newtheorem{Definition}[Theorem]{Definition}

\newtheorem{Example}[Theorem]{Example}
\newtheorem{Remark}[Theorem]{Remark} }

\usepackage{booktabs}
\usepackage{verbatim,amscd,mathrsfs,stmaryrd}
\usepackage{tikz-cd}

\usepackage{mathtools}

\DeclarePairedDelimiter\floor{\lfloor}{\rfloor}

\usepackage{bbm}
\usepackage{bm}
\usetikzlibrary{positioning,calc,arrows,decorations.pathreplacing,decorations.markings,patterns}
\usepackage{ifthen}
\usepackage{hhline}

\usepackage{mathtools}
\newcommand{\myarrow}[1][-45]{%
	\mathrel{%
		\text{$
			\begin{tikzpicture}[baseline = -0.5ex]
				\node[inner sep=0pt,outer sep=0pt,rotate = #1] (a) at (0,0) {$\xrightarrow{}$};
			\end{tikzpicture}
			$}%
	}%
}%

\makeatletter
\newcommand{\doublewidetilde}[1]{{%
		\mathpalette\double@widetilde{#1}%
}}
\newcommand{\double@widetilde}[2]{%
	\sbox\z@{$\m@th#1\widetilde{#2}$}%
	\ht\z@=.9\ht\z@
	\widetilde{\box\z@}%
}
\makeatother

\usepackage{blkarray,stmaryrd}

\newcommand{\old}[1]{}

\newcommand{\bcg}{{balanced cylinder graph}}

\newcommand{\Z}{{\mathbb Z}}
\newcommand{\N}{{\mathbb N}}
\newcommand{\R}{{\mathbb R}}
\newcommand{\C}{{\mathbb C}}
\newcommand{\CP}{{\mathbb C\mathrm P}}

\newcommand{\T}{{\mathbb T}}
\newcommand{\A}{{\mathbb A}}

\newcommand{\disp}{{\rm disp}}
\renewcommand{\P}{{\mathbb P}}
\newcommand{\PGL}{\operatorname{PGL}}

\newcommand{\arv}{\stackrel{\vec \alpha}{\sim}}
\makeatletter
\newcommand{\extp}{\@ifnextchar^\@extp{\@extp^{\,}}}
\def\@extp^#1{\mathop{\bigwedge\nolimits^{\!#1}}}
\makeatother
\newcommand\restr[2]{{
		\left.\kern-\nulldelimiterspace
		#1
		\vphantom{\big|}
		\right|_{#2}
}}
\definecolor{calpolypomonagreen}{rgb}{0, 0.6, 0.2}

\newcommand{\ra}{\rightarrow}
\newcommand{\be}{\begin{equation}}
	\newcommand{\ee}{\end{equation}}
\newcommand{\bt}{\begin{Theorem}}
	
		\newcommand{\et}{\end{Theorem}}
	\newcommand{\bd}{\begin{Definition}}
		\newcommand{\ed}{\end{Definition}}
	\newcommand{\bp}{\begin{Proposition}}
		\newcommand{\ep}{\end{Proposition}}
	
\newcommand{\bl}{\begin{Lemma}}
	\newcommand{\el}{\end{Lemma}}
\newcommand{\bc}{\begin{Corollary}}
	\newcommand{\ec}{\end{Corollary}}
\newcommand{\bcon}{\begin{conjecture}}
	\newcommand{\econ}{\end{conjecture}}
\newcommand{\la}{\label}
\newcommand{\w}{{\rm w}}
\newcommand{\bw}{{\rm b}}
\newcommand{\wt}{{\rm wt}}

\def\cro{\operatorname{cr}}
\def\mr{\operatorname{mr}}

\DeclareMathOperator{\coker}{coker}
\DeclareMathOperator{\Hom}{Hom}
\DeclareMathOperator{\tr}{tr}
\DeclareMathOperator{\im}{im}

\tikzset{mid arrow/.style={postaction={decorate,decoration={
				markings,
				mark=at position .5 with {\arrow{latex}}
	}}},
	mid rarrow/.style={postaction={decorate,decoration={
				markings,
				mark=at position .5 with {\arrow{latex reversed}}
	}}},
}

\tikzset{qvert/.style={draw,black,circle,fill=gray,minimum size=5pt,inner sep=0pt} }
\tikzset{bvert/.style={draw,circle,fill=black,minimum size=5pt,inner sep=0pt} }
\tikzset{wvert/.style={draw,circle,fill=white,minimum size=5pt,inner sep=0pt} }
\tikzset{fvert/.style={text=blue} }
\tikzset{sqvert/.style={draw,black,rectangle,fill=black,minimum size=5pt,inner sep=0pt} }
\tikzset{lvert/.style={draw,circle,fill=black,minimum size=4pt,inner sep=0pt} }

\begin{document}

\allowdisplaybreaks

\newcommand{\arXivNumber}{2108.12692}

\renewcommand{\PaperNumber}{040}

\FirstPageHeading

\ShortArticleName{Integrable Dynamics in Projective Geometry}

\ArticleName{Integrable Dynamics in Projective Geometry\\ via Dimers and Triple Crossing Diagram Maps\\ on the Cylinder}

\Author{Niklas Christoph AFFOLTER~$^{\rm abc}$, Terrence GEORGE~$^{\rm d}$ and Sanjay RAMASSAMY~$^{\rm e}$}

\AuthorNameForHeading{N.~Affolter, T.~George and S.~Ramassamy}

\Address{$^{\rm a)}$~Technische Universit\"at Berlin, Institute of Mathematics,\\
\hphantom{$^{\rm a)}$}~Strasse des 17. Juni 136, 10623 Berlin, Germany}

\Address{$^{\rm b)}$~D\'epartement de math\'ematiques et applications, \'Ecole Normale Sup\'erieure,\\
\hphantom{$^{\rm b)}$}~CNRS, PSL University, 45 rue d'Ulm, 75005 Paris, France}

\Address{$^{\rm c)}$~Institute of Discrete Mathematics and Geometry, TU Wien,\\
\hphantom{$^{\rm c)}$}~Wiedner Hauptstra{\ss}e 8–10/104, 1040 Wien, Austria}
\EmailD{\href{mailto:niklas.affolter@tuwien.ac.at}{niklas.affolter@tuwien.ac.at}}
\URLaddressD{\url{https://sites.google.com/view/niklasaffolter/}}

\Address{$^{\rm d)}$~Department of Mathematics, UCLA, 520 Portola Plaza, Los Angeles, CA 90095, USA}
\EmailD{\href{tegeorge@math.ucla.edu}{tegeorge@math.ucla.edu}}
\URLaddressD{\url{https://terrencegeorge.github.io}}

\Address{$^{\rm e)}$~Universit\'e Paris-Saclay, CNRS, CEA, Institut de Physique Th\'eorique,\\
\hphantom{$^{\rm e)}$}~91191 Gif-sur-Yvette, France}
\EmailD{\href{mailto:sanjay.ramassamy@ipht.fr}{sanjay.ramassamy@ipht.fr}}
\URLaddressD{\url{https://www.normalesup.org/~ramassamy/}}

\ArticleDates{Received December 23, 2024, in final form May 20, 2025; Published online June 03, 2025}

\Abstract{We introduce twisted triple crossing diagram maps, collections of points in projective space associated to bipartite graphs on the cylinder, and use them to provide geometric realizations of the cluster integrable systems of Goncharov and Kenyon constructed from toric dimer models. Using this notion, we provide geometric proofs that the pentagram map and the cross-ratio dynamics integrable systems are cluster integrable systems. We show that in appropriate coordinates, cross-ratio dynamics is described by geometric $R$-matrices, which solves the open question of finding a cluster algebra structure describing cross-ratio dynamics.}


\Classification{37J70; 82B20; 13F60}

\section{Introduction}
	
A widely studied class of discrete systems which are integrable both in the algebro-geometric sense and in the Liouville sense is the cluster integrable systems of Goncharov and Kenyon \cite{GK13}. Cluster integrable systems are constructed from the dimer model on bipartite graphs on the torus coming from statistical mechanics. The class of cluster integrable systems was shown to contain many other integrable systems \cite{FM16}, several of which have geometric origins as moduli spaces of points in projective space. The most prominent example of such an integrable system is the pentagram map, discovered by Schwartz \cite{Schwartz1}. However, proofs of the coincidence of such a~geometric integrable system with some cluster integrable system are essentially algebraic; they involve constructing isomorphisms between the two integrable systems using coordinates.
	
	The main idea of this paper is that one can actually provide a more geometric identification by making use of a generalization of the recently introduced triple crossing diagram maps (TCD maps) \cite{AGPR,AGR}, which we call \textit{twisted TCD maps}. We develop the general theory of twisted TCD maps and show that any cluster integrable system can be realized as a geometric integrable system. A key role is played by the monodromy matrix, which provides a robust and systematic way to compute the conserved quantities (Hamiltonians and Casimirs) of these geometric integrable systems and relate them to the conserved quantities of cluster integrable systems.
	
	We then use the theory to show that the pentagram map and the cross-ratio dynamics integrable system of \cite{AFIT} are cluster integrable systems. For cross-ratio dynamics, this is a completely new result and this answers an open question asked in \cite{AFIT}. For the pentagram map, its cluster algebra structure was identified with that of a dimer model in \cite{AGPR}. In this paper, we illustrate the power of the framework of twisted TCD maps by providing a new derivation of the Hamiltonians and Casimirs of the pentagram map. This new derivation is faster than the classical ones~\mbox{\cite{OST1,Soloviev}} once we have the theory of twisted TCD maps. We expect that the geometric integrable systems arising from twisted TCD maps should encompass a large part (if not all) of the class of Y-meshes of \cite{GP}, which is one of the broadest generalizations of the pentagram~map.
	
	\subsection{Twisted TCD maps}
	
	The phase space of the cluster integrable system is an $\mathcal X$ cluster variety constructed from bipartite graphs in the torus as follows (see Section~\ref{sec:gk} for more details).
	Let $\Gamma$ be a bipartite graph in the torus $\T:= \R^2/\Z^2$ satisfying a certain minimality condition. Associated to $\Gamma$ is the space $\mathcal L_\Gamma:=H^1(\Gamma,\C^\times)$ of weights on it, where a weight $[\wt]$ is a cohomology class assigning to every homology class $[L] \in H_1(\Gamma,\Z)$ a nonzero complex number $[\wt]([L])$. Two such graphs are \textit{move-equivalent} if they are related by the two moves shown on the left and middle of Figure~\ref{fig:bigon}. Move-equivalence classes of (minimal) bipartite graphs are classified by convex integral polygons~$N$ in the plane; if $\Gamma$ is in the move-equivalence class of $N$, we write $N(\Gamma)=N$. Each move~${\Gamma \ra \Gamma'}$ induces a birational map of weights $\mathcal L_\Gamma \rightarrow \mathcal L_{\Gamma'}$; for the spider move, the map is given by the mutation formula for cluster $X$-variables \cite{FG} or $y$-variables \cite{FZ} depending on the authors. Gluing all the $\mathcal L_\Gamma$ in the move-equivalence class, we get the $\mathcal X$ cluster variety $\mathcal X_N$. Goncharov and Kenyon identified a Poisson structure on $\mathcal X_N$, generalizing the cluster Poisson structure of~\cite{GSTV}. Associated to $(\Gamma,[\wt])$ is a periodic finite-difference operator $K(z,w)$ called the Kasteleyn matrix whose determinant $P(z,w)$ is a bivariate polynomial whose coefficients are weighted enumerations of dimer covers with prescribed homology. The curve defined by this polynomial is called the \textit{spectral curve}. Goncharov and Kenyon showed that $\mathcal X_N$ is a Liouville integrable system whose Hamiltonians and Casimirs are coefficients of the polynomial defining the spectral curve. We will henceforth call them the GK Poisson structures, Casimirs and Hamiltonians.
	
	A twisted TCD map lives in the infinite cylinder \smash{$\widehat \A:= \R^2/\Z(0,1)$}. By using the contraction-uncontraction move, we may assume that every black vertex in $\Gamma$ has degree $3$. Let \smash{$\Gamma_{\widehat \A}$} denote the infinite periodic graph that is the preimage of $\Gamma$ under the quotient map \smash{$\widehat \A\ra \T$}. A \textit{TCD map} is a function \smash{$P\colon W\bigl(\Gamma_{\widehat \A}\bigr) \ra \CP^d$} assigning to every white vertex in \smash{$\Gamma_{\widehat \A}$} a point in $\CP^d$ such that for every black vertex $\bw$ incident to white vertices $\w_1$, $\w_2$, $\w_3$, the points~$P_{\w_1}$,~$P_{\w_2}$ and~$P_{\w_3}$ are contained in a line $\CP^1 \subseteq \CP^d$ \cite{AGR}. These maps are a version of the vector-relation configurations of \cite{AGPR} better adapted to geometric dynamics. A TCD map $P$ is called \textit{twisted} if there is an $M \in \PGL_{d+1}$, called the \textit{monodromy matrix}, such that $P_{\w+(1,0)} = M(P_\w)$, where $\w+(1,0)$ denotes the translate of $\w$.
	A twisted TCD map defines a weight $[\wt]$ on $\Gamma$ as follows. Let $L$ be a closed loop in $\Gamma$ such that $[L] \in \{0\} \times \Z$, i.e., which has zero homology in the horizontal direction. Let $\widetilde L={\rm w}_1 \ra {\rm b}_1 \ra {\rm w}_2 \ra {\rm b}_2 \ra \cdots \ra {\rm w}_n \ra {\rm b}_n \ra {\rm w}_1$ denote any lift of $L$ to \smash{$\Gamma_{\widehat \A}$} such that ${\rm w}_i \neq {\rm w}_{i+1}$ for every $i$. Let ${\rm w}_i'$ denote the third white vertex incident to ${\rm b}_i$ that is not in $\{{\rm w}_i,{\rm w}_{i+1}\}$. Then,
	\[
	[{\rm wt}]([L]) := \pm \mr{\big(P_{{\rm w}_1}, P_{{\rm w}_1'},P_{{\rm w}_2},P_{{\rm w}_2'},\dots, P_{{\rm w}_n},P_{{\rm w}_n'}\big)},
	\]
	where $\pm$ denotes an explicit sign, and
	\begin{equation*}
		\mr(P_1,\dots,P_{2m}):=\frac{\prod\limits_{i=1}^m (P_{2i-1}-P_{2i})}{\prod\limits_{i=1}^m (P_{2i}-P_{2i+1})}
	\end{equation*}
	is a $\PGL_{d+1}$-invariant generalizing the cross-ratio called the \textit{multi-ratio}. Note that $\PGL_{d+1}$-invariance of the multi-ratio and the twisted condition imply that $[{\rm wt}]([L])$ is independent of the choice of lift $\widetilde L$. The two moves in Figure~\ref{fig:bigon}
	induce transformations of TCD maps that give rise to the birational maps of weights described above; see Figure~\ref{fig:localmoves}.
	
	In Section~\ref{subsec:Kasteleyncylinder}, we construct from a weighted bipartite graph on the torus $(\Gamma,[\wt])$ a matrix~$\Pi(w)$. In Section~\ref{subsec:tcdmaps}, we use Kasteleyn theory in the cylinder to construct from $(\Gamma,[\wt])$ and a~choice of zig-zag path a TCD map $P$ on \smash{$\widehat \A$}. Our first main result is the following.
	
	\begin{Theorem}[cf.\ Theorem~\ref{prop:mon}]
		The TCD map $P$ is a twisted TCD map with monodro\-my~$-\Pi(1)$.	
	\end{Theorem}
	In other words, the construction of $[\wt]$ from a twisted TCD map by taking multi-ratios has a (left) inverse. Hamiltonians and Casimirs in geometric integrable systems are typically constructed as invariants of the monodromy matrix $M$ \cite{GSTV, KhesinSoloviev, OST1, Soloviev}. By definition, the GK Hamiltonians and Casimirs are coefficients of the spectral curve. Therefore, the following theorem turns out to be the key to proving that Hamiltonians and Casimirs in geometric integrable systems coincide with their GK counterparts.
	\begin{Theorem}[cf.\ Theorem~\ref{p:sc}]
		\label{thm:spectralbm}
		The spectral curve of the dimer model on an arbitrary minimal graph $\Gamma$ on the torus is given by
		\[
		\bigl\{(z,w)\in(\C^\times)^2 \mid \det (zI+\Pi(w))=0\bigr\}.
		\]
	\end{Theorem}
	The matrix $-\Pi(w)$ is closely related to the boundary measurement matrix for networks on cylinders of \cite{GSTV, GSV}, as detailed in Remark~\ref{rem:bmm}. In a recent preprint of which we learned during the completion of this work, Izosimov \cite{Izosimov1} made precise the connection between the integrable systems of \cite{GK13} and \cite{GSTV}. In particular, his main result provides a representation of the dimer spectral curve similar to Theorem~\ref{thm:spectralbm}. We note that analogous representations have appeared in physics in the special case of the periodic Toda chain \cite{EFS} and in mathematical physics via representation-theoretic arguments \cite{FM16}.

	\subsection{The pentagram map}
	
	\begin{figure}[ht]
		\centering
		\begin{tikzpicture}[scale=0.3,xslant=-2,xscale=4,yscale=4.5]
			\draw[dashed, gray] (0,0)--(8,0)--(11,1)--(3,1)--(0,0);
			
			\foreach[evaluate={\yy=int(\xx+1)}] \xx in {1,2,3,4} {
				\pgfmathtruncatemacro{\label}{\xx}
				\coordinate[wvert,label=above:$p_{\label}$] (p\xx) at (2*\xx+1,1);
				\coordinate[wvert,label=above:$\widetilde p_{\label}$] (q\xx) at (2*\xx+2,1);
				\coordinate[wvert,label=below:$p_{\label}$] (pp\xx) at (2*\xx-2,0);
				\coordinate[wvert,label=below:$\widetilde p_{\label}$] (qq\xx) at (2*\xx-1,0);
				\node[bvert] (b\xx) at (2*\xx+0.3,0.333*2.3) {};
				\node[bvert] (bb\xx) at (2*\xx+1-0.3,0.333*0.7) {};
			}	
			\node[bvert] (bb0) at (1-0.3,0.333*0.7) {};
			\node[bvert] (b5) at (2*5+0.3,0.333*2.3) {};
			\coordinate[wvert,label=right:${p_5=M(p_1)}$] (p5) at (2*5+1,1);
			\coordinate[wvert,label=right:${p_5=M(p_1)}$] (pp5) at (2*5-2,0);
			\foreach[evaluate={\yy=int(\xx+1)}] \xx in {1,2,3,4} {
				\pgfmathtruncatemacro{\label}{2*\xx+1}
				\pgfmathtruncatemacro{\u}{\xx+1}
				\draw[-,text=red,inner sep=1]
				(p\xx)--(b\xx)--(pp\yy)--(bb\xx)--(p\xx);
			}
			\foreach[evaluate={\yy=int(\xx+1)}] \xx in {1,2,3} {
				\pgfmathtruncatemacro{\label}{2*\xx+1}
				
			}
			
			\foreach[evaluate={\yy=int(\xx+1)}] \xx in {1,2,3,4} {
				\pgfmathtruncatemacro{\label}{2*\xx}
				
			}
			
			\foreach[evaluate={\yy=int(\xx-1)}] \xx in {1,2,3,4} {
				\draw[-] (qq\xx)--(bb\yy)
				;
			}
			\foreach[evaluate={\yy=int(\xx+1)}] \xx in {1,2,3,4} {
				\draw[-] (q\xx)--(b\yy)
				;
			}
		\end{tikzpicture}
		\caption{Twisted TCD map for the pentagram map with $n=4$, where $\widetilde p_i:=\overline{p_{i-1}p_{i}} \cap \overline{p_{i+1} p_{i+2}}$.}
		\label{fig:pentagramtcdintro}
	\end{figure}
	
	Our first application is to the pentagram map, a discrete dynamical system discovered by Richard Schwartz \cite{Schwartz1}, proved to be Liouville and discrete integrable in \cite{OST1,OST2} and algebro-geometric integrable in \cite{Soloviev}. Glick \cite{Glick} discovered an underlying cluster algebra structure (see also \cite{Schwartz}). Gekhtman, Shapiro, Tabachnikov and Vainstein \cite{GSTV} generalized the pentagram map and related it to the integrable systems associated to weighted networks in a torus \cite{GSV}. The pentagram map was further generalized to the noncommutative setting \cite{Ovenhouse} and to general algebraically closed fields \cite{Weinreich}.
	
	A \emph{twisted $n$-gon} in $\CP^2$ is a pair $(p,M)$ where $p\colon \Z \ra \CP^2$ and $M \in \PGL_{3}$ is a projective transformation called \emph{monodromy} such that $p_{i+n}=M(p_i)$ for all $i \in \Z$. The phase space of the pentagram map is the moduli space $\mathcal T_n$ of twisted $n$-gons satisfying a nondegeneracy condition. The rational map $T\colon \mathcal T_n \ra \mathcal T_n$ defined by $(p,M) \mapsto (q,M)$, where $q_i=\overline{p_{i-1}p_{i+1}}\cap \overline{p_{i}p_{i+2}}$ is called the \textit{pentagram map}.

	In Section~\ref{sec:pentagram}, we construct a twisted TCD map on a bipartite torus graph denoted $\Xi_n$ for the pentagram map (see Figure~\ref{fig:pentagramtcdintro} for the twisted TCD map when $n=4$). The coordinates of Schwartz \cite{Schwartz} and Glick \cite{Glick} can be obtained from the twisted TCD map by taking weights of appropriate cycles. Using the correspondence between twisted TCD maps and the $\mathcal X$ cluster variety $\mathcal X_{\Xi_n}$ associated to the graph $\Xi_n$, we show the following.

\begin{Theorem} [cf.\ Proposition~\ref{prop:pentagramtcd} and Theorem~\ref{mainthm:pentagram}] \label{thm:pentintro}
The map sending the twisted TCD map~$P$ to the twisted $n$-gon $p$ induces a Poisson birational map $\pi_n\colon \mathcal X_{N_{\Xi_n}}^\lambda \ra \mathcal T_n$ that restricts to a~birational isomorphism between symplectic leaves on the two sides. The GK Hamiltonians are the pullbacks of the pentagram map Hamiltonians by $\pi_n$.
	\end{Theorem}

	Here, \smash{$\mathcal X_{N_{\Xi_n}}^\lambda$} is a level set of \smash{$\mathcal X_{N_{\Xi_n}}$} where we set a GK Casimir equal to $\lambda \in \C^\times$ and by a birational isomorphism, we mean a birational map that preserves the symplectic structure. Almost all of Theorem~\ref{thm:pentintro} is well known; see \cite{FM16,GSTV,GR}. However, the precise correspondence between the OST Hamiltonians and GK Hamiltonians for general $n$ has not been clarified before; the case $n=5$ appears in \cite{GR}. Our motivation for including it is that twisted TCD maps give a~quick proof, and the same type of argument will be used to prove the analogous results for the cross-ratio dynamics integrable system for which these results are new.

	\subsection{Cross-ratio dynamics}
	\label{subsec:crdynintro}

	\begin{figure}[ht]
		\centering
		\def\scl{0.7}
		\begin{tikzpicture}[scale=\scl,yscale=1.8]
			
			\node[wvert,label=below:$p_2$] (p2) at (2,0) {};
			\node[wvert,label=below:$p_3$] (p4) at (4,0) {};
			\node[wvert,label=below:$p_4$] (px0) at (6,0) {};
			
			\node[wvert,label=above:$p_1$] (pp2) at (2,1.73) {};
			\node[wvert,label=above:$p_2$] (pp4) at (4,1.73) {};
			\node[wvert,label=above:$p_3$] (ppx0) at (6,1.73) {};
			\node[wvert,label=left:$q_1$] (p1) at (1,0.87) {};
			\node[wvert,label=above:$q_2$] (p3) at (3,0.87) {};
			\node[wvert,label=above:$q_3$] (p5) at (5,0.87) {};
			\node[wvert,label=above:$q_4$] (pz1) at (7,0.87) {};
			\node[wvert,label=above:$p_4$] (pz2) at (8,1.73) {};
			
			\node[wvert,label=below:$p_1$] (pz02) at (0,0) {};

			\node[wvert,label=above:$M(p_1)$] (qz2) at (10,1.73) {};
			\node[wvert,label=right:$M(q_1)$] (qz1) at (9,0.87) {};
			\node[wvert,label=below:$M(p_1)$] (qx0) at (8,0) {};

			\draw[dashed,gray] (qx0)-- (qz1) -- (qz2) --(pz2)-- (ppx0)--(pp4) -- (pp2) -- (p1)--(pz02)--(p2)--(p4)--(px0) -- (qx0);	
			
			\node[bvert] (b1) at (1,0.37) {};
			\node[bvert] (b2) at (2,1.24) {};
			\node[bvert] (b3) at (3,0.37) {};
			\node[bvert] (b4) at (4,1.24) {};
			\node[bvert] (b5) at (5,0.37) {};
			\node[bvert] (b7) at (7,0.37) {};
			\node[bvert] (b8) at (8,1.24) {};
			
			\node[bvert] (bx0) at (6,1.24) {};
			\draw[-,text=red,inner sep=1]
			(b1) edge (pz02) edge (p2) edge (p1)
			(b3) edge (p2) edge (p3) edge (p4)
			(b5) edge (p4) edge (p5) edge (px0)
			(bx0) edge (pz1)
			(b2) edge (p1) edge (pp2) edge (p3)
			(b4) edge (p3) edge (pp4) edge (p5)
			(bx0) edge (p5) edge (ppx0)		
			(b8) edge (qz1)	 edge (pz1) edge (pz2)
			(b7) edge (qx0) edge (pz1) edge (px0)
			;
			
		\end{tikzpicture}\hspace{10mm}
		\begin{tikzpicture}[scale=\scl]
			\def\nnp{5};
			\def\nn{4};
			\def\nnm{3};
			
			\draw[dashed,gray] (1,4)--(1,0)--(5,0)--(5,4)--(1,4);
			\coordinate[wvert,label=left:$p_{1}$] (p1) at (1,3);
			\coordinate[wvert,label=left:$q_{1}$] (q1) at (1,1);
			\coordinate[bvert] (b1) at (1+0.4,3);		
			\coordinate[bvert] (c1) at (1+0.4,1);
			\foreach[evaluate={\yy=int(\xx+1)}] \xx in { 3, ..., \nnm} {
				\pgfmathtruncatemacro{\label}{\xx}
				\coordinate[wvert,label=above:$p_{\label}$] (p\xx) at (\xx,3);
				\coordinate[wvert,label=above:$q_{\label}$] (q\xx) at (\xx,1);		
				\coordinate[bvert] (b\xx) at (\xx+0.4,3);		
				\coordinate[bvert] (c\xx) at (\xx+0.4,1);	
			}
			\foreach[evaluate={\yy=int(\xx+1)}] \xx in {2, 4, ..., \nn} {
				\pgfmathtruncatemacro{\label}{\xx}
				\coordinate[wvert,label=above:$q_{\label}$] (p\xx) at (\xx,4);
				\coordinate[wvert,label=below:$q_{\label}$] (pp\xx) at (\xx,0);
				\coordinate[wvert,label=left:$p_{\label}$] (q\xx) at (\xx,2);
			}
			\foreach[evaluate={\yy=int(\xx+1)}] \xx in {1, 3, ..., \nnm} {
				\draw[-]
				(b\xx) edge (p\xx) edge (p\yy) edge (q\yy)
				(c\xx) edge (q\xx) edge (q\yy) edge (pp\yy)
				;
			}
			
			\coordinate[wvert,label=right:$M(p_{1})$] (p5) at (\nn+1,3);
			\coordinate[wvert,label=right:$M(q_{1})$] (q5) at (\nn+1,1);
			\foreach[evaluate={\yy=int(\xx-1)}] \xx in {3, 5} {
				\coordinate[bvert] (b\xx) at (\xx-0.4,3);		
				\coordinate[bvert] (c\xx) at (\xx-0.4,1);		
				\draw[-]
				(b\xx) edge (p\xx) edge (p\yy) edge (q\yy)
				(c\xx) edge (q\xx) edge (pp\yy) edge (q\yy)
				;			}

		\end{tikzpicture}
		
		\caption{Twisted TCD maps for cross-ratio dynamics when $n=4$. The graph on the left is $\Delta_4$ and the one on the right is $\Gamma_4$.}
		\label{fig:crtcdintro}
	\end{figure}
	
	Let $(\alpha_j)_{j\in\Z}$ be a bi-infinite sequence of elements of $ \C^\times$ and consider maps $f\colon \Z^2 \rightarrow \CP^1$ such that for every $(i,j)\in\Z^2$, we have
	\begin{align}
		\frac{(f_{i,j}-f_{i+1,j})(f_{i+1,j+1}-f_{i,j+1}) }{(f_{i+1,j}-f_{i+1,j+1})(f_{i,j+1}-f_{i,j})} = \alpha_j,
		\label{eq:disconf}
	\end{align}
	where we denote by $f_{i,j}$ the value taken by $f$ at $(i,j)\in\Z^2$. Such a map is a special case of a~discrete version of the Schwarzian KdV equation \cite{NC}, or a special case of a \emph{discrete isothermic surface} \cite{BP} restricted to the sphere $S^2$. In both cases it was shown that these maps are a discrete integrable system in the sense that they admit a discrete Lax representation. An interesting question is: Given all the $\alpha_j$, what is the space of solutions of \eqref{eq:disconf}? To answer this, consider each column $i$ of $\Z^2$ as a \emph{discrete curve} $f_i\colon \Z\rightarrow \CP^1$. The discrete curve corresponding to two adjacent columns $f_i$ and $f_{i+1}$ are called $\vec\alpha$-related, where $\vec\alpha$ is the vector of all $\alpha_j$. In the case when $\alpha_j$ is independent of $j$, the curve $f_{i+1}$ is called a \emph{Darboux transform} \cite{hjbook} of $f_i$ in the discrete differential geometry community, see also \cite{bsddgbook}. If we know $f_i$ and one point of $f_{i+1}$, then equation~\eqref{eq:disconf} determines all of $f_{i+1}$. As a consequence, there is a complex one-parameter freedom for each additional column of $\Z^2$.
	
	However, this changes if one considers \emph{periodic} maps, that is maps $\Z \times \Z/n\Z\rightarrow \CP^1$ that satisfy equation~\eqref{eq:disconf} for some $n\in \N$. These periodic discrete curves can be seen as \emph{closed $n$-gons} in $\CP^1$. In this case, if we know $f_i$ then there are only two possible solutions for $f_{i+1}$, because $f_{i+1}$ has to be periodic as well. Thus if we know both $f_{i-1}$ as well as $f_i$ and assume that $f_{i+1} \neq f_{i-1}$, then $f_{i+1}$ is determined \emph{uniquely}. Special attention has been paid to the case that~$\alpha_j$ does not depend on $j$. In this case, periodic solutions to equation \eqref{eq:disconf} have been studied as \emph{periodic discrete conformal maps} \cite{HMNP} with respect to algebro-geometric integrability. Also in this case, the map $(f_{i-1},f_i) \mapsto (f_i,f_{i+1})$ is called \emph{cross-ratio dynamics} \cite{AFIT}. Cross-ratio dynamics can also be generalized from closed $n$-gons to \emph{twisted} $n$-gons, that is curves $f_i\colon \Z \rightarrow \CP^1$ such that $f_i(j+n) = M(f_i(j))$ for all $j\in \Z$ and for some {length} $n\in \N$ and monodromy $M\in \PGL_2$.
	
	We now transition to the notation of \cite{AFIT}. Let $p_i := f_0(i)$ and $q_i := f_1(i)$ for $i \in \Z$.
	The phase space of the cross-ratio dynamics integrable system is the moduli space ${\mathcal U}_{n,\vec\alpha}$ of pairs~$(p,q)$ of $\vec\alpha$-related nondegenerate twisted $n$-gons modulo $\PGL_2$. Arnold, Fuchs, Izmestiev and Tabachnikov~\cite{AFIT} identified a Poisson structure on ${\mathcal U}_{n,\vec\alpha}$ and proved integrability in the sense of Liouville. Indeed, they provide Poisson brackets that are preserved by the dynamics as well as integrals of motion that are Casimirs and Hamiltonians. We will henceforth call them the AFIT Poisson structures, Casimirs and Hamiltonians.
	
	In Section~\ref{sec:crtcd12}, we give two different constructions of twisted TCD maps for cross-ratio dynamics, one on a hexagonal lattice denoted $\Delta_n$ and the other on a square lattice denoted $\Gamma_n$ (see Figure~\ref{fig:crtcdintro} for the case $n=4$). The map sending the twisted TCD map to the pair of twisted $n$-gons $(p,q)$ induces a birational map \smash{$\pi_{\vec{\alpha}}\colon \mathcal X_{N_{\Theta_n},\vec \alpha}^\lambda \ra {\mathcal U}_{n,\vec\alpha}$}, where \smash{$\mathcal X_{N_{\Theta_n},\vec \alpha}^\lambda$} denotes a closed subvariety of the space of dimer weights for $\Theta_n$, where $\Theta \in \{\Gamma,\Delta\}$. We summarize several results of Section~\ref{sec:crtcd12} in the following theorem.
	\begin{Theorem}
		\label{thm:main1}
		Let $n\geq2$, let $\vec{\alpha}\in(\C\setminus\{0\})^n$, and let $\Theta \in \{\Gamma,\Delta\}$. The map $\pi_{\vec{\alpha}}$ is a Poisson birational map from \smash{$\mathcal X_{N_{\Theta_n},\vec\alpha}^\lambda$} to ${\mathcal U}_{n,\vec\alpha}$ that restricts to a birational isomorphism between symplectic leaves on the two sides. The GK Hamiltonians are related to the pullbacks of the AFIT Hamiltonians by $\pi_{\vec{\alpha}}$ by an invertible linear transformation.
	\end{Theorem}	
	
	An explicit geometric bridge between cross-ratio dynamics and the dimer model is given by the following result.
	
	\begin{Theorem}[cf.\ Theorems~\ref{thm:crdynamicslocalmoves} and~\ref{thm:sqseq}] \label{thm:main2}
		Let $n\geq2$, let $\vec{\alpha}\in(\C\setminus\{0\})^n$, and let $\Theta \in \{\Gamma,\Delta\}$. Pairs of $\vec{\alpha}$-related twisted polygons of length $n$ arise as twisted TCD maps on $\Theta_{n,\A}$ taking values in $\CP^1$ and cross-ratio dynamics arises as an explicit sequence of local moves on these twisted TCD maps.
	\end{Theorem}
	
	An important note regarding Theorem~\ref{thm:main2} is that the first local move in the sequence depends on a parameter and that parameter depends globally on the initial pair of $\vec{\alpha}$-related twisted polygons. {In this sense, the sequence of transformations could be termed a \emph{semi-local} transformation.} Furthermore, combining Theorem~\ref{thm:main1} with Theorem~\ref{thm:main2}, we obtain an alternative proof of the conservation of the AFIT Hamiltonians (stated as Corollary~\ref{cor:alternative}), since the dynamics on TCD maps is conjugated to the dimer integrable dynamics of~\cite{GK13}. A more explicit statement of Theorem~\ref{thm:main2} is given by Theorem~\ref{thm:crdynamicslocalmoves}.
	
	As a first corollary of Theorem~\ref{thm:main2}, we find that the evolution of certain coordinates under cross-ratio dynamics is given by a so-called \emph{geometric $R$-matrix transformation}. Geometric $R$-matrices have been introduced in representation theory in relation with geometric crystals~\mbox{\cite{BK,Etingof,KNY,KNO}} and are so named because they are birational maps that tropicalize to combinatorial $R$-matrices \cite{MR1187560}. They first received an interpretation in terms of semi-local transformations of electrical networks~\cite{LP3, LP1,LP2} then in terms of semi-local transformations for dimer models~\mbox{\cite{Chepuri,GeRa, ILP1,ILP2}}. Transforming our graphs $\Gamma_n$, we recover the graphs of \cite{ILP1} whose semi-local transformation is described by a geometric $R$-matrix transformation, hence the following result.
	
	\begin{Corollary}[cf.\ Proposition~\ref{prop:uevolution}]
		\label{cor:rmatrix}
		The evolution of some coordinates under cross-ratio dynamics is given by a geometric $R$-matrix transformation.
	\end{Corollary}
	
	Recently, it was observed that another geometric dynamics, polygon recutting, was also governed by geometric $R$-matrices~\cite{Izosimov2}.
	
	As a second corollary, we answer an open question of~\cite{AFIT} asking for an interpretation of cross-ratio dynamics in terms of cluster algebras. Indeed, all but the first and the last operations for TCD maps of Theorem~\ref{thm:main2} have a cluster algebra interpretation~\cite{AGPR,AGR}. Actually, Inoue--Lam--Pylyavskyy showed in~\cite{ILP2} that this sequence of operations, including the first and the last one, could be interpreted as cluster algebra mutations provided one considers a decorated version of the bipartite graph.
	
	\begin{Corollary}
		The evolution of some coordinates under cross-ratio dynamics can be written as an explicit composition of cluster algebra mutations.
	\end{Corollary}
	
	Geometric $R$-matrix transformations give rise to the class of generalized cluster transformations that were systematically studied in~\cite{GeRa}. In Section~\ref{subsec:otherdynamics}, we describe explicitly the group of all generalized cluster transformations associated with the Newton polygon $\Delta_n$.
	
	As noted by~\cite{AFIT}, cross-ratio dynamics bears a lot of resemblances with the pentagram map. There is however a notable difference with cross-ratio dynamics. For the pentagram map and its generalizations, the dynamics is local in the sense than one can construct a point of the twisted $n$-gon $q$ knowing only a bounded number of points of the twisted $n$-gon $p$. For cross-ratio dynamics the dynamics is global, one needs to know all the points of $p$ to construct any given point of $q$.
	
	We end the introduction by remarking that it is mysterious to us that cross-ratio dynamics can be realized as a cluster integrable system in at least two different ways. The two realizations have different Casimirs and reveal different symmetries of the system. We believe this phenomenon deserves further study.

	\subsection*{Organization of the paper}
	In Section~\ref{sec:gk}, we recall the Goncharov-Kenyon integrable system \cite{GK13} associated with the dimer model on the torus. In Section~\ref{sec:cyl}, we consider the dimer model on the cylinder, construct the matrix $\Pi(w)$ and prove Theorem~\ref{thm:spectralbm}. We introduce in Section~\ref{sec:tcdtorus} the notion of twisted TCD maps associated to a bipartite graph on the cylinder. In Section~\ref{sec:pentagram}, we realize the pentagram map as a~twisted TCD map and show that it coincides with a cluster integrable system. In~Section~\ref{sec:crdynamics}, we provide the necessary background on cross-ratio dynamics and its integrability following mostly~\cite{AFIT}. In~Section~\ref{sec:crtcd12}, we realize the cross-ratio dynamics integrable system as a~cluster integrable system in two different ways, and describe the sequence of local moves for twisted TCD maps that realize cross-ratio dynamics, as stated in Theorem~\ref{thm:main2}. Section~\ref{subsec:localmovestcd} shows Corollary~\ref{cor:rmatrix} on the relation with geometric $R$-matrices. Finally Appendix~\ref{sec:schur} presents some results used in Sections~\ref{sec:cyl} and \ref{sec:tcdtorus} related to the classical notion of Schur complement.

	\section{The cluster integrable system}
	\label{sec:gk}
	
	In this section, we recall the integrable system associated with the dimer model on a weighted graph on a torus. For further details, see \cite{GK13}.
	
	\subsection{The dimer model in a torus}
	
	Let $\Gamma=(B \cup W,E)$ be a bipartite graph embedded in a torus $\T$ such that $|B|=|W|$ and such that the faces of $\Gamma$, that is, the connected components of the complement of $\Gamma$, are topological disks. We denote by $F$ the set of faces of $\Gamma$. An \emph{edge weight} on $\Gamma$ is a function ${\rm wt}\colon E \ra \C^\times$. Two edge weights ${\rm wt}_1$ and ${\rm wt}_2$ are said to be \emph{gauge equivalent} if there is a function $g\colon B \cup W \ra \C^\times$ such that for every edge $e=\bw \w$ with ${\rm b} \in B$, ${\rm w}\in W$, we have ${\rm wt}_2(e)={\rm wt}_1(e) g({\rm w}) g({\rm b})^{-1}$. Let~$\mathcal L_\Gamma$ denote the space of edge weights modulo gauge equivalence {and denote by $[{\rm wt}]$ the gauge equivalence class of the weight ${\rm wt}$}.
	
	To rephrase the above in the language of algebraic topology, we consider the graph $\Gamma$ to be a~cell complex whose $0$- and $1$-cells are $B \cup W$ and $E$, respectively. Considering each edge $e={\rm b w}$ to be oriented from ${\rm b}$ to ${\rm w}$, we have the nonzero cellular chain groups
	\[
	C_0(\Gamma,\Z)=\Z B \oplus \Z W, \qquad C_1(\Gamma,\Z)=\Z E,
	\]
	with boundary homomorphism $\partial\colon C_1(\Gamma,\Z) \ra C_0(\Gamma,\Z)$ given by $\partial(e)={\rm w}-{\rm b}$, so that $H_1(\Gamma,\Z)= \ker \partial$. Dually, we have cellular cochain groups
	$
	C^q(\Gamma,\C^\times):= \Hom _\Z(C_q(\Gamma,\Z),\C^\times),
	$
	for $q \in \{0,1\}$, with coboundary homomorphism $\delta\colon C^0(\Gamma,\C^\times) \ra C^1(\Gamma,\C^\times)$ given by $\delta(g)(e)=\frac{g({\rm w})}{g({\rm b})}$.
	Since an edge weight is a $1$-cochain and two edge weights are gauge equivalent if and only if they differ by a~$1$-coboundary, we have
	\[
		\mathcal L_\Gamma=H^1\big(\Gamma,\C^\times\big):=C^1\big(\Gamma,\C^\times\big)/\delta\bigl(C^0\big(\Gamma,\C^\times\big)\bigr).
	\]
	{Then}, $[{\rm wt}]$ {is} the cohomology class represented by the cochain ${\rm wt}$.
	
	For $[L] \in H_1(\Gamma,\Z)$, we denote by ${[\rm wt]}([L])$ the result of evaluating the cohomology class $[{\rm wt}]$ on the homology class $[L]$ yielding an alternating product of edge weights around $L$ (the product is alternating due to our choice of orientation of edges from $\bw$ to $\w$). Explicitly, if the $L$ is the $1$-cycle ${\rm w}_1 \xrightarrow[]{e_1} {\rm b}_1 \xrightarrow[]{e_2} {\rm w}_2 \xrightarrow[]{e_3} {\rm b}_2 \xrightarrow[]{e_4} \cdots \xrightarrow[]{e_{2n-2}} {\rm w}_n \xrightarrow[]{e_{2n-1}} {\rm b}_n \xrightarrow[]{e_{2n}} {\rm w}_1 \in H_1(\Gamma,\Z)$, we have
	\[
	[{\rm wt}]([L])=\prod_{i=1}^n \frac{\wt(e_{2i})}{\wt(e_{2i-1})}.
	\]

	Since $\mathcal L_\Gamma=\Hom _\Z(H_1(\Gamma,\Z),\C^\times)$ is an algebraic torus, the algebra $\mathcal O_{\mathcal L_\Gamma}$ of regular functions on $\mathcal L_\Gamma$ is generated by the characters $\chi_{[L]}$ for $[L] \in H_1(\Gamma,\Z)$ defined by $\chi_{[L]}([{\rm wt}]):=[{\rm wt}]([L])$.
	
	We now give a description of $\mathcal O_{\mathcal L_\Gamma}$ in terms of a basis. For a face $f$ of $\Gamma$, let $\partial f$ denote the counterclockwise oriented cycle given by the walk along the boundary of $f$ and {define the \emph{face weight}} $X_f :=\chi_{[\partial f]}$. Let $a$ and $b$ denote two cycles in $\Gamma$ {such that their homology classes $[a]$ and~$[b]$ generate} $H_1(\T,\Z)$. Then
	\[
		\mathcal O_{\mathcal L_\Gamma}=\C\bigl[X_f^{\pm 1},\chi_{[a]}^{\pm 1},\chi_{[b]}^{\pm 1}\bigr]/\big(1-\mbox{$\prod_{f \in F}$} X_f\big),
	\]
	where the relation \smash{$\prod_{f \in F} X_f=1$} comes from the relation \smash{$\sum_{f \in F} [\partial f]=0$} in $H_1(\Gamma,\Z)$. {While this set of generators is natural, both from the point of view of cluster algebras and topology, we will see in the examples of the pentagram map and cross-ratio dynamics that other generators are often more convenient to work with.}
	
{\bf Zig-zag paths and the Newton polygon.}
	A \emph{zig-zag} path in $\Gamma$ is a path that turns maximally left at white vertices and maximally right at black vertices. Let $\mathcal Z$ denote the set of zig-zag paths of $\Gamma$. Each zig-zag path $\beta \in \mathcal Z$ defines a homology class $[\beta]\in H_1(\T,\Z)$. Label the zig-zag paths $\beta_1, \beta_2,\dots,\beta_{|\mathcal Z|}$ so that the $[\beta_i]$ regarded as vectors in $H_1(\T,\Z) \otimes \R \cong \R^2$ are in counterclockwise order. We construct a closed convex integral polygon $N(\Gamma)$ (or just $N$ when $\Gamma$ is clear from the context) by placing the $[\beta_i]$ such that the head of $[\beta_i]$ is the tail of $[\beta_{i+1}]$. Each edge of $\Gamma$ is contained in two zig-zag paths that traverse the edge in opposite directions, so we have \smash{$\sum_{\beta \in \mathcal Z} [\beta]=0$}, which shows that $N$ constructed as above is a closed polygon. $N$ is unique up to translation and is called the Newton polygon of $\Gamma$. The name Newton polygon will be justified at the end of this section by the fact that this polygon arises as the Newton polygon of the characteristic polynomial of the dimer model on $\Gamma$.
	
A graph $\Gamma$ is said to be \emph{minimal} if any lift of a zig-zag path to the universal cover of $\T$ has no self intersections and any lifts of two zig-zag paths to the universal cover of $\T$ do not form parallel bigons (pairs of zig-zag paths oriented the same way intersecting twice). Hereafter, when considering a graph $\Gamma$ in $\T$, we assume that it is minimal unless stated otherwise. By construction, the set of primitive edge vectors of the Newton polygon of a minimal graph $\Gamma$ is in bijection with $\mathcal Z$, but this bijection is not canonical when there is more than one zig-zag path with a given homology class.

	\begin{figure}[ht]\centering
			\begin{tikzpicture}
				\begin{scope}
					\def\lw{1}
					\draw[-latex,blue,line width=\lw] (-1,0.05)--(0,0.05)--(30:1);
					\draw[-latex,red,line width=\lw] (-1,-0.05)--(0,-0.05)--(-30:1);
					\node[](no) at (0,0.3) {$v$};
					\node[](no) at (0,-1) {$\frac 1 2$};
				\end{scope}
				
				\begin{scope}[shift={(3,0)}]
					\def\lw{1}
					\draw[-latex,red,line width=\lw] (180-45:1)--(-45:1);
					\draw[-latex,blue,line width=\lw] (180+45:1)--(45:1);
					\node[](no) at (0,0.3) {$v$};
					\node[](no) at (0,-1) {$1 $};
				\end{scope}

				\begin{scope}[shift={(6,0)}]
					\def\lw{1}
					\draw[-latex,blue,line width=\lw] (180-45:1)--(0,0)--(45:1);
					\draw[-latex,red,line width=\lw] (180+45:1)--(0,0)--(-45:1);
					\node[](no) at (0,0.3) {$v$};
					\node[](no) at (0,-1) {$0 $};
				\end{scope}

				\begin{scope}[shift={(9,0)}]
					\def\lw{1}
					\draw[-latex,blue,line width=\lw] (-1,0)--(1,0);
					\draw[-latex,red,line width=\lw] (-1,-0.1)--(1,-0.1);
					\node[](no) at (0,0.3) {$v$};
					\node[](no) at (0,-1) {$0 $};
				\end{scope}
				
			\end{tikzpicture}
		\caption{Local rules for computing $\epsilon_\Gamma$. $L_1$ and $L_2$ are the blue and red cycles, respectively.}\label{fig:localpairing}

	\end{figure}

{\bf Conjugate surface and Poisson structure.}
	Thickening the edges of $\Gamma$, we obtain a~ribbon graph. Equivalently a ribbon graph is a graph along with the data of a cyclic order of edges around each vertex. The ribbon graph obtained from $\Gamma$ has the cyclic order induced from the embedding in $\T$. Let \smash{$\widehat \Gamma$} be the ribbon graph obtained from $\Gamma$ by reversing the cyclic order at all black vertices. The boundary components of \smash{$\widehat \Gamma$} are in bijection with the zig-zag paths of~$\Gamma$. Gluing in disks along these boundary components of \smash{$\widehat \Gamma$}, we obtain a surface $\widehat S$, called the \emph{conjugate surface}, along with an embedding of $\Gamma$ in $\widehat S$. Let \smash{$\epsilon_{\widehat S}$} denote the intersection form on~$H_1\big(\widehat S,\Z\big)$ defined as follows. If $L_1$ and $L_2$ are two cycles on $\widehat S$ intersecting transversely, then
	\[
		\epsilon_{\widehat S }([L_1],[L_2]):=\sum_{p \in L_1 \cap L_2} \epsilon_p(L_1,L_2),
	\]
	where $\epsilon_p(L_1,L_2)$ is the local intersection index, with sign chosen so that it is positive if $L_2$ crosses $L_1$ at $p$ from its right side to its left side. {Note that the definition of \smash{$\epsilon_{\widehat S }([L_1],[L_2])$} is~independent of the choice of cycles representing $[L_1]$ and $[L_2]$.} The embedding $\iota\colon \Gamma \hookrightarrow \widehat S$~induces the homomorphism of homology groups $\iota_*\colon H_1(\Gamma,\Z) \ra H_1(\widehat S,\Z)$. We define the alternating form $\epsilon_\Gamma$ on~$H_1(\Gamma,\Z)$ by $\epsilon_\Gamma([L_1],[L_2]):=\epsilon_{\widehat S}(\iota_*[L_1],\iota_*[L_2])$. The pairing $\epsilon_\Gamma$ has the following local description which is useful for computations (see \cite[Appendix]{GK13}): $\epsilon_\Gamma([L_1],[L_2])=\sum_{v \in B} \epsilon_v(L_1,L_2)-\sum_{v \in W} \epsilon_v(L_1,L_2)$, where $\epsilon_v(L_1,L_2)$ is defined in Figure~\ref{fig:localpairing}. In particular, if $f$ and $f'$ are two faces having a single edge in common and $f$ lies to the left of that edge when traversed from its black endpoint to its white endpoint, then $\epsilon_\Gamma([\partial f],[\partial f'])=1$.

	For ${[L_1]},{[L_2]} \in H_1(\Gamma,\Z)$, define the Poisson bracket
	\[
		\{\chi_{[L_1]},\chi_{[L_2]}\}_\Gamma:=\epsilon_\Gamma([L_1],[L_2])\chi_{[L_1]}\chi_{[L_2]}.
	\]
	By linearity and Leibniz's rule, we obtain a Poisson bracket on $\mathcal O_{\mathcal L_\Gamma}$. The faces of $\Gamma$ in $\widehat S$ become the zig-zag paths of $\Gamma$ in $\T$, so we have $\{\chi_{[L_1]},\chi_{[L_2]}\}=0$ for all $[L_2] \in H_1(\Gamma,\Z)$ if and only if~$[L_1] \in \bigoplus_{\beta \in \mathcal Z} \Z \cdot [\beta]$. Therefore, the center of the Poisson algebra $\mathcal O_{\mathcal L_\Gamma}$ is the subalgebra
	\[
	\C\bigl[C_\beta^{\pm 1}\bigr]/\biggl(1-\prod_{\beta \in \mathcal Z} C_\beta\biggr),
	\]
	generated by the functions $C_\beta:=\chi_{[\beta]}, \beta \in \mathcal Z$. Elements of the center of a Poisson algebra are called \emph{Casimirs}.
	
\subsection{Local and semi-local transformations}
	
There are two local modifications of bipartite graphs called \emph{elementary transformations}. An~elementary transformation $s\colon \Gamma \ra \Gamma'$ induces a unique up to isotopy homeomorphism of conjugate surfaces {$\hat s \colon \widehat S_\Gamma \ra \widehat S_{\Gamma'}$} \cite[Lemma 4.1]{GK13}, which in turn induces an isomorphism $\hat s_*\colon H_1(\Gamma,\Z) \ra H_1(\Gamma',\Z)$. For $[L'] \in H_1(\Gamma',\Z)$, let $[L]=(\hat s_*)^{-1}([L'])$. Associated to the elementary transformation $s$ is a Poisson birational map of weights $\mu_s\colon \mathcal L_\Gamma \ra \mathcal L_{\Gamma'}$:
	\begin{enumerate}\itemsep=0pt
		\item \emph{The spider move $s$ at face $f$}: We define \[\mu_{s}^*(\chi_{[L']})=\begin{cases}
			X_f^{-1} &\text{if $[L]=[\partial f]$},\\
			\chi_{[L]}\big(1+X_f^{-\text{sign }\epsilon_\Gamma([L],\partial f)}\big)^{-\epsilon_\Gamma([L],\partial f)} & \text{otherwise.}
		\end{cases}
		\]
		This is illustrated on the left side of Figure~\ref{fig:bigon}.
		\item \emph{Contracting/expanding degree two vertices}: We define $\mu_s^*(\chi_{[L']})=\chi_{[L]}$. This is illustrated in the middle of Figure~\ref{fig:bigon}.
	\end{enumerate}
	In other words, the spider move at $f$ inverts the face weight at $f$ and multiplies the face weights of a face $f'$ adjacent to $f$ by some power of $1+X_f$ or of \smash{$\big(1+X_f^{-1}\big)^{-1}$}. Such a transformation on face weights corresponds to the mutation rule for coefficient variables in cluster algebras \cite{FG, FZ} and indeed one can associate a cluster algebra to a dimer model on a torus \cite{GK13}. Contracting/expanding degree two vertices does not change the face weights.

	Elementary transformations do not change homology classes of zig-zag paths, and therefore the Newton polygon. Gluing the Poisson affine varieties $\mathcal L_\Gamma$ for all $\Gamma$ minimal with $N(\Gamma)=N$ using these Poisson birational maps, we obtain the Poisson space $\mathcal X_N$ called the \emph{dimer cluster Poisson variety} associated to $N$. $\mathcal X_N$ is a cluster Poisson variety as defined by Fock and Goncharov \cite{FG}, and will be the phase space of the cluster integrable system. {Each $\mathcal L_\Gamma$ such that $\Gamma$ is minimal with $N(\Gamma)=N$ is Zariski-dense inside $\mathcal X_N$.}
	
	\begin{figure}[ht]\centering
			\def\shf{10mm}
			\def\scl{0.6}
			\begin{tikzpicture}[scale=\scl,rotate=45,baseline={([yshift=-.7ex]current bounding box.center)},rotate=90]
				\coordinate (ws) at (-2,-1);
				\coordinate (wn) at (-2,1);
				\coordinate (es) at (2,-1);
				\coordinate (en) at (2,1);
				\coordinate (nw) at (-1,2);
				\coordinate (ne) at (1,2);
				\coordinate (sw) at (-1,-2);
				\coordinate (se) at (1,-2);
				\coordinate (n) at (0,1);
				\coordinate (s) at (0,-1);			
				\node[bvert] (n) at (n) {};
				\node[bvert] (s) at (s) {};
				\node[wvert] (nn) at ($(nw)!.5!(ne)$) {};
				\node[wvert] (ss) at ($(sw)!.5!(se)$) {};
				\node[wvert] (ww) at ($(wn)!.5!(ws)$) {};
				\node[wvert] (ee) at ($(en)!.5!(es)$) {};
				\draw[-]
				(n) edge (nn) edge (ww) edge (ee)
				(s) edge (ss) edge (ww) edge (ee)
				;			
			\end{tikzpicture}\hspace{3mm}$\longleftrightarrow$\hspace{1mm}
			\begin{tikzpicture}[scale=\scl,rotate=45,baseline={([yshift=-.7ex]current bounding box.center)}]
				\coordinate (ws) at (-2,-1);
				\coordinate (wn) at (-2,1);
				\coordinate (es) at (2,-1);
				\coordinate (en) at (2,1);
				\coordinate (nw) at (-1,2);
				\coordinate (ne) at (1,2);
				\coordinate (sw) at (-1,-2);
				\coordinate (se) at (1,-2);
				\coordinate (n) at (0,1);
				\coordinate (s) at (0,-1);		
				
				\node[bvert] (n) at (n) {};
				\node[bvert] (s) at (s) {};
				\node[wvert] (nn) at ($(nw)!.5!(ne)$) {};
				\node[wvert] (ss) at ($(sw)!.5!(se)$) {};
				\node[wvert] (ww) at ($(wn)!.5!(ws)$) {};
				\node[wvert] (ee) at ($(en)!.5!(es)$) {};
				\draw[-]
				(n) edge (nn) edge (ww) edge (ee)
				(s) edge (ss) edge (ww) edge (ee)
				;			
			\end{tikzpicture}\hspace{\shf}
			\begin{tikzpicture}[scale=2*\scl,baseline={([yshift=-1ex]current bounding box.center)}]
				\node[wvert] (v) at (0,0) {};
				\node (arr) at (0.4,0) {$\longleftrightarrow$};
				\node[wvert] (w1) at (1,-0.5) {};
				\node[bvert] (b) at (1,0) {};
				\node[wvert] (w2) at (1,0.5) {};
				\draw[-]
				(w1) -- (b) -- (w2)
				;
			\end{tikzpicture}\hspace{\shf}
			\begin{tikzpicture}[scale=2*\scl,baseline={([yshift=-1ex]current bounding box.center)}]
				\node[wvert] (v1) at (30:1) {};
				\node[bvert, label=above:$\mathrm b$] (v2) at (90:1) {};
				\node[wvert] (v3) at (150:1) {};
				\node[bvert] (v4) at (210:1) {};
				\node[wvert, label=below:$\mathrm w$] (v5) at (270:1) {};
				\node[bvert] (v6) at (330:1) {};
				\node[blue] (f) at (0,0) {$f$};
				\draw[-]
				(v1) -- (v2) -- (v3) -- (v4) -- (v5) -- (v6) -- (v1)
				;
			\end{tikzpicture}\hspace{6mm}$\longleftrightarrow$\hspace{5mm}
			\begin{tikzpicture}[scale=2*\scl,baseline={([yshift=-1ex]current bounding box.center)}]
				\node[wvert] (v1) at (30:1) {};
				\node[bvert, label=above:$\mathrm b$] (v2) at (90:1) {};
				\node[wvert] (v3) at (150:1) {};
				\node[bvert] (v4) at (210:1) {};
				\node[wvert, label=below:$\mathrm w$] (v5) at (270:1) {};
				\node[bvert] (v6) at (330:1) {};
				\node[blue] (f) at (0,0) {$f_b$};
				\node[blue] (fr) at (0.5,0) {$f_r$};
				\node[blue] (fl) at (-0.5,0) {$f_l$};
				\draw[-]
				(v1) -- (v2) -- (v3) -- (v4) -- (v5) -- (v6) -- (v1)
				(v2) edge [bend left=25] (v5) edge [bend right=25] (v5)
				;
			\end{tikzpicture}	
		\caption{On the left, the spider move. In the middle, contracting/expanding a degree two vertex. On the right, adding/removing a bigon inside a face.}
		\label{fig:bigon}
	\end{figure}

	{\bf Inserting/removing a bigon.}
	The right side of Figure~\ref{fig:bigon} shows the insertion of a bigon between vertices $\w \in W$ and $\bw \in B$ belonging to a common face $f$, with parameter $u$. This divides $f$ into three new faces, the bigon $f_b$ and the face $f_l$ (resp.\ $f_r$) to the left (resp.\ right) of~$f_b$ when traversing the bigon from $\rm w$ to $\rm b$. Let $\Gamma_b$ denote the graph obtained. The embedding $i_b\colon \Gamma \hookrightarrow \Gamma_b$ induces a homomorphism $(i_b)_*\colon H_1(\Gamma,\Z) \ra H_1(\Gamma_b,\Z)$. We define the induced map of weights $\mu_u\colon \mathcal L_\Gamma \ra \mathcal L_{\Gamma_b}$ on a basis as follows: If $[L]$ is topologically nontrivial in $H_1(\T,\Z)$ or is the boundary of a face of $\Gamma$ set $\mu_u^*(\chi_{(i_b)_*[L]})=\chi_{[L]}$. Define also $\mu_u^*(X_{f_l})=u$ and $\mu_u^*(X_{f_b})=-1$. Note that the second equation implies that in any cocycle, the weights of the two edges of the bigon sum to zero. On the other hand, if we have a bigon with $X_{f_b}=-1$, we may remove~it. This induces a map of weights $\mu_b'\colon \{X_{f_b}=-1\} \ra \mathcal L_\Gamma$ given by $\big(\mu_b'\big)^*(\chi_{[L]})=\chi_{(i_b)_*[L]}$, where $\{X_{f_b}=-1\}$ denotes the subvariety in $ \mathcal L_{\Gamma_b}$. In other words, the insertion of a bigon with parameter $u$ inside a face $f$ assigns to the faces $f_l$, $f_b$ and $f_r$ the respective weights $u$, $-1$ and~\smash{$-\tfrac{X_f}{u}$}, while the deletion of a bigon $f_b$ with face weight $-1$ assigns to the resulting face the product of the weights of the three faces that got merged.
	
	\begin{figure}[ht]
		\centering
		\begin{tikzpicture}[scale=0.9,yscale=1.8,baseline={(current bounding box.center)}]
			\begin{scope}
				\node[wvert] (w1) at (1,0.87) {};
				\node[wvert] (w2) at (3,0.87) {};
				\node[wvert] (w3) at (5,0.87) {};
				\node[wvert] (w4) at (7,0.87) {};
				\node[wvert] (w5) at (2,0) {};
				\node[wvert] (w6) at (4,0) {};
				\node[wvert] (w7) at (6,0) {};
				\node[bvert] (b1) at (2,1.24) {};
				\node[bvert] (b2) at (4,1.24) {};
				\node[bvert] (bb3) at (6,1.24) {};
				\node[bvert] (b4) at (1,0.37) {};
				\node[bvert] (b5) at (3,0.37) {};
				\node[bvert] (b6) at (5,0.37) {};
				\node[bvert] (b7) at (7,0.37) {};
				\draw[-] (w1) -- (b1) -- (w2) -- (b2) -- (w3) -- (bb3)--(w4);
				\draw[-] (b4) -- (w5) -- (b5) -- (w6) -- (b6)--(w7) -- (b7);
				\draw[-] (w1) -- (b4)
				(w2) -- (b5)
				(w3) --(b6)
				(w4) -- (b7);
				\node[blue] (no) at (4,0.62) {$a_2$};
				\node[blue] (no) at (2,0.62) {$a_1$};
				\node[blue] (no) at (6,0.62) {$a_3$};
				\draw[->] (4,-0.2) -- (4,-0.55);
				\draw[->,dashed] (8,0.62) --node[above] {$\Phi$} (9.03,0.62);
			\end{scope}
			
			\begin{scope}[shift={(0,-2)}]
				\node[wvert] (w1) at (1,0.87) {};
				\node[wvert] (w2) at (3,0.87) {};
				\node[wvert] (w3) at (5,0.87) {};
				\node[wvert] (w4) at (7,0.87) {};
				\node[wvert] (w5) at (2,0) {};
				\node[wvert] (w6) at (4,0) {};
				\node[wvert] (w7) at (6,0) {};
				\node[bvert] (b1) at (2,1.24) {};
				\node[bvert] (b2) at (4,1.24) {};
				\node[bvert] (bb3) at (6,1.24) {};
				\node[bvert] (b4) at (1,0.37) {};
				\node[bvert] (b5) at (3,0.37) {};
				\node[bvert] (b6) at (5,0.37) {};
				\node[bvert] (b7) at (7,0.37) {};
				\draw[-] (w1) -- (b1) -- (w2) -- (b2) -- (w3) -- (bb3)--(w4);
				\draw[-] (b4) -- (w5) -- (b5) -- (w6) -- (b6)--(w7) -- (b7);
				\draw[-] (w1) -- (b4)
				(w2) -- (b5)
				(w3) --(b6)
				(w4) -- (b7);
				\draw[-]
				(b2) edge [bend left=60] (w6) edge [bend right=60] (w6);
				\node[blue] (no) at (4,0.62) {$-1$};
				\node[blue] (no) at (3.35,0.62) {$f_l$};
				\node[blue] (no) at (4.65,0.62) {$f_r$};
				\node[red] (no) at (4.6,0.85) {$\times$};
				\draw[->] (4,-0.2) -- (4,-0.55);
			\end{scope}
			
			\begin{scope}[shift={(0,-4)}]
				\node[wvert] (w1) at (1,0.87) {};
				\node[wvert] (w2) at (3,0.87) {};
				\node[wvert] (w3) at (5,0.87) {};
				\node[wvert] (w4) at (7,0.87) {};
				\node[wvert] (w5) at (2,0) {};
				\node[wvert] (w6) at (4,0) {};
				\node[wvert] (w7) at (6,0) {};
				\node[bvert] (b1) at (2,1.24) {};
				\node[bvert] (b2) at (4,1.24) {};
				\node[bvert] (bb3) at (6,1.24) {};
				\node[bvert] (b4) at (1,0.37) {};
				\node[bvert] (b5) at (3,0.37) {};
				\node[bvert] (b6) at (5,0.37) {};
				\node[bvert] (b7) at (7,0.37) {};
				\draw[-] (w1) -- (b1) -- (w2) -- (b2) -- (w3) -- (bb3)--(w4);
				\draw[-] (b4) -- (w5) -- (b5) -- (w6) -- (b6)--(w7) -- (b7);
				\draw[-] (w1) -- (b4)
				(w2) -- (b5)
				(w3) --(b6)
				(w4) -- (b7);
				\draw[-]
				(b2) -- (w6)
				(bb3) -- (w7);
				\node[red] (no) at (6.5,0.62) {$\times$};
				\draw[->] (8,0.62) -- (9.03,0.62);
			\end{scope}
			
			\begin{scope}[shift={(9,0)}]
				\node[wvert] (w1) at (1,0.87) {};
				\node[wvert] (w2) at (3,0.87) {};
				\node[wvert] (w3) at (5,0.87) {};
				\node[wvert] (w4) at (7,0.87) {};
				\node[wvert] (w5) at (2,0) {};
				\node[wvert] (w6) at (4,0) {};
				\node[wvert] (w7) at (6,0) {};
				\node[bvert] (b1) at (2,1.24) {};
				\node[bvert] (b2) at (4,1.24) {};
				\node[bvert] (bb3) at (6,1.24) {};
				\node[bvert] (b4) at (1,0.37) {};
				\node[bvert] (b5) at (3,0.37) {};
				\node[bvert] (b6) at (5,0.37) {};
				\node[bvert] (b7) at (7,0.37) {};
				\draw[-] (w1) -- (b1) -- (w2) -- (b2) -- (w3) -- (bb3)--(w4);
				\draw[-] (b4) -- (w5) -- (b5) -- (w6) -- (b6)--(w7) -- (b7);
				\draw[-] (w1) -- (b4)
				(w2) -- (b5)
				(w3) --(b6)
				(w4) -- (b7);
				\node[blue] (no) at (4,0.62) {$a_2'$};
				\node[blue] (no) at (2,0.62) {$a_1'$};
				\node[blue] (no) at (6,0.62) {$a_3'$};
				\draw[<-] (4,-0.2) -- (4,-0.55);
			\end{scope}
			
			\begin{scope}[shift={(9,-2)}]
				\node[wvert] (w1) at (1,0.87) {};
				\node[wvert] (w2) at (3,0.87) {};
				\node[wvert] (w3) at (5,0.87) {};
				\node[wvert] (w4) at (7,0.87) {};
				\node[wvert] (w5) at (2,0) {};
				\node[wvert] (w6) at (4,0) {};
				\node[wvert] (w7) at (6,0) {};
				\node[bvert] (b1) at (2,1.24) {};
				\node[bvert] (b2) at (4,1.24) {};
				\node[bvert] (bb3) at (6,1.24) {};
				\node[bvert] (b4) at (1,0.37) {};
				\node[bvert] (b5) at (3,0.37) {};
				\node[bvert] (b6) at (5,0.37) {};
				\node[bvert] (b7) at (7,0.37) {};
				\draw[-] (w1) -- (b1) -- (w2) -- (b2) -- (w3) -- (bb3)--(w4);
				\draw[-] (b4) -- (w5) -- (b5) -- (w6) -- (b6)--(w7) -- (b7);
				\draw[-] (w1) -- (b4)
				(w2) -- (b5)
				(w3) --(b6)
				(w4) -- (b7);
				\draw[-]
				(b2) edge [bend left=60] (w6) edge [bend right=60] (w6);
				\node[blue] (no) at (4,0.62) {$-1$};
				
				\draw[<-] (4,-0.2) -- (4,-0.55);
			\end{scope}
			
			\begin{scope}[shift={(9,-4)}]
				\node[wvert] (w1) at (1,0.87) {};
				\node[wvert] (w2) at (3,0.87) {};
				\node[wvert] (w3) at (5,0.87) {};
				\node[wvert] (w4) at (7,0.87) {};
				\node[wvert] (w5) at (2,0) {};
				\node[wvert] (w6) at (4,0) {};
				\node[wvert] (w7) at (6,0) {};
				\node[bvert] (b1) at (2,1.24) {};
				\node[bvert] (b2) at (4,1.24) {};
				\node[bvert] (bb3) at (6,1.24) {};
				\node[bvert] (b4) at (1,0.37) {};
				\node[bvert] (b5) at (3,0.37) {};
				\node[bvert] (b6) at (5,0.37) {};
				\node[bvert] (b7) at (7,0.37) {};
				\draw[-] (w1) -- (b1) -- (w2) -- (b2) -- (w3) -- (bb3)--(w4);
				\draw[-] (b4) -- (w5) -- (b5) -- (w6) -- (b6)--(w7) -- (b7);
				\draw[-] (w1) -- (b4)
				(w2) -- (b5)
				(w3) --(b6)
				(w4) -- (b7);
				\draw[-]
				(b2) -- (w6)
				(b1) -- (w5);
				\node[red] (no) at (2.5,0.62) {$\times$};
				
			\end{scope}

		\end{tikzpicture}
		\caption{The sequence of operations corresponding to the geometric $R$-matrix transformation when $n=3$. For each of the six pictures, the left and right sides are identified. Here we have $k=2$, so that we add the bigon in the second hexagon. The red crosses indicate the faces at which we perform the next spider move. The face weights are depicted in blue.}
		\label{fig:Rmatrixsequence}
	\end{figure}
	
	{\bf Geometric $\boldsymbol{R}$-matrices.}
	We now recall the dimer interpretation of geometric $R$-matrix transformations given in \cite[Section 11]{ILP1} as a composition of bigon insertion/removal and spider moves. We call this a semi-local move, because the choice of the parameter $u$ associated with the bigon insertion is a function of weights of faces that may be arbitrarily far away from the bigon. Consider a bipartite graph embedded on a surface, which possesses a cyclic chain of $n$ hexagons (see the first picture of Figure~\ref{fig:Rmatrixsequence} for an example with $n=3$). The two edges of each hexagon which separate it from the neighboring hexagons must form an opposite pair of edges. Denote by $a_1,\dots,a_n$ the face weights of the hexagons. Fix $k$ between $1$ and $n$ and add a bigon between the two vertices of the $k$th hexagon that are not part of the neighboring two hexagons. We impose the weight of the bigon to be $-1$ and we denote by $f_r$ and \smash{$f_l=-\tfrac{a_k}{f_r}$} the weight of the two newly created quadrilaterals, as on the right picture of Figure~\ref{fig:bigon}. For now $f_r$ is an unknown. We then perform a sequence of $n$ spider moves, starting at the face of weight $f_r$ and moving in the direction of increasing values of $k$. At the end of this sequence of spider moves, we come back to a situation where the $k$th hexagon has a bigon. It follows from~\cite{ILP1} that the closing condition for the weight of this bigon to be $-1$ is linear in $f_r$. We set $f_r$ to be the unique solution of this equation. We finally delete the bigon, obtaining again a cyclic chain of $n$ hexagons, the weights of which we denote by $a'_1,\dots,a'_n$. Formulas (11.1) and (11.2) of \cite{ILP1} lead to the following result.
	
	\begin{Theorem}[\cite{ILP1}]
		\label{thm:ILP}
		With the setting defined above, the values $a'_1,\dots,a'_n$ are independent of the choice of the starting position $k$ and are given for every $1\leq i \leq n$ by
		\begin{equation}
			\label{eq:Rmatrix}
			a'_i=\frac{\sum\limits_{t=0}^{n-1}\prod\limits_{s=0}^{t-1}a_{i+s}}{\sum\limits_{t=1}^{n}\prod\limits_{s=1}^{t}a_{i+s}},
		\end{equation}
		where indices are taken modulo $n$.
	\end{Theorem}

	We point out that the formula we stated above slightly differs from the one obtained from \cite{ILP1} in that our indices are increasing while theirs are decreasing ($i+s$ instead of $i-s$). This comes from having the opposite convention for mutation rules, which results from different convention in the definitions of zig-zag paths and of dimer face weights. We also note that the framework of \cite{ILP1} was a bit more restrictive than the one we are considering, since they were assuming that on each side of the cyclic chain of hexagons there were other chains of hexagons. Here we are only assuming that no face immediately above or below the cyclic chain of $n$ hexagons may be one of these $n$ hexagons. The proof of \cite{ILP1} holds verbatim in this framework. Finally, we point out that the map sending $(a_1,\dots,a_n)$ to $\big(a'_1,\dots,a'_n\big)$ is an involution \cite{ILP1}. We denote this map by $\Phi$.
	
	\subsection{Dimer covers and Kasteleyn theory}
	
	\begin{figure}[ht]
		\centering
		\begin{tikzpicture}[scale=1.5]
			\def\nnp{3};
			\def\nn{2};
			\def\nnm{1};
			\draw[dashed, gray] (1,0) rectangle (\nnp,4);
			\draw[->,densely dotted] (2.8,0) -- (2.8,4);
			\draw[->,densely dotted] (1,3.75) -- (3,3.75);
			\node (no) at (1-0.25,3.75) {$\gamma_z$};
			\node (no) at (2.75,-0.25) {$\gamma_w$};

			\foreach[evaluate={\yy=int(\xx+1)}] \xx in {1} {
				\pgfmathtruncatemacro{\label}{\nnm-\xx+3}
				\pgfmathtruncatemacro{\uu}{\nnm-\xx+2}
				\coordinate[wvert,label=left:${\rm w}_{2}$] (p\xx) at (\xx,3);
				\coordinate[wvert,label=left:${\rm w}_{1}$] (q\xx) at (\xx,1);		
				\coordinate[bvert,label={[label distance=-0.5mm]right:${\rm b}_{2}$}] (b\xx) at (\xx+0.6,3);		
				\coordinate[bvert,label={[label distance=-0.5mm]right:${\rm b}_{1}$}] (c\xx) at (\xx+0.6,1);	
			}

			\foreach[evaluate={\yy=int(\xx+1)}] \xx in {3} {
				\pgfmathtruncatemacro{\label}{\nnm-\xx+3}
				\pgfmathtruncatemacro{\uu}{\nnm-\xx+2}
				\coordinate[wvert] (p\xx) at (\xx,3);
				\coordinate[wvert] (q\xx) at (\xx,1);		
			}

			\foreach[evaluate={\yy=int(\xx+1)}] \xx in {2} {
				\pgfmathtruncatemacro{\label}{\nnm-\xx+3}
				\pgfmathtruncatemacro{\uu}{\nnm-\xx+2}
				\coordinate[wvert,label=above:${\rm w}_{3}$] (p\xx) at (\xx,4);
				\coordinate[wvert,label=below:${\rm w}_{3}$] (pp\xx) at (\xx,0);
				\coordinate[wvert,label=right:${\rm w}_{4}$] (q\xx) at (\xx,2);
			}
			
			\coordinate[wvert,label=right:${\rm w}_{2}$] (p11) at (\nn+1,3);
			\coordinate[wvert,label=right:${\rm w}_{1}$] (q11) at (\nn+1,1);
			\foreach[evaluate={\yy=int(\xx-1)}] \xx in {3} {
				\coordinate[bvert,label={[label distance=-0.5mm]left:${\rm b}_{4}$}] (b\xx) at (\xx-0.6,3);		
				\coordinate[bvert,label={[label distance=-0.5mm]left:${\rm b}_{3}$}] (c\xx) at (\xx-0.6,1);

				\draw[-,text=red]
				(b\xx) edge node[below] {$z$} (p\xx) edge node[right] {$a w$} (p\yy) edge node[right] {$d$} (q\yy)
				(c\xx) edge node[below] {$z$} (q\xx) edge node[right] {$h$} (pp\yy) edge node[right] {$e$} (q\yy)
				;
				\draw[line width=1.8pt] (b3) -- (p11);
				\draw[line width=1.8pt] (c3) -- (q11);
				\draw[line width=1.8pt] (b1) -- (p2);
				\draw[line width=1.8pt] (c1) -- (q2);		
			}
			
			\foreach[evaluate={\yy=int(\xx+1)}] \xx in {1} {
				\draw[-,text=red]
				(b\xx) edge node[below] {\ifthenelse{\xx>1}{$1$}{$1$}} (p\xx) edge node[left] {$- b w$} (p\yy) edge node[left] {$c$} (q\yy)
				(c\xx) edge node[below] {\ifthenelse{\xx>1}{$1$}{$1$}} (q\xx) edge node[left] {$-f$} (q\yy) edge node[left] {$g$} (pp\yy)
				;
			}			
			
		\end{tikzpicture}
		\caption{Edge weight, Kasteleyn sign, cochain $\phi$ and $M_0$ (thick edges) on $\Gamma_2$.}
		\label{fig:giew}
	\end{figure}

	A \emph{dimer cover} $M$ of $\Gamma$ is a subset of $E$ such that each vertex of $\Gamma$ is incident to exactly one edge in $M$. Let $\mathcal M$ denote the set of dimer covers of $\Gamma$. If we fix a reference dimer cover $M_0$, then we can associate to each dimer cover a homology class
	\[
	M \mapsto [M-M_0] \in H_1(\T,\Z),
	\]
	where, as before, we orient $e=\bw \w$ from ${\rm b}$ to ${\rm w}$. Given $[\rm wt] \in \mathcal L_\Gamma$, each dimer cover also gets a~\emph{weight} $[{\rm wt}](M-M_0)$. If $\Gamma$ is minimal, we can describe the Newton polygon in terms of dimer covers.
	
	\begin{Proposition}
		[\protect{\cite[Theorem 3.12]{GK13}}]
		For a minimal bipartite graph $\Gamma$ in $\T$, we have
		\[
		N(\Gamma)=\operatorname{Convex-hull}\{[M-M_0] \mid M \text{ is a dimer cover of }\Gamma\},
		\]
		up to a translation.
	\end{Proposition}
	Let $R$ be a fundamental rectangle of $\T$ and let $\gamma_z$, $\gamma_w$ be cycles in $\T$ such that~$[\gamma_z]$,~$[\gamma_w]$ generate $H_1(\T,\Z)$. We choose $\gamma_z$, $\gamma_w$ parallel to the sides of $R$ as shown in Figure~\ref{fig:giew}. Isotoping if necessary, we assume that the edges of $\Gamma$ intersect $\gamma_z$, $\gamma_w$ transversely. Applying $\Hom _\Z(\cdot,\C^\times)$ to the surjection $H_1(\Gamma,\Z) \ra H_1(\T,\Z)$, we get an inclusion $H^1(\T,\C^\times) \hookrightarrow H^1(\Gamma,\C^\times)$. Let~${[\phi'] \in H^1(\Gamma,\C^\times)}$ be in the image of $H^1(\T,\C^\times)$. In other words, $X_f([\phi'])=1$ for all $f \in F$. We choose a cochain~$\phi$ representing $[\phi']$ as follows: Let $z:=[\phi']([\gamma_z])$, $w=[\phi']([\gamma_w])$, and define
	\[
	\phi(e):=z^{(e,\gamma_w)} w^{(e,-\gamma_z)},
	\]
	where $(\cdot,\cdot)$ is the intersection index, i.e., the sum of the local intersection indices defined in the previous subsection.
	
	The cohomology class $[\kappa] \in H^1(\Gamma,\C^\times)$ is called a \emph{Kasteleyn sign} if the following conditions hold:
	\begin{enumerate}\itemsep=0pt
		\item[(1)] $X_{[L]}([\kappa])=\pm 1$ for all $[L] \in H_1(\Gamma,\Z)$;
		\item[(2)] \smash{$X_f([\kappa])=(-1)^{\frac{|\partial f|}{2}+1}$} for all $f \in F$, where $|\partial f|$ is the number of edges in $\partial f$.
	\end{enumerate}
	Let $\kappa\colon E \ra \C^\times$ be a cochain representing the Kasteleyn sign $[\kappa]$. We define the \emph{Kasteleyn matrix}
	\[
	K(z,w)\colon \ \C\bigl[z^{\pm 1},w^{\pm 1}\bigr]^B \ra \C\bigl[z^{\pm 1},w^{\pm 1}\bigr]^W
	\]
	by
	\[
	K{(z,w)}_{{\rm w},{\rm b}}:=\sum_{e=\bw \w\in E} {\rm wt}(e) \kappa(e) \phi(e),
	\]
	where the sum is over edges between $\bw$ and $\w$.
	\begin{Theorem}[\cite{Kas}]
		We have
		\[
		\frac{1}{{\rm wt}(M_0) \kappa(M_0) \phi(M_0) } \det K(z,w)= \sum_{M \in \mathcal M} \operatorname{sign}([M-M_0]) [{\rm wt}]([M-M_0]) [\phi]([M-M_0]),
		\]
		where $\operatorname{sign}([M-M_0])$ is a sign that depends on $[\kappa]$ and on the homology class $[M-M_0]$ in~$H_1(\T,\Z)$ and that is irrelevant for our purposes.
	\end{Theorem}
	Moreover,
	\[
	P(z,w):=\frac{1}{{\rm wt}(M_0)\kappa(M_0) \phi(M_0) } \det K(z,w)
	\]
	is called the \emph{characteristic polynomial} and $\Sigma:=\bigl\{(z,w) \in (\C^\times)^2 \mid P(z,w)=0\bigr\}$ is called the \emph{spectral curve} of $(\Gamma,[\wt])$. Although $K(z,w)$ depends on the choice of cochains representing~$[{\rm wt}]$ and $[\kappa]$, the spectral curve is independent of these choices. Moreover $N(\Gamma)$ is the Newton polygon of $P(z,w)$, that is, the convex hull of the pairs $(i,j)\in\Z^2$ such that $z^iw^j$ has a nonzero coefficient in $P(z,w)$.
	
	\begin{Example}
		Consider the graph $\Gamma_2$ with edge weights, Kasteleyn sign, $\phi$ and reference dimer cover $M_0$ chosen as in Figure~\ref{fig:giew}. The Kasteleyn matrix is
		\[
		K(z,w)=\begin{blockarray}{ccccc}
			{\rm b}_1 & {\rm b}_2 & {\rm b}_3 & {\rm b}_4 \\
			\begin{block}{[cccc]c}
				1 & 0 & z & 0 & {\rm w}_1\\
				0 & 1 & 0 & z & {\rm w}_2\\
				g & -bw & h & aw & \w_3\\
				-f & c & e & d & \w_4\\
			\end{block}
		\end{blockarray},
		\]
		and the spectral curve is
		\be \la{eq:scg2}
		P(z,w)=\frac{1}{b f z^2 w }
		\big(-dh + (dg+ch)z -cg z^2+a e w + (b e+a f) zw + bf z^2 w\big).
		\ee
		
	\end{Example}

	{\bf Hamiltonians.}
	Let $N^\circ$ denote the interior of $N$. For $[\gamma] \in N^\circ \cap H_1(\T,\Z)$, let
	\[
	H_{[\gamma]}:=\sum_{M \in \mathcal M: [M-M_0]=[\gamma]} {[\rm wt]}([M-M_0])
	\]
	denote the coefficient of $[\phi]([\gamma])$ (up to a sign) in $P(z,w)$.
	
	A space equipped with a Poisson bracket is a \emph{Liouville integrable system} if the generic level sets of the Casimirs are symplectic leaves of some dimension $2m$, which possess $m$ mutually Poisson-commuting \emph{Hamiltonians} which are functionally independent.
	
	\begin{Proposition}[\protect{\cite[Theorem 1.2]{GK13}}]
		The generic level sets of the Casimirs are symplectic leaves of $\mathcal X_N$. The quantities $H_{[\gamma]}$ for $[\gamma] \in H_1(\T,\Z) \cap N^\circ$ mutually Poisson-commute, making these symplectic leaves into Liouville integrable systems with Hamiltonians $H_{[\gamma]}$.
	\end{Proposition}

	\section{The dimer model in a cylinder}
	\la{sec:cyl}
	
	In this section, we consider balanced cylinder graphs, which are bipartite graphs on a cylinder satisfying certain conditions. In Section~\ref{subsec:Kasteleyncylinder}, we construct a matrix $\Pi(w)$ from a dimer model on a balanced cylinder graph. Then in Section~\ref{subsec:spectralcurve}, we prove Theorem~\ref{thm:spectralbm} relating the spectrum of $\Pi(w)$ to the spectral curve of the dimer model on the torus graph obtained by gluing the two boundaries of the balanced cylinder graph. Finally, in Section~\ref{subsec:t2c}, we show that this result holds for a large class of torus graphs, namely minimal graphs.
	
	Let $\Gamma_{\A}=(B \cup W,E)$ be a bipartite graph embedded in a cylinder $\A$ satisfying the following conditions:
	\begin{enumerate}\itemsep=0pt
\item Every vertex on the boundary of $\A$ is white and of degree $1$.
		\item Let $S$ and $T$ denote the boundary white vertices on the two components of the boundary of $\A$, called the \emph{source} and \emph{target} vertices, respectively. Let $W_{\text{int}}=W \setminus(S \cup T)$ denote the set of internal white vertices. We assume that $|S|=|T|$ and $|B|=|W|-|S|$.
		\item $\Gamma_\A$ has a dimer cover $M_0$ that uses all the vertices in $S$ and none of the vertices in $T$.
		\item The faces of $\Gamma_\A$ (including boundary faces) are topological disks.
	\end{enumerate}
	Here by a dimer cover of $\Gamma_\A$, we mean a matching that uses all the vertices in $B$ and in \mbox{a~$|B|$-element} subset of $W$ exactly once. Note that assumptions 1 and 3 imply that the black vertices incident to the white vertices in $S$ are all different.

	We call graphs $\Gamma_\A$ satisfying these conditions \emph{{\bcg }s}.

	An edge weight on $\Gamma_\A$ is a function $\wt\colon E \ra \C^\times$. Two edge weights ${\rm wt}_1$ and ${\rm wt}_2$ are gauge equivalent if there is a function $g\colon B \cup W \ra \C^\times$ satisfying $g(\w)=1$ for all $\w \in S \cup T$ such that for every edge $e=\bw \w$ with ${\rm b} \in B$, ${\rm w}\in W$, we have ${\rm wt}_2(e)=g({\rm b})^{-1} {\rm wt}_1(e) g({\rm w}) $. In other words we only allow gauge transformations at interior vertices. The space of edge weights modulo gauge transformations is the relative cohomology group $H^1(\Gamma_\A, S \cup T,\C^\times)$. {This relative cohomology group is generated by functions of cycles in $\Gamma_\A$ and of paths in $\Gamma_\A$ starting and ending at $S\cup T$.}
	As before, we denote by $[{\rm wt}]$ the cohomology class represented by ${\rm wt}$.
	
	Let $\mathcal M$ denote the set of dimer covers of $\Gamma_\A$. For $M \in \mathcal M$, we define its {weight} to be $\wt(M)=\prod_{e \in M} \wt(e)$. For $M \in \mathcal M$, let $\partial M$ denote the set of boundary white vertices incident to $M$. For example, $\partial M_0=S$.
	
\subsection[Kasteleyn theory in A]{Kasteleyn theory in $\boldsymbol{\A}$}
	\label{subsec:Kasteleyncylinder}
	
Suppose $\Gamma_\A$ is a \bcg. Let $\gamma_z$ be a simple path connecting the two boundaries of the cylinder and directed from the boundary containing $T$ towards the boundary containing~$S$. Write $h=|S|=|T|$. Denote by $\w_1,\dots,\w_h$ the vertices of $T$ labelled consecutively and by $\w'_1,\dots,\w'_h$ the vertices of $S$ labelled consecutively. The orientations of the boundaries induced by these labelings are prescribed to be compatible with the orientation of the cylinder. We also prescribe that $\gamma_z$ starts between $\w_h$ and $\w_1$ and ends between $\w'_h$ and $\w'_1$. For any face $f\in F$, denote by $T_f$ (resp.\ $S_f$) the subset of $i\in \{1,\dots,h\}$ such that the boundary segment $\w_i\w_{i+1}$ \big(resp.\ $\w'_i \w'_{i+1}$\big) is adjacent to $f$.
	
	An element $[\kappa] \in H^1(\Gamma_\A,S \cup T,\C^\times)$ is called a Kasteleyn sign if the following conditions hold:
	\begin{enumerate}\itemsep=0pt
		\item[(1)] $\chi_{[L]}([\kappa])=\pm 1$ for all $[L] \in H_1(\Gamma_\A,S \cup T,\Z)$;
		\item[(2)] there exists $(\sigma_1,\dots,\sigma_h)\in\{0,1\}^h$ such that for every $f\in F$,
		\[
			X_f([\kappa])=(-1)^{\frac{|\partial f|}{2}+1+\sum\limits_{i\in T_f} \sigma_i + \sum\limits_{i\in S_f} 1-\sigma_i}.
		\]
	\end{enumerate}
	As an example of the second condition, if all the $\sigma_i$ are zero and $|S_f|\leq1$ for every $f\in F$ (see, for example, Figure~\ref{fig:giew}), then \smash{$X_f([\kappa])=(-1)^{\frac{|\partial f|}{2}}$} if $f$ is adjacent to the boundary containing~$S$, otherwise \smash{$X_f([\kappa])=(-1)^{\frac{|\partial f|}{2}+1}$}.
	The existence of a Kasteleyn sign is shown in \cite[Proposition~2.1]{CR2}. Note that this definition of Kasteleyn signs for balanced cylinder graphs makes them compatible with concatenation or with gluing the two boundaries to obtain a torus graph. {The signs are admittedly complicated but they can mostly be ignored for the purposes of this paper.}
	
	We define the \emph{Kasteleyn matrix} of $\Gamma_\A$:
	\begin{align*}
		K_\A(w)\colon \ \C[w^{\pm 1}]^B \ra \C[w^{\pm 1}]^W, \qquad
		K_\A(w)_{{\rm w},{\rm b}} :=\sum_{e=\bw \w\in E} {\rm wt}(e) \kappa(e) w^{(e,-\gamma_z)}.
	\end{align*}
	
	We have the following version of Kasteleyn's theorem.
	\begin{Theorem}[\protect{\cite[Theorem 2.4]{CR2}}]\la{thm:cimres}
		Let $I \subset S \cup T$ such that $|I|=|S|$, and let $K_{\A,I}(w)$ denote the submatrix of the Kasteleyn matrix with rows indexed by white vertices in $I \cup W_{\text{int}}$ and columns indexed by black vertices in $B$. Then we have
		\begin{gather*}
		\frac{1}{\wt(M_0) \kappa(M_0)w^{(M_0,-\gamma_z)}}\det K_{\A,I}(w)=\!\sum_{\partial M=I} \operatorname{sign}([M-M_0]) [\wt]([M-M_0]) w^{([M-M_0],-\gamma_z)},
		\end{gather*}
		where $[M-M_0]$ is the relative homology class in $H_1(\A, \partial \A,\Z)$ defined by the relative cycle $M-M_0$ and $\operatorname{sign}([M-M_0])$ is a sign that depends on $[\kappa]$ and on the relative homology class $[M-M_0]$ and that is irrelevant for our purposes.
	\end{Theorem}

	Let $B_S$ denote the set of black vertices incident to $S$. We denote by $\bw_i$ the black vertex in~$B_S$ that is connected to the white vertex $\w'_i$ in $S$. $B_S$ is matched to $S$ by $M_0$. Up to performing gauge transformations at $B_S$ to ensure that the edges connecting $B_S$ to $S$ have $\kappa=1$, the~Kasteleyn matrix $K_\A(w)$ of $\Gamma_\A$ has the block matrix form
	\be
	K_\A(w)=\begin{blockarray}{ccc}
		B_S & B \setminus B_S \\
		\begin{block}{[cc]c}
			I & 0 & S\\
			K_1 & K_2 & T\\
			K_3 & K_4 & W_{\text{int}}\\
		\end{block}
	\end{blockarray}. \la{k4}
	\ee

	For generic $[{\rm wt}]$ and for generic $w \in \C^\times$, the submatrix $K_4$ is invertible using Theorem~\ref{thm:cimres} with $I=S$, since $M_0$ is a dimer cover with $\partial M_0=S$ that will appear as a summand in $\det K_4=\det K_{\A,S}(w)$. {We will now resort to the notion of Schur complement and we refer the reader to Appendix~\ref{sec:schur} for some background on this.} Define the Schur complement
 \begin{align}
		L:=K_\A(w)/K_4 =\begin{bmatrix}I\\K_1 \end{bmatrix}
		-\begin{bmatrix}0\\K_2 \end{bmatrix} K_4^{-1} K_3
		= \begin{bmatrix}
			I \\ \Pi(w)
		\end{bmatrix}, \la{eq:pi}
	\end{align}
	where $\Pi(w):= K_1-K_2 K_4^{-1} K_3$. To get an explicit formula for the entries of $\Pi(w)$, notice that for~$\w'_i \in S$, $\w_j \in T$, the square submatrix $L_{S \setminus \{\w'_i\} \cup \{\w_j\}}$ of $L$ with rows indexed by ${S \setminus \{\w'_i\} \cup \{\w_j\}}$ is the Schur complement $K_{\A, S \setminus \{\w'_i\} \cup \{\w_j\}}(w)/K_4$. Using $\det K_4=\det K_{\A,S}(w)$ and Theorem \ref{thm::schur1}, we have
	\begin{equation}
		\la{eq:pientry}
		\Pi(w)_{\w_j,\bw_i}=(-1)^{|S|-i} \det L_{S \setminus \{\w'_i\} \cup \{\w_j\}}=(-1)^{|S|-i} \frac{ \det K_{\A,S \setminus \{ \w'_i \} \cup \{ \w_j \}}(w)}{\det K_{\A,S}(w)}.
	\end{equation}
	
	The $\Pi(w)$ matrix has the following multiplicativity property which will be very useful later for computations.
	
	\begin{Proposition}\la{prop:prod}
		Suppose $\Gamma_\A$ is a \bcg\ obtained by gluing \bcg s $\Gamma_i$ for $i=1, \dots, n$ from left to right, so that $S(\Gamma_i)$ is identified with $T(\Gamma_{i+1})$. Assume that the Kasteleyn signs on the $\Gamma_i$ induce a Kasteleyn sign on $\Gamma_\A$ $($which is the case if they use the same $(\sigma_1,\dots,\sigma_h)$ for their definition$)$. Then,
		\[
		\Pi(\Gamma_\A)(w)=(-1)^{n-1}\Pi(\Gamma_1)(w) \Pi(\Gamma_{2})(w) \cdots \Pi(\Gamma_n)(w).
		\]
		Here, $S(\Gamma_i)$ and $T(\Gamma_{i+1})$ denote the source vertices of $\Gamma_i$ and target vertices of $\Gamma_{i+1}$, respectively.
	\end{Proposition}

	\begin{proof}
		We may assume that $n=2$, the general case will follow by induction. For $i\in\{1,2\}$, denote by $B_S^i$, $B^i$, $S^i$, $T^i$ and $W_{\text{int}}^i$ the sets of vertices associated with $\Gamma_i$. Then we have the following form for the Kasteleyn matrices of $\Gamma_1$ and $\Gamma_2$:
		\[
		K(\Gamma_1)(w)=\begin{blockarray}{ccc}
			B_S^1 & B^1 \setminus B_S^1 \\
			\begin{block}{[cc]c}
				I & 0 & S^1\\
				K_1 & K_2 & T^1\\
				K_3 & K_4 & W_{\text{int}}^1\\
			\end{block}
		\end{blockarray},\qquad
		K(\Gamma_2)(w)=\begin{blockarray}{ccc}
			B_S^2 & B^2 \setminus B_S^2 \\
			\begin{block}{[cc]c}
				I & 0 & S^2\\
				K_1' & K_2' & T^2\\
				K_3' & K_4' & W_{\text{int}}^2\\
			\end{block}
		\end{blockarray}.
		\]
		
		Observing that $S^1=T^2$, we have the following Kasteleyn matrix for $\Gamma_\A$:
		
		\[
		K(\Gamma_\A)(w)=\begin{blockarray}{ccccc}
			B_S^2 & B_S^1 & B^2 \setminus B_S^2 & B^1 \setminus B_S^1 \\
			\begin{block}{[cccc]c}
				I & 0 & 0 & 0 & S^2\\
				0 & K_1 & 0 & K_2 & T^1\\
				K_1' & I & K_2' & 0 & T^2\\
				K_3' & 0 & K_4' & 0 & W_{\text{int}}^2\\
				0 & K_3 & 0 & K_4 & W_{\text{int}}^1\\
			\end{block}
		\end{blockarray}.
		\]
		Thus,
		\[
		-\Pi(\Gamma_\A)(w)=
		\begin{bmatrix}
			K_1 & 0 & K_2
		\end{bmatrix}
		\begin{bmatrix}
			I & K_2' & 0 \\
			0 & K_4' & 0 \\
			K_3 & 0 & K_4
		\end{bmatrix}^{-1}
		\begin{bmatrix}
			K_1' \\
			K_3' \\
			0
		\end{bmatrix}.
		\]
		Since
		\[
		\begin{bmatrix}
			I & K_2' & 0 \\
			0 & K_4' & 0 \\
			K_3 & 0 & K_4
		\end{bmatrix}^{-1}
		=
		\begin{bmatrix}
			I & -K_2'K_4'^{-1} & 0 \\
			0 & K_4'^{-1} & 0 \\
			-K_4^{-1}K_3 & K_4^{-1}K_3K_2'K_4'^{-1} & K_4^{-1}
		\end{bmatrix},
		\]
		we conclude that $-\Pi(\Gamma_\A)(w)=\Pi(\Gamma_1)(w)\Pi(\Gamma_2)(w).$
	\end{proof}
	
	When $w=1$, we abbreviate $K_\A(1)$ (resp.\ $\Pi(1)$) to $K_\A$ (resp.\ $\Pi$).
	
	\begin{Remark}
		\label{rem:bmm}
		The matrix $\Pi(w)$ is related to the boundary measurement matrix of \cite{GSTV} constructed from networks on cylinders.
		The reference dimer cover $M_0$ makes $\Gamma_\A$ into a directed network $\mathcal N$ as follows. Orient each edge $e=\bw \w$ contained in $M_0$ from $\w$ to $\bw$, and assign it weight $\frac 1 {\wt(e)}$, and each edge not contained in $M_0$ from $\bw$ to $\w$ and assign it weight $\wt(e)$. Each directed path in $\mathcal N$ gets a weight that is the product of weights of all edges appearing in it. Then using Theorem \ref{thm:cimres} and formula~\eqref{eq:pientry}, we have for $\w'_i \in S$, $\w_j \in T$,
		\begin{align}
			\Pi(w)_{\w_j,\bw_i}&=(-1)^{|S|-i} \frac{ \det K_{\A,S \setminus \{ \w'_i \} \cup \{ \w_j \}}(w)}{ \det K_{\A,S}(w)}\nonumber\\
			&{}=(-1)^{|S|-i} \frac{\wt(M_0) \kappa(M_0) w^{(M_0,-\gamma_z)}}{ \det K_{\A,S}(w)} \frac{ \det K_{\A,S \setminus \{ \w'_i \} \cup \{ \w_j \}}(w)}{\wt(M_0) \kappa(M_0) w^{(M_0,-\gamma_z)}}\nonumber\\
			&{}=(-1)^{|S|-i} \frac{\wt(M_0) \kappa(M_0) w^{(M_0,-\gamma_z)}}{ \det K_{\A,S}(w)} \nonumber \\
			&\quad{} \times\biggl( \sum_{\partial M=S \setminus \{ \w'_i \}\cup \{\w_j\}} \operatorname{sign}([M-M_0]) [\wt]([M-M_0])w^{([M-M_0],-\gamma_z)}\biggr). \la{eq:path}
		\end{align}
		Notice that if $M$ is a dimer cover with $\partial M =S \setminus \{ \w'_i \}\cup \{\w_j\}$, then $M-M_0$ is the union of some directed cycles and a single directed simple path in $\mathcal N$ from $\w'_i$ to $\w_j$. The multiplicative factor is $\pm1$ if $M_0$ is the only dimer cover $M'$ with $\partial M'=S$. If there is another dimer cover~$M'$ with $\partial M'=S$, then $M'-M_0$ is a collection of directed cycles in $\mathcal N$ that can be attached to any directed path from $\w'_i$ to $\w_j$ to get a new directed path from $\w'_i$ to $\w_j$. Expanding the multiplicative factor
		\begin{gather*}
		\frac{\wt(M_0) \kappa(M_0) w^{(M_0,-\gamma_z)}}{ \det K_{\A,S}(w)}\\
\qquad{}=\frac{ \operatorname{sign}([0])}{ 1+\Biggl(\sum\limits_{\substack{M' \neq M_0 \\ \partial M'=S}} \operatorname{sign}([M'-M_0]) \operatorname{sign}([0]) [\wt]([M'-M_0])w^{([M'-M_0],-\gamma_z)}\Biggr)}
		\end{gather*}
		as a geometric series, we see that on the right-hand side of equation \eqref{eq:path}, we have a (signed) partition function for (not necessarily simple) directed paths from $\w'_i$ to $\w_j$ along with collections of directed cycles. The boundary measurement matrix of \cite{GSTV} is also a matrix whose entries are signed partition functions for directed paths from $\w'_i$ to $\w_j$. Therefore, the matrix $\Pi(w)$ is the boundary measurement matrix of \cite{GSTV} up to signs. A careful choice of $\kappa$ is required to make the signs match up; this was worked out recently in \cite{Izosimov1}. This is the reason for calling $S$ and $T$ the source and target vertices, respectively.
	\end{Remark}
	
	\begin{figure}[ht]
		\centering
		\begin{tikzpicture}[scale=1.5] 		\draw[dashed, gray] (1,0) rectangle (3,2);
			
			\draw[->,densely dotted] (1,.25) -- (3,.25);
			\node (no) at (1-0.25,.25) {$\gamma_z$};
			\coordinate[wvert,label=left:${\rm w}_{1}$] (w2) at (1,1);
			\coordinate[wvert,label=right:${\rm w}'_{1}$] (w1) at (3,1);
			\coordinate[wvert,label=above:${\rm w}$] (w3) at (2,2);
			\coordinate[wvert,label=below:${\rm w}$] (w4) at (2,0);
			\coordinate[bvert,label=below:${\rm b}_{1}$] (b1) at (2.5,1);
			\coordinate[bvert,label=right:${\rm b}_{2}$] (b2) at (1.5,1);
			
			\draw[-,text=red]
			(b1) edge node[below] {$1$} (w1)
			edge node[right] {$-b $} (w3)
			;
			\draw[-,text=red]
			(b2) edge node[below] {$1$} (w2)
			edge node[left] {$c/w$} (w4) edge node[left] {$a$} (w3)
			;
			\draw[line width=1.8pt] (b1) -- (w1);
			\draw[line width=1.8pt] (b2) -- (w3);

		\end{tikzpicture} \hspace{5mm}
		\begin{tikzpicture}[scale=1.5]
			
			\draw[dashed, gray] (1,0) rectangle (3,2);
			
			\draw[->,densely dotted] (1,.25) -- (3,.25);
			\node (no) at (1-0.25,.25) {$\gamma_z$};
			\coordinate[wvert,label=left:${\rm w}_{1}$] (w2) at (1,1);
			\coordinate[wvert,label=right:${\rm w}'_{1}$] (w1) at (3,1);
			\coordinate[wvert,label=above:${\rm w}$] (w3) at (2,2);
			\coordinate[wvert,label=below:${\rm w}$] (w4) at (2,0);
			\coordinate[bvert,label=below:${\rm b}_{1}$] (b1) at (2.5,1);
			\coordinate[bvert,label=right:${\rm b}_{2}$] (b2) at (1.5,1);
			
			\draw[->,text=red]
			(b1)
			edge node[right] {$b $} (w3)
			;
			\draw[->,text=red]
			(b2) edge node[below] {$1$} (w2)
			edge node[left] {$c/w$} (w4)
			;
			\draw[->,text=red] (w3) edge node[left] {$\frac 1 a$} (b2);
			\draw[->,text=red]
			(w1) edge node[below] {$1$} (b1);

		\end{tikzpicture}
		\caption{On the left, a bipartite graph $\Gamma_{\A}$ on a cylinder with the dimer cover $M_0$ in bold. On the right, the associated network $\mathcal N$.}
		\label{fig:network}
	\end{figure}
	\begin{Example}
		Consider the bipartite graph shown in Figure~\ref{fig:network}. We compute
		\[
		K_\A(w)=\begin{blockarray}{ccc}
			{\rm b}_1 & {\rm b}_2 \\
			\begin{block}{[cc]c}
				1&0 &{\rm w}'_1\\
				0&1&{\rm w}_1\\
				-b & a+\frac c w & \w\\
			\end{block}
		\end{blockarray}, \qquad \Pi(w)=\begin{bmatrix}
			-\frac b {a+\frac c w}
		\end{bmatrix}.
		\]
		
		There are two dimer covers with $\partial M=\{\w'_1\}$ with weights $a$ and \smash{$\frac c w$}, respectively. Notice that $\w'_1 \ra \bw_1 \ra \w \ra \bw_2 \ra \w_1$ is a directed simple path from $\w'_1$ to $\w_1$ with weight $\frac b a$ and that $\w \ra \bw_2 \ra \w$ is a directed cycle in $\mathcal N$ with weight $\frac c {aw}$. Moreover, any directed path from $\w'_1$ to $\w_1$ is obtained by attaching a finite number of copies of the cycle to the simple path at $\bw_2$. Therefore, we see that
		\[
		\Pi_{\w_1, \bw_1}(w)=- \biggl(\frac 1 {1+ \frac c {aw}}\biggr) \frac b a =-\frac b a \sum_{k \geq 0} (-1)^k \biggl(\frac c {aw} \biggr)^k
		\]
		is the (signed) partition function for all paths from $\w'_1$ to $\w_1$.
	\end{Example}
	
	\begin{Example}
		\label{ex:bcg}
		Let $\Gamma_{2,\A}$ denote the balanced cylinder graph obtained by gluing the top and bottom sides of the rectangle on Figure~\ref{fig:giew} (here we should take $z=1$ on the picture). Gluing also the left and right sides yields the torus graph $\Gamma_2$. The targets are the two white vertices~$\w_1$ and $\w_2$ on the left boundary, while the sources are their copies on the right boundary that we denote by $\w'_1$ and $\w'_2$. The set $B_S$ is $\{\bw_3,\bw_4\}$. The Kasteleyn matrix (with blocks as in \eqref{k4} and with $w=1$) is
		\[
		\aboverulesep=0pt \belowrulesep=0pt
		K_\A=\begin{blockarray}{ccccc}
			{\rm b}_3 & {\rm b}_4 & {\rm b}_1 & {\rm b}_2 \\
			\begin{block}{[cc|cc]c}
				1& 0 & 0&0 &{\rm w}'_1\\
				0 & 1 & 0&0&{\rm w}'_2\\
				\cmidrule(lr){1-4}
				0 & 0 & 1&0 &{\rm w}_1\\
				0 & 0 & 0&1&{\rm w}_2\\
				\cmidrule(lr){1-4}
				h & a & g & -b & \w_3\\
				e & d & -f & c & \w_4\\
			\end{block}
		\end{blockarray},
		\]
		and \be
		\Pi=\frac{1}{bf-cg}\begin{bmatrix}
			ch+be & ac+bd\\
			eg+fh & dg+af
		\end{bmatrix}. \la{eq:pmg2}
		\ee
	\end{Example}

	\subsection{Spectral curve of a torus graph constructed from a cylinder graph}
	\label{subsec:spectralcurve}

	We now show how the spectral curve of the torus graph obtained by gluing the two boundaries of a balanced cylinder graph can be obtained from the $\Pi$ matrix of the cylinder graph.

	\begin{Theorem} \la{p:sc}
		Let $\Gamma_\A$ be a balanced cylinder graph and assume that by gluing the two boundaries of $\Gamma_\A$ we obtain a torus graph $\Gamma$. Write $\Pi(w)$ for $\Pi(\Gamma_\A)(w)$. Then $\Sigma=\bigl\{(z,w) \in (\C^\times)^2 \mid \det (z I+\Pi(w))=0\bigr\}$ is the spectral curve.
	\end{Theorem}
	
	Due to the construction of Section~\ref{subsec:t2c} this result holds in particular when $\Gamma_\A$ is obtained from an arbitrary minimal torus graph $\Gamma$ cut along a zig-zag path.
	
	\begin{proof}
		In $\Gamma$, split the white vertices that are in the image of $S$ and $T$ under the projection of $\Gamma_{\A}$ to $\T$, so that we now have two copies of these white vertices which we identify with $S$ and $T$, respectively, connected by degree two black vertices. Let $\Gamma_{ST}$ denote the torus graph obtained. Let $B_{ST}$ denote the newly created degree two black vertices. Perturb $\gamma_w$ so that it goes transversely through all the edges connecting $B_{ST}$ with $T$ (see the right picture of Figure~\ref{fig:cfroma}). We extend the Kasteleyn sign $\kappa$ on $\Gamma$ to $\Gamma_{ST}$ by defining $\kappa(e)=1$ if $e$ is an edge between $B_{ST}$ and $T$ and $\kappa(e)=-1$ if $e$ is an edge between $B_{ST}$ and $S$.
		The Kasteleyn matrix $K(z,w)$ of $\Gamma_{ST}$ has the block matrix form
		\[
		K(z,w)=\begin{blockarray}{ccc}
			B(\Gamma_{\A}) & B_{ST} \\
			\begin{block}{[c|c]c}
				& -I & S\\
				K_{\A}(w) & z I & T\\
				& 0 & W_{\text{int}} \\
			\end{block}
		\end{blockarray}.
		\]
		Defining $K_4(w)$ to be the square submatrix of $K(z,w)$ with rows indexed by $W_{\text{int}}$ and columns indexed by $B(\Gamma_\A)\setminus B_S$, we have the Schur complement
		\[
		K(z,w)/K_4(w)=\begin{blockarray}{ccc}
			B_S & B_{ST} \\
			\begin{block}{[c|c]c}
				I & -I & S\\
				\Pi(w) & z I & T\\
			\end{block}
		\end{blockarray}.
		\]
		By Theorem \ref{thm::schur1}, we get $\det {K(z,w)} =\det K_4(w) \det{(zI+\Pi(w))}$.
	\end{proof}
	
	\begin{Example}
		Consider again the graph $\Gamma_2$ from Figure~\ref{fig:giew} for which we have
		\[
		\Pi(w)=\frac{1}{-cg+bf w}\begin{bmatrix}
			ch+be w & (ac+bd)w\\
			eg+fh & dg+af w
		\end{bmatrix}
		\]
		from \eqref{eq:pmg2}. We have
		\[
		\det (zI+\Pi(w))=\frac{1}{-cg+bf w}\big(-dh + (dg+ch)z -cg z^2+a e w + (b e+a f) zw + bf z^2 w\big),
		\]
		which agrees with \eqref{eq:scg2}.
	\end{Example}
	
	\subsection{Torus to cylinder}
	\la{subsec:t2c}
	
	\begin{figure}[ht]
		\centering
		\def\scl{0.8}
		\begin{tikzpicture}[scale=\scl]
			\node[bvert] (b1) at (0,0) {};
			\node[wvert] (w1) at (1,1) {};
			\node[bvert] (b2) at (0,2) {};
			\node[wvert] (w2) at (1,3) {};
			\node[wvert] (w0) at (1,-1) {};
			\draw[->] (w0) -- (b1);
			\draw (b1) -- (-0.5,-0.25);
			\draw (b1) -- (-0.5,0.25);
			\draw (b2) -- (-0.5,2-0.25);
			\draw (b2) -- (-0.5,2+0.25);
			\draw (w1) -- (1+0.5,1-0.25);
			\draw (w1) -- (1.5,1+0.25);
			\draw (w2) -- (1.5,3-0.25);
			\draw (w2) -- (1.5,3+0.25);
			\draw (w0) -- (1+0.5,-1-0.25);
			\draw (w0) -- (1.5,-1+0.25);
			
			\draw[->] (w1) -- (b2);
			\draw[->] (b1) -- (w1);
			\draw[->] (b2) -- (w2);
			\draw[->,dashed] (w2) -- (0,4);
			\draw[->,dashed] (0,-2) -- (w0);
			\begin{scope}[shift={(-1.1,0)}]
				\node[bvert] (bb1) at (6,0) {};
				\node[bvert] (bb3) at (4,2) {};
				\node[bvert] (bb4) at (4,0) {};
				\node[wvert] (ww1) at (6+1,1) {};
				\node[bvert] (bb2) at (6+0,2) {};
				\node[wvert] (ww2) at (6+1,3) {};
				\node[wvert] (ww0) at (6+1,-1) {};
				\node[wvert] (ww4) at (5,2) {};
				\node[wvert] (ww5) at (5,0) {};
				
				\draw (bb4) -- (4-0.5,-0.25);
				\draw (bb4) -- (4-0.5,0.25);
				\draw (bb3) -- (4-0.5,2-0.25);
				\draw (bb3) -- (4-0.5,2+0.25);
				\draw (ww1) -- (7+0.5,1-0.25);
				\draw (ww1) -- (7.5,1+0.25);
				\draw (ww2) -- (7.5,3-0.25);
				\draw (ww2) -- (7.5,3+0.25);
				\draw (ww0) -- (7+0.5,-1-0.25);
				\draw (ww0) -- (7.5,-1+0.25);
				
				\draw[->] (ww0) -- (bb1);
				\draw[->] (ww1) -- (bb2);
				\draw[->] (bb1) -- (ww1);
				\draw[->] (bb2) -- (ww2);
				\draw[->,dashed] (ww2) -- (6+0,4);
				\draw[->,dashed] (6+0,-2) -- (ww0);
				\draw (ww4) -- (bb3);
				\draw (ww5) -- (bb4);
				\draw (ww4) -- (bb2);
				\draw (ww5) -- (bb1);
				\draw[->,densely dotted] (5,-2) -- (ww5) -- (ww4) -- (5,4);
				\node (no) at (0.5,-2.5) {$\beta$};
				\node (no) at (5,-2.5) {$\gamma_w$};
				\node (no) at (6+0.5,-2.5) {$\beta$};
				\node (no) at (5+0.5,-0.5) {$T$};
				\node (no) at (5-0.5,-0.5) {$S$};
			\end{scope}
		\end{tikzpicture}\hspace{12mm}
		\begin{tikzpicture}[scale=\scl]
			\fill [gray!50] (3,-2) rectangle (5,4);
			\node[bvert] (bb1) at (6,0) {};
			\node[bvert] (bb3) at (4,2) {};
			\node[bvert] (bb4) at (4,0) {};
			\node[bvert] (bb5) at (2,2) {};
			\node[bvert] (bb6) at (2,0) {};
			\node[wvert] (ww1) at (6+1,1) {};
			\node[bvert] (bb2) at (6+0,2) {};
			\node[wvert] (ww2) at (6+1,3) {};
			\node[wvert] (ww0) at (6+1,-1) {};
			\node[wvert] (ww4) at (5,2) {};
			\node[wvert] (ww5) at (5,0) {};
			\node[wvert] (ww6) at (3,2) {};
			\node[wvert] (ww7) at (3,0) {};
			\draw (bb6) -- (2-0.5,-0.25);
			\draw (bb6) -- (2-0.5,0.25);
			\draw (bb5) -- (2-0.5,2-0.25);
			\draw (bb5) -- (2-0.5,2+0.25);
			\draw (ww1) -- (7+0.5,1-0.25);
			\draw (ww1) -- (7.5,1+0.25);
			\draw (ww2) -- (7.5,3-0.25);
			\draw (ww2) -- (7.5,3+0.25);
			\draw (ww0) -- (7+0.5,-1-0.25);
			\draw (ww0) -- (7.5,-1+0.25);
			\draw[->] (ww0) -- (bb1);
			\draw[->] (ww1) -- (bb2);
			\draw[->] (bb1) -- (ww1);
			\draw[->] (bb2) -- (ww2);
			\draw[->,dashed] (ww2) -- (6+0,4);
			\draw[->,dashed] (6+0,-2) -- (ww0);
			\draw (ww4) -- (bb3);
			\draw (ww5) -- (bb4);
			\draw (ww4) -- (bb2);
			\draw (ww5) -- (bb1);
			\draw (ww6) -- (bb5);
			\draw (ww6) -- (bb3);
			\draw (ww7) -- (bb6);
			\draw (ww7) -- (bb4);
			\draw[->,densely dotted] (4.5,-2) -- (4.5,4);
			
			\node (no) at (4.5,-2.5) {$\gamma_w$};
			\node (no) at (6+0.5,-2.5) {$\beta$};
			\node (no) at (5,-0.5) {$T$};
			\node (no) at (3,-0.5) {$S$};
			\node (no) at (4,-0.5) {$B_{ST}$};
			\node (no) at (2,-0.5) {$B_{S}$};
		\end{tikzpicture}
		\caption{On the left is a zig-zag path $\beta$ on a torus graph $\Gamma$. In the middle is the graph resulting from splitting the black vertices on $\beta$. On the right is the torus graph $\Gamma_{ST}$ obtained from further splitting the white vertices lying on $\gamma_w$. Removing the shaded region from $\Gamma_{ST}$, we get the cylinder graph $\Gamma_{\A}$.} \label{fig:cyl}	\label{fig:cfroma}
	\end{figure}
	
In this subsection, we show that Theorem~\ref{p:sc} actually applies to any minimal bipartite graph on the torus. We outline a general procedure to construct a balanced cylinder graph $\Gamma_\A$ from a~minimal graph $\Gamma$ in a torus $\T$. This procedure is a generalization of Example~\ref{ex:bcg}, in which the balanced cylinder graph $\Gamma_{2,\A}$ is obtained from the minimal torus graph $\Gamma_2$ by cutting along the vertical side of the fundamental rectangle, which is parallel to the zig-zag path $\zeta_2$.
	
	Let $\beta$ be a zig-zag path in $\Gamma$. Without loss of generality, we assume that there are no $2$-valent black vertices in $\beta$. Changing the fundamental domain if necessary, we can assume that~${[\beta]=[\gamma_w]}$. Split each black vertex in $\beta$ to create a $2$-valent white vertex, in such a way that one of the newly created black vertices is trivalent, having as neighbors the newly created white vertex as well as the two white vertices on $\beta$ that were adjacent to the black vertex before the split. See the left and middle pictures of Figure~\ref{fig:cyl}. The homology class $[\gamma_w]$ has a representative cycle $\gamma_w$ in $\T$ that goes through each of the newly created $2$-valent white vertices and does not intersect $\Gamma$ anywhere else. Cutting $\T$ along $\gamma_w$, we obtain a cylinder $\A$ and a graph $\Gamma_\A$ embedded in it. The $2$-valent white vertices become $S$ and $T$, where $T$ is connected to $\beta$ (see the middle picture of Figure~\ref{fig:cyl}). We label the vertices of $T$ in clockwise order as $\w_1,\dots,\w_h$, where~${h=|T|}$. Since $|B(\Gamma)|=|W(\Gamma)|$, we have $|B(\Gamma_\A)|=|W(\Gamma_\A)|-|S|$. Since $\Gamma$ is minimal, there is a dimer cover $M_0$ in $\Gamma$ that contains half the edges in $\beta$ (see, for example, \cite[Theorem~3.12]{GK13}), which becomes a dimer cover in $\Gamma_\A$ such that $\partial M_0=S$. For every $1\leq i\leq h$ let $f_i$ be the face of $\Gamma_\A$ adjacent to the boundary segment $w_i w_{i+1}$ on the $T$ side. The Kasteleyn signs $\kappa$ on $\Gamma$ induce Kasteleyn signs on $\Gamma_\A$ provided we set $\sigma_i=1$ if and only if \smash{$X_{f_i}([\kappa])=(-1)^{\frac{|\partial f_i|}{2}}$} for every~${1\leq i\leq h}$.
	
\section{TCD maps on cylinders}
	\label{sec:tcdtorus}
	
	In this section, we describe the cokernel of the Kasteleyn matrix $K$ from a projective point of view in terms of \emph{triple crossing diagram maps}, which we abbreviate to TCD maps. Our presentation of TCD maps is self-contained but we refer to \cite{AGPR,AGR} for more details on TCD maps. We then define the notion of twisted TCD maps on a cylinder and compute the monodromy of such twisted TCD maps.
	
	TCD maps are introduced in \cite{AGR, athesis} as a special case of the vector-relation configurations of~\cite{AGPR}. The spider move for TCD maps first appeared in \cite{AGPR} while the resplit move is introduced in~\cite{AGR, athesis}. The geometric $R$-matrix move for TCD maps is a novel contribution of the present paper.
	
	\subsection{TCD maps}
	\label{subsec:tcdmaps}
	
	Let $\Gamma_\A$ be a balanced cylinder graph. Assume that the black vertices of $\Gamma_\A$ are all of degree~$2$ or~$3$, which we can always do using expansion moves. A TCD map is a collection of points \smash{$(P_{\rm w})_{{\rm w}\in W} \in \CP^{|W|-|B|-1}$} such that the following two conditions hold:
	\begin{itemize}\itemsep=0pt
		\item for each ${\rm b} \in B$ of degree $3$, the three points $P_{\rm w}$ for ${\rm w}$ incident to ${\rm b}$ are distinct and are all contained in a line;
		\item for each ${\rm b} \in B$ of degree $2$, the two points $P_{\rm w}$ for ${\rm w}$ incident to ${\rm b}$ are equal.
	\end{itemize}
	Strictly speaking, the black vertices of a TCD should all be of degree $3$ \cite{thurston}, but for the purposes of this article, it will be convenient to also allow black vertices of degree $2$. The above definition of a TCD map was given in \cite{AGR} for any bipartite graph with black vertices of degree $2$ or $3$, not necessarily a balanced cylinder graph. However starting in the next paragraph we use the Kasteleyn matrix hence we have to restrict the level of generality to consider only balanced cylinder graphs.

	Recall that for a linear map $f$ between two vector spaces $E$ and $F$, the \emph{cokernel} of $f$ is defined as $F/\im f$. Given a generic weight $\wt$ on $\Gamma_\A$, we obtain a TCD map as follows: consider the exact sequence
	\[
	0 \ra \C^{B} \xrightarrow[]{K_\A} \C^{W} \ra \coker K_\A \ra 0,
	\]
	where $\coker K_\A$ is ${(|W|-|B|)}$-dimensional. Let $e_\w$ be the unit basis vector corresponding to~$\w$ in $\C^{W(\Gamma_{\A})}$ and let $v_{\rm w} \in \coker K_\A$ be the image of $e_{\rm w}$. Then the projectivizations $P_{\rm w}$ of the vectors $v_{\rm w}$ define a TCD map. Clearly the definition is invariant under gauge equivalence. On~the~other hand, given a TCD map, we recover the edge weights modulo gauge transformations from the equations of the lines associated to the black vertices (see Lemma \ref{lem:mrcyl} below).
	
	\begin{Remark}
		Note that the cokernel is only defined up to isomorphism. Different choices for a representative of the isomorphism class of the cokernel give different TCD maps related by projective transformations.
	\end{Remark}
	
	\begin{figure}[ht]
		\centering
		\begin{tikzpicture}[scale=.75,rotate=45,baseline={([yshift=-.7ex]current bounding box.center)},rotate=90]
			\coordinate (ws) at (-2,-1);
			\coordinate (wn) at (-2,1);
			\coordinate (es) at (2,-1);
			\coordinate (en) at (2,1);
			\coordinate (nw) at (-1,2);
			\coordinate (ne) at (1,2);
			\coordinate (sw) at (-1,-2);
			\coordinate (se) at (1,-2);
			\coordinate (n) at (0,1);
			\coordinate (s) at (0,-1);			
			\node[bvert] (n) at (n) {};
			\node[bvert] (s) at (s) {};
			\node[wvert,label=below:$\mathrm w_1$] (nn) at ($(nw)!.5!(ne)$) {};
			\node[wvert,label=above:$\mathrm w_3$] (ss) at ($(sw)!.5!(se)$) {};
			\node[wvert,label=below:$\mathrm w_2$] (ww) at ($(wn)!.5!(ws)$) {};
			\node[wvert,label=above:$\mathrm w_4$] (ee) at ($(en)!.5!(es)$) {};
			\draw[-]
			(n) edge (nn) edge (ww) edge (ee)
			(s) edge (ss) edge (ww) edge (ee)
			;			
		\end{tikzpicture}\hspace{3mm}$\leftrightarrow$\hspace{1mm}
		\begin{tikzpicture}[scale=.75,rotate=45,baseline={([yshift=-.7ex]current bounding box.center)}]
			\coordinate (ws) at (-2,-1);
			\coordinate (wn) at (-2,1);
			\coordinate (es) at (2,-1);
			\coordinate (en) at (2,1);
			\coordinate (nw) at (-1,2);
			\coordinate (ne) at (1,2);
			\coordinate (sw) at (-1,-2);
			\coordinate (se) at (1,-2);
			\coordinate (n) at (0,1);
			\coordinate (s) at (0,-1);		
			
			\node[bvert] (n) at (n) {};
			\node[bvert] (s) at (s) {};
			\node[wvert,label=above:$\mathrm w_4$] (nn) at ($(nw)!.5!(ne)$) {};
			\node[wvert,label=below:$\mathrm w_2$] (ss) at ($(sw)!.5!(se)$) {};
			\node[wvert,label=below:$\mathrm w_1$] (ww) at ($(wn)!.5!(ws)$) {};
			\node[wvert,label=above:$\mathrm w_3$] (ee) at ($(en)!.5!(es)$) {};
			\draw[-]
			(n) edge (nn) edge (ww) edge (ee)
			(s) edge (ss) edge (ww) edge (ee)
			;			
		\end{tikzpicture}\hspace{16mm}
		\begin{tikzpicture}[scale=.75,baseline={([yshift=-.7ex]current bounding box.center)},rotate=90]
			\coordinate (ws) at (-2,-1);
			\coordinate (wn) at (-2,1);
			\coordinate (es) at (2,-1);
			\coordinate (en) at (2,1);
			\coordinate (nw) at (-1,2);
			\coordinate (ne) at (1,2);
			\coordinate (sw) at (-1,-2);
			\coordinate (se) at (1,-2);
			\coordinate (n) at (0,1);
			\coordinate (s) at (0,-1);		
			
			\node[bvert] (n) at (n) {};
			\node[bvert] (s) at (s) {};
			\node[wvert,label=below:$\mathrm w_1$] (vnw) at ($(nw)!.5!(wn)$) {};
			\node[wvert,label=above:$\mathrm w_4$] (vne) at ($(ne)!.5!(en)$) {};
			\node[wvert,label=below:$\mathrm w_2$] (vsw) at ($(sw)!.5!(ws)$) {};
			\node[wvert,label=above:$\mathrm w_3$] (vse) at ($(se)!.5!(es)$) {};
			\node[wvert,label=above:$\mathrm w$] (c) at (0,0) {};
			\draw[-]
			(n) edge (vnw) edge (vne) edge (c)
			(s) edge (vsw) edge (vse) edge (c)
			;			
		\end{tikzpicture}\hspace{3mm}$\leftrightarrow$\hspace{1mm}
		\begin{tikzpicture}[scale=.75,baseline={([yshift=-.7ex]current bounding box.center)}]
			\coordinate (ws) at (-2,-1);
			\coordinate (wn) at (-2,1);
			\coordinate (es) at (2,-1);
			\coordinate (en) at (2,1);
			\coordinate (nw) at (-1,2);
			\coordinate (ne) at (1,2);
			\coordinate (sw) at (-1,-2);
			\coordinate (se) at (1,-2);
			\coordinate (n) at (0,1);
			\coordinate (s) at (0,-1);		
			
			\node[bvert] (n) at (n) {};
			\node[bvert] (s) at (s) {};
			\node[wvert,label=above:$\mathrm w_4$] (vnw) at ($(nw)!.5!(wn)$) {};
			\node[wvert,label=above:$\mathrm w_3$] (vne) at ($(ne)!.5!(en)$) {};
			\node[wvert,label=below:$\mathrm w_1$] (vsw) at ($(sw)!.5!(ws)$) {};
			\node[wvert,label=below:$\mathrm w_2$] (vse) at ($(se)!.5!(es)$) {};
			\node[wvert,label=right:$\mathrm w'$] (c) at (0,0) {};
			\draw[-]
			(n) edge (vnw) edge (vne) edge (c)
			(s) edge (vsw) edge (vse) edge (c)
			;			
		\end{tikzpicture}\\
		\begin{tikzpicture}[scale=.7,baseline={([yshift=-.7ex]current bounding box.center)}]
			\node[wvert,label=left:$P_{\mathrm w_1}$] (t1) at (-1,-1) {};
			\node[wvert,label=left:$P_{\mathrm w_4}$] (t4) at (1,2) {};
			\draw[-]
			(t1) -- (t4)
			;			
			\node[wvert,label=left:$P_{\mathrm w_3}$] (t3) at ($(t1)!.25!(t4)$) {};
			\node[wvert,label=left:$P_{\mathrm w_2}$] (t2) at ($(t1)!.7!(t4)$) {};
		\end{tikzpicture}\hspace{3mm}$\leftrightarrow$\hspace{2mm}
		\begin{tikzpicture}[scale=.7,baseline={([yshift=-.7ex]current bounding box.center)}]
			\node[wvert,label=left:$P_{\mathrm w_1}$] (t1) at (-1,-1) {};
			\node[wvert,label=left:$P_{\mathrm w_4}$] (t4) at (1,2) {};
			\draw[-]
			(t1) -- (t4)
			;			
			\node[wvert,label=left:$P_{\mathrm w_3}$] (t3) at ($(t1)!.25!(t4)$) {};
			\node[wvert,label=left:$P_{\mathrm w_2}$] (t2) at ($(t1)!.7!(t4)$) {};
		\end{tikzpicture}\hspace{14mm}
		\begin{tikzpicture}[scale=.7,baseline={([yshift=-.7ex]current bounding box.center)}]
			\node[wvert,label=left:$P_{\mathrm w_1}$] (t1) at (-1,-1) {};
			\node[wvert,label=left:$P_{\mathrm w_4}$] (t4) at (-0.2,2) {};
			\coordinate (t) at (2,1) {};
			\node[wvert,label=below:$\qquad \ P_{\mathrm w_2}$] (t2) at ($(t1)!.75!(t)$) {};
			\node[wvert,label=right:$P_{\mathrm w_3}$] (t3) at ($(t4)!.4!(t)$) {};
			\node[wvert,label=right:$P_{\mathrm w}$] (tt) at (intersection of t1--t4 and t2--t3) {};
			\draw[-]
			(t1) -- (t4) -- (tt) -- (t3) -- (t2)
			;						
		\end{tikzpicture}\hspace{5mm}$\leftrightarrow$\hspace{-4mm}	
		\begin{tikzpicture}[scale=.7,baseline={([yshift=-.7ex]current bounding box.center)}]
			\node[wvert,label=left:$P_{\mathrm w_1}$] (t1) at (-1,-1) {};
			\node[wvert,label=left:$P_{\mathrm w_4}$] (t4) at (-0.2,2) {};
			\node[wvert,label=above:$P_{\mathrm w'}$] (t) at (2,1) {};
			\draw[-]
			(t1) -- (t) -- (t4)
			;			
			\node[wvert,label=below:$\qquad \ P_{\mathrm w_2}$] (t2) at ($(t1)!.75!(t)$) {};
			\node[wvert,label=above:$P_{\mathrm w_3}$] (t3) at ($(t4)!.4!(t)$) {};
			\node[wvert,white,label=right:$\phantom{P_{\mathrm w}}$] (tt) at (intersection of t1--t4 and t2--t3) {};
			\draw[-]
			;						
		\end{tikzpicture}
		
		\caption{Elementary transformations allowed in TCD maps: spider move (left) and resplit (right).}\label{fig:localmoves}
	\end{figure}
	
	The advantage of working with a graph $\Gamma_\A$ having trivalent black vertices is that we can keep track of both the geometric dynamics and the invariants while performing local moves. For TCD maps, there are two allowed elementary transformations, the spider move and the resplit, see Figure~\ref{fig:localmoves}. In a generic situation, the points associated to white vertices after one of these two moves are determined by the combinatorics. Indeed, points do not change when performing the spider move. Furthermore, if the ambient projective space is of dimension at least~$2$, the new point appearing in the resplit is determined as the intersection of the two lines represented by the two black vertices. One can give a~formula for it using multi-ratios.
	
	The \emph{multi-ratio} of $2m$ points $P_1,P_2,\dots,P_{2m}\in \CP^n$ with $n\geq1$ is defined by
	\[
\mr(P_1,\dots,P_{2m})=\frac{\prod\limits_{i=1}^m (P_{2i-1}-P_{2i})}{\prod\limits_{i=1}^m (P_{2i}-P_{2i+1})}.
	\]
	Such a definition makes sense by taking an affine chart of $\CP^n$ (and is independent of the~choice of such a chart) by pairing up each term in the numerator with a collinear term in the~denominator, which is possible whenever one of the following two conditions is satisfied:
	\begin{enumerate}\itemsep=0pt
		\item[(1)] for every $1\leq i\leq m$ the points $P_{2i-1}$, $P_{2i}$ and $P_{2i+1}$ are aligned;
		\item[(2)] for every $1\leq i\leq m$ the points $P_{2i}$, $P_{2i+1}$ and $P_{2i+2}$ are aligned.
	\end{enumerate}
	The \emph{cross-ratio} of four aligned points $P_1,P_2,P_3,P_4\in \CP^n$ with $n\geq1$ is defined by
	\[
		\cro(P_1,P_2,P_3,P_4)=\mr(P_1,P_2,P_3,P_4).
	\]
	
	If the ambient dimension is at least $2$, then the points involved in a resplit satisfy the classical Menelaus' theorem (see for, e.g., \cite{Lester}):
	\begin{align}
		\label{eq:menelaus}
		\mr(P_{\mathrm w_1},P_{\mathrm w},P_{\mathrm w_2},P_{\mathrm w_3},P_{\mathrm w'},P_{\mathrm w_4}) = -1,
	\end{align}
	where the white vertices are labeled as on the right-hand side of Figure~\ref{fig:localmoves}. In $\CP^1$ however, there is no incidence geometry. In this case, we \emph{define} the new white vertex in the resplit via equation \eqref{eq:menelaus}. Equation \eqref{eq:menelaus} has the symmetries of the octahedron.
	
	\begin{Lemma}
		Let $n\geq1$ and let $P_1,\dots, P_6$ be six points in $\CP^n$. For every permutation $\sigma$ of $\{1,\dots,6\}$ such that $\sigma(i+3)=\sigma(i)+3 \mod 6$ for every $1\leq i \leq 6$, we have
		\[
		\mr(P_1,P_2,P_3,P_4,P_5,P_6)=-1 \quad\Leftrightarrow\quad \mr\big(P_{\sigma(1)},P_{\sigma(2)},P_{\sigma(3)},P_{\sigma(4)},P_{\sigma(5)},P_{\sigma(6)}\big)=-1.
		\]
	\end{Lemma}
	
	\begin{proof}
		The permutations $\sigma$ such that $\sigma(i+3)=\sigma(i)+3 \mod 6$ for every $i$ form the symmetry group of the octahedron, namely they leave invariant the collection of pairs of opposite points $\{\{P_1,P_4\},\{P_2,P_5\},\{P_3,P_6\}\}$. This subgroup is generated by $\sigma_1$, $\sigma_2$ and $\sigma_3$, where $\sigma_1(i)=i+1\mod 6$ for every $i$, $\sigma_2(i)=7-i$ for every $i$ and $\sigma_3$ is the transposition $(1,4)$.

		Observe that $\sigma_1$ changes a multi-ratio to its inverse, while~$\sigma_2$ leaves it invariant. In the case of~$\sigma_3$, solving the linear equation $\mr(P_1,\dots,P_6)=-1$ for $P_1$ and reinserting it in $\mr(P_4,P_2,P_3,\allowbreak P_1,P_5,P_6)$ yields the value $-1$.
	\end{proof}

	Another useful property of TCD maps is that the face weights can be recovered as multi-ratios as stated in the following lemma.
	
	\begin{Lemma}[\protect{\cite[Proposition 2.6]{AGPR}}]\la{lem:mrcyl}
		For a loop
\[
L = {\rm w}_1 \ra {\rm b}_1 \ra {\rm w}_2 \ra {\rm b}_2 \ra \cdots \ra {\rm w}_n \ra {\rm b}_n \ra {\rm w}_1
\] such that ${\rm w}_i \neq {\rm w}_{i+1}$ for every $i$, let ${\rm w}_i'$ denote the third white vertex incident to ${\rm b}_i$ that is not in~${\{{\rm w}_i,{\rm w}_{i+1}\}}$. Then we have
		\[
		[{\rm wt}]([L]) =[\kappa]([L])^{-1} \mr \big(P_{{\rm w}_1}, P_{{\rm w}_1'},P_{{\rm w}_2},P_{{\rm w}_2'},\dots, P_{{\rm w}_n},P_{{\rm w}_n'}\big) .
		\]
	\end{Lemma}

	\subsection[TCD map on Gamma\_\{widehat A\} from Gamma]{TCD map on $\boldsymbol{\Gamma_{\widehat \A}}$ from $\boldsymbol{\Gamma}$}
	\la{subsec:tcd}
	
	Let $\Gamma_\A$ be a balanced cylinder graph.
	
	\begin{Lemma} \la{lem:m1}
		Suppose there exists a dimer cover $M_1$ {of $\Gamma_\A$} such that $\partial M_1=T$. Then the~matrix~$\Pi$ in \eqref{eq:pi} is invertible for generic weights on $\Gamma_\A$.
	\end{Lemma}
	\begin{proof}
		Since $\det K_4=\det K_{\A,S}$, using Theorem \ref{thm::schur1}, we get
		\[
		\det \Pi=\frac{\det K_{\A,T}}{\det K_{\A,S}}.
		\]
		The existence of $M_1$ with $\partial M_1=T$ and Theorem \ref{thm:cimres} with $I=T$ gives $\det K_{\A,T} \neq 0$ for generic weights.
	\end{proof}

	Hereafter, we assume that there is a dimer cover $M_1$ of $\Gamma_\A$ such that $\partial M_1=T$ and that weights on $\Gamma_\A$ are generic.

	Let $\T$ be the torus obtained by gluing together the two boundary components of $\A$ in such a~way that the two endpoints of $\gamma_z$ are identified and that each vertex in $S$ is identified with a~vertex in $T$. Let $\gamma_w$ be the image in $\T$ of the boundaries of $\A$ and let $\Gamma$ be the image in $\T$ of $\Gamma_\A$. Let \smash{$\widehat \A:=H_1(\T,\R)/\Z [\gamma_w]$} denote the infinite cylinder covering $\A$. Note that \smash{$\widehat \A = \bigcup_{k \in \Z} \A+k [\gamma_z]$}, that is \smash{$\widehat \A$} is obtained by gluing together infinitely many copies of $\A$. Let
	\[
		\Gamma_{\widehat \A}=\big(B\bigl(\Gamma_{\widehat \A}\bigr)\cup W\bigl(\Gamma_{\widehat \A}\bigr),E\bigl(\Gamma_{\widehat \A}\bigr)\big)
	\]
	denote the preimage of $\Gamma$ under the covering map \smash{$\widehat \A \ra \T$}. Fix a white vertex ${\rm w} \in W\bigl(\Gamma_{\widehat \A}\bigr)$ and choose a large enough $m$ so that ${\rm w}$ is in \smash{$\A_m:=\bigcup_{k \in [-m,m] \cap \Z} (\A+k\gamma_z) \subset \widehat \A$. Let $\Gamma_{\A_m}:=\Gamma_{\widehat \A} \cap \A_m$}.
	\begin{Lemma}\la{lem:inftcd}
		We have $\coker K_{{{\A_m}}} \cong \coker K_\A \cong \C^{|W(\Gamma_{\A})|-|B(\Gamma_{\A})|}$ for all $m \geq 0$. Moreover with these identifications, the image of $e_{\w}$ in \smash{$\C^{|W(\Gamma_{\A})|-|B(\Gamma_{\A})|}$} under the cokernel map of $K_{\A_m}$ is independent of $m$, where $e_\w$ is the unit basis vector corresponding to $\w$ in \smash{$\C^{W(\Gamma_{\A_m})}$}.
	\end{Lemma}
	\begin{proof}
		Suppose $m>{m'}$. $K_{{{\A_m}}}$ has the block form
		\[
		K_{{{\A_m}}}=\begin{blockarray}{ccc}
			B\bigl(\Gamma_{{\A_{m'}}}\bigr)& B(\Gamma_{{\A_m}})\setminus B\bigl(\Gamma_{{\A_{m'}}}\bigr)& \\
			\begin{block}{[cc] c}
				K_{{\A_{m'}}} & *& W\bigl(\Gamma_{{\A_{m'}}}\bigr)\\
				0 & K' & W(\Gamma_{{\A_m}})\setminus W\bigl(\Gamma_{{\A_{m'}}}\bigr)\\
			\end{block}
		\end{blockarray},
		\]
		where $K'$ is invertible by existence of $M_0$ and $M_1$, and Theorem \ref{thm:cimres}. Therefore, $K_{{{\A_m}}}/K'=K_{{\A_{m'}}}$ and so the second statement follows from Theorem \ref{thm::schur}. By Theorem \ref{thm::schur} with ${m'}=0$, we get $\coker K_{{\A_m}} \cong \coker K_\A \cong \C^{|W(\Gamma_{\A})|-|B(\Gamma_{\A})|}$ for all $m \geq 0$.
	\end{proof}

	We define a TCD map \smash{$P\colon W\bigl(\Gamma_{\widehat \A}\bigr) \!\ra\! \CP^{|W(\Gamma_{\A})|-|B(\Gamma_{\A})|-1}$} in the following way. For \smash{${\rm w} \!\in\! W\bigl(\Gamma_{\widehat \A}\bigr)$}, we choose $m$ sufficiently large so that ${\rm w} \in W(\Gamma_{\A_m})$. Let $v_\w$ denote the image of $e_\w$ in $\C^{|W(\Gamma_{\A})|-|B(\Gamma_{\A})|}$ as in Lemma~\ref{lem:inftcd}, and define $P_{\rm w} \in \CP^{|W(\Gamma_{\A})|-|B(\Gamma_{\A})|-1}$ to be the projectivization of $v_{\rm w}$. Lemma~\ref{lem:inftcd} guarantees that the definition is independent of the choice of $m$.
	
\subsection[Monodromy of a TCD map on Gamma\_\{widehat A\}]{Monodromy of a TCD map on $\boldsymbol{\Gamma_{\widehat \A}}$}
	
	A pair $(P,M)$ where $P\colon W\bigl(\Gamma_{\widehat \A}\bigr) \ra \CP^{|W(\Gamma_{\A})|-|B(\Gamma_{\A})|-1}$ is a TCD map and the operator $M \in \PGL_{|W(\Gamma_{\A})|-|B(\Gamma_{\A})|}$ is called a \emph{twisted TCD map} if $M(P_{{\rm w}})=P_{{\rm w}+\gamma_z}$ for all ${\rm w} \in W\bigl(\Gamma_{\widehat \A}\bigr).$ $M$ is called the \emph{monodromy} of~$P$.
	
	Now suppose $P\colon W\bigl(\Gamma_{\widehat \A}\bigr) \ra \CP^{|W(\Gamma_{\A})|-|B(\Gamma_{\A})|-1}$ is a TCD map on \smash{$\Gamma_{\widehat \A}$} as in Section~\ref{subsec:tcd}. Let~$[\Pi]$ denote the class of $\Pi$ in $\PGL_{|W(\Gamma_{\A})|-|B(\Gamma_{\A})|}$, where $\Pi$ is the matrix in \eqref{eq:pi}.
	\begin{Theorem} \la{prop:mon}
		The map $P$ is a twisted TCD map. The $\PGL_{|W(\Gamma_{\A})|-|B(\Gamma_{\A})|}$ matrix {class} $[-\Pi]$ is the monodromy of $P$ in the basis $\{v_\w\}_{\w \in T}$.
	\end{Theorem}
	\begin{proof}
		The existence of the dimer cover $M_1$ implies that the matrix $\Pi$ is invertible (Lemma \ref{lem:m1}). Let $T=\{{\rm w}_1,\dots,{\rm w}_h\}$ be the set of target vertices in $W(\Gamma_{\A})$ so that $S=\{{\rm w}_1+\gamma_z,\dots,{\rm w}_h+\gamma_z\}$ is the set of source vertices. Since each boundary white vertex of $\Gamma_\A$ has degree $1$, we have a~unique black vertex $\bw_i \in B_S$ incident to $\w_i+\gamma_z \in S$ in $\Gamma_\A$. Therefore, we have a canonical isomorphism $\C^{S} \cong \C^{B_S}$, $e_{\w_i+\gamma_z} \mapsto e_{{\mathrm b}_i}$.
		
		By Theorem \ref{thm::schur}, we have $\coker \bigl[\begin{smallmatrix}
			I\\
			\Pi
		\end{smallmatrix}\bigr] \cong \coker K_\A$ such that the cokernel map $\C^S \oplus \C^T \ra \coker \bigl[\begin{smallmatrix}
			I\\
			\Pi
		\end{smallmatrix}\bigr] \cong \coker K_\A$ is $e_\w \mapsto v_\w$. Consider the following exact sequence:
		\[
		0 \ra \C^{B_S} \xrightarrow[]{\bigl[\begin{smallmatrix}
				I\\
				\Pi
		\end{smallmatrix}\bigr]} \C^S \oplus \C^T \xrightarrow[]{[\begin{smallmatrix}
				-\Pi & I
		\end{smallmatrix}]} \C^T \ra 0.
		\]
		When we write $-\Pi$ above the third arrow, we are abusing notation and mean the composition \smash{$\C^S \cong \C^{B_S} \xrightarrow[]{-\Pi} \C^T$}. By the universal property of the cokernel, we have a canonical isomorphism of $\coker K_\A$ with $\C^T$ such that $v_{\w_i} = e_{\w_i} $ for $\w_i \in T$ and $v_{\w_i+\gamma_z} = -\Pi e_{\w_i+\gamma_z}$ for ${\w_i+\gamma_z \in S}$. Translation by $\gamma_z$ gives us an isomorphism $\C^T \cong \C^S$ sending $e_{\w_i}$ to $e_{\w_i + \gamma_z}$. Therefore, ${v_{{\rm w}_i+\gamma_z}=-\Pi{}v_{{\rm w}_i}}$ for $i=1,2,\dots,h$, where again we are abusing notation by calling~$-\Pi$ the~composition \smash{$\C^T \cong \C^S \xrightarrow[]{-\Pi} \C^T$} which sends $v_{\w _i}$ to $v_{\w_i + \gamma_z}$.
		
		The same argument applied to the translated graph $\Gamma_{\A+k \gamma_z}$ gives \smash{$v_{{\rm w}_i+(k+1)\gamma_z}=-\Pi{}v_{{\rm w}_i+k \gamma_z}$}. This implies $v_{{\rm w}_i+k\gamma_z}=(-\Pi{})^k v_{{\rm w}_i}$ for all $k \in \Z$.
		For ${\rm w} \in W(\Gamma_{ \A})$, since $\{v_{\w_i}\}$ is a basis {of $\coker K_\A$}, there exist $a_i\in\C$ such that \smash{$v_{\rm w}=\sum_{i=1}^h a_i v_{{\rm w}_i}$}. Then we have
  \begin{align*}(-\Pi{})^k v_{{\rm w}}&=\sum_{i=1}^h a_i (-\Pi{})^k v_{{\rm w}_i} = \sum_{i=1}^h a_i v_{{\rm w}_i+k\gamma_z} =v_{{\rm w}+k\gamma_z},
		\end{align*}
		where the last equality follows from $K_{\A}{}=K_{{\A+ k \gamma_z}}{}$. Any white \smash{${\rm w}'\in W\bigl(\Gamma_{\widehat \A}\bigr)$} is of the form~${\rm w}+m\gamma_z$ for some ${\rm w} \in W(\Gamma_{\A})$. Then ${\rm w}'+\gamma_z={\rm w}+(m+1) \gamma_z$, so that we have
		\[
		v_{{\rm w}'}=(-\Pi{})^m v_{{\rm w}}, \qquad v_{{\rm w}'+\gamma_z}=(-\Pi{})^{m+1}v_{\rm w},
		\]
		and therefore $v_{{\rm w}'+\gamma_z}=-\Pi{}v_{{\rm w}'}$. Projectivizing, we get the statement of the proposition.
	\end{proof}
	
	In the second half of the paper, the setting of twisted TCD maps will be used to study dynamical systems on spaces of twisted polygons.
	
	\begin{Definition}
		Let $d\geq1$ and let $n \geq 3$. A \emph{twisted $n$-gon in dimension $d$} is a pair $(p,M)$ where $p\colon \Z \ra \CP^d$ and $M \in \PGL_{d+1}$ is a projective transformation called \emph{monodromy} such that $p_{i+n}=M(p_i)$ for all $i \in \Z$.
		
		The group $\PGL_{d+1}$ acts on the space of twisted $n$-gons in dimension $d$ by
		\begin{equation}
			\label{eq:pglaction}
			A \cdot (p_1,p_2,\dots,p_n,M) = \big(A(p_1),A(p_2),\dots,A(p_n),AMA^{-1}\big).
		\end{equation}
	\end{Definition}
	
	For each dynamical system considered, we will impose some additional nondegeneracy conditions for the twisted $n$-gons, which will be specific to each dynamics.
	
\section{The pentagram map}\label{sec:pentagram}
	
	Our exposition follows \cite{OST1,Weinreich}.
	Let $n \geq 4$. A twisted $n$-gon in dimension $2$ is said to be \emph{nondegenerate} if in each $5$-tuple $(p_i,p_{i+1},p_{i+2},p_{i+3},p_{i+4})$ of consecutive points, no three points are collinear, except possibly $p_i$, $p_{i+2}$, $p_{i+4}$.
	
	Let \smash{$\widetilde{\mathcal T}_n$} denote the space of nondegenerate twisted $n$-gons. \smash{$\widetilde {\mathcal T}_n$} is an open subvariety of $\big(\CP^2\big)^n \times \PGL_3$, and $\PGL_3$ acts on it by \eqref{eq:pglaction}. The quotient \smash{$\mathcal T_n:=\widetilde{\mathcal T}_n / \PGL_3$} is the moduli space parameterizing projective equivalence classes of nondegenerate twisted $n$-gons. For $(p,M) \in \widetilde{\mathcal T}_n$, the \textit{left and right corner invariants} are defined by
	\begin{align*}
		&x_i:= \cro(p_{i-2},p_{i-1},\overline{p_{i-2} p_{i-1}} \cap \overline{p_{i+1} p_{i+2}} ,\overline{p_{i-2} p_{i-1}}\cap \overline{p_i p_{i+1}} ),\\
		&y_i:= \cro( \overline{p_{i-2} p_{i-1}} \cap \overline{p_{i+1} p_{i+2}},\overline{p_{i-1} p_{i}}\cap \overline{p_{i+1} p_{i+2}},p_{i+2},p_{i+1}),
	\end{align*}
	where $\overline{ab}$ denotes the projective line spanned by $a,b \in \CP^2$. The corner invariants define a~morphism
	\begin{align}
		(x,y)\colon \ \widetilde{\mathcal T}_n \ra (\C \setminus\{0,1\})^{2n} , \qquad
		(p,M) \mapsto (x_1,\dots,x_n,y_1,\dots,y_n).\la{eq:xy}
	\end{align}
	Since the $x_i,y_i$'s are cross-ratios, they are $\PGL_3$-invariant. Therefore, the morphism \eqref{eq:xy} descends to a morphism $(x,y)\colon \mathcal T_n \ra (\C \setminus\{0,1\})^{2n}$, which is an isomorphism \cite{Schwartz}. The Poisson brackets
	\be \la{pent:pb}
	\{x_i,x_{i + 1}\}_{\mathcal T_n}:= - x_i x_{i+1}, \qquad \{y_i,y_{i + 1}\}_{\mathcal T_n} := y_i y_{i+1},
	\ee
	make $\mathcal T_n$ a Poisson variety. Here, we only give the nonzero values obtained by pairing two coordinate functions.

	The rational map $T\colon \mathcal T_n \ra \mathcal T_n$ defined by $(p,M) \mapsto (q,M)$, where $q_i=\overline{p_{i-1}p_{i+1}}\cap \overline{p_{i}p_{i+2}}$ is called the \textit{pentagram map}. In the coordinates $(x,y)$, the rational map $T$ is given by {\cite[Lemma~2.4]{OST1}}
	\[
	T^* x_i = x_i \frac{1-x_{i-1}y_{i-1}}{1-x_{i+1}y_{i+1}}, \qquad T^* y_i = y_{i+1} \frac{1-x_{i+2}y_{i+2}}{1-x_{i}y_{i}}.
	\]

	\subsection{Integrability of the pentagram map}
	
	We recall the following well known result from the theory of algebraic groups (see, for example,~\mbox{\cite[Theorem 3.2.3]{Springer}}). If we have an action of a torus $\C^\times$ on a vector space $V$, then $V=\bigoplus_{k \in \Z} V_k$ can be decomposed into the \textit{weight subspaces} corresponding to the \textit{characters} of~$\C^\times$, where a~character $\chi\colon \C^\times \ra \C^\times$ is just a Laurent monomial $\chi(w)=w^{k}$, $k \in \Z$, and the weight subspace of $\chi(w)=w^k$ is defined as
	\[
	V_k:=\{v \in V \mid w \cdot v = w^k v \text{ for }w \in \C^\times\}.
	\]
	If $v \in V_k$, then we say that $v$ has \textit{weight} $k$.
	
	Now consider the (rational) action of $\C^\times$ on $\mathcal T_n$: $R_w(x,y)=\big(w^{-1}x,wy\big)$, $x=(x_1,\dots,x_n)$, ${y=(y_1,\dots,y_n)}$,
	and let $R_w^*$ denote the induced action on the vector space $\C[x,y]$:
	\[
	R_w^*( f(x,y))= f(R_{w^{-1}}(x,y)).
	\]
	Define $O_n:=\prod_{i=1}^n x_i$ and $E_n:=\prod_{i=1}^n y_i$. Then, $O_n$ and $E_n$ have weights $n$ and $-n$, respectively. For $(p,M) \in \mathcal T_n$, let $M(w)$ denote the monodromy matrix of $R_w(p,M)$, and let
	\[
	P(z,w):= \det(zI-M(w)).
	\]
	In \cite{Schwartz}, it is shown that
	\begin{gather}
	\biggl(\frac{w^n O_n^2 E_n}{\det M(w)^{-1}}\biggr)^{\frac 1 3}\tr M(w)^{-1} = 1+ \sum_{k=1}^{\floor{\frac n 2}} O_k w^k, \nonumber\\
 \biggl(\frac{O_n E_n^2}{w^{n} \det M(w)}\biggr)^{\frac 1 3}\tr M(w) = 1+ \sum_{k=1}^{\floor{\frac n 2}} E_k w^{-k},\la{schinv}
	\end{gather}
	where $O_k$, $E_k$, $k=1,\dots,\floor{\frac n 2}$, are polynomials in $\C[x,y]$.
	\begin{Theorem}[\cite{OST1}]\la{thm:ost}
		We have
		\begin{enumerate}\itemsep=0pt
			\item[$1.$] For $n$ even, $\{\cdot,\cdot \}_{\mathcal T_n}$ has corank $4$, and the subalgebra of Casimirs is generated by \smash{$O_{\frac n 2}$}, \smash{$E_{\frac n 2}$}, $O_n$, $E_n$.
			\item[$2.$] For $n$ odd, $\{\cdot,\cdot \}_{\mathcal T_n}$ has corank $2$, and the subalgebra of Casimirs is generated by $O_n$, $E_n$.
			\item[$3.$] For \smash{$k=1,2,\dots, \floor{\frac{n-1}{2}}$}, the functions $O_k$ and $E_k$ mutually commute and form a maximal set of functionally independent Hamiltonians, making the Poisson variety $\mathcal T_n$ a Liouville integrable system.
		\end{enumerate}
		Moreover, the pentagram map $T$ is discrete integrable in the following sense:
		\begin{enumerate}\itemsep=0pt
			\item[$4.$] $T$ is Poisson.
			\item[$5.$] The Hamiltonians and the Casimirs are invariant under $T$.
		\end{enumerate}
	\end{Theorem}
	
	\subsection{Pentagram TCD maps} \label{sec:pentagramtcd}
	Let $n \geq 4$. Let $\Xi_n$ denote the torus graph obtained by gluing in cyclic order the graphs $H_i$ in Figure~\ref{fig:pentagram} for $i=1,2,\dots,n$ and contracting two-valent black vertices (see Figure~\ref{fig:pentagramtcd} for a picture of $\Xi_4$).
	Let $N_{\Xi_n}$ denote the Newton polygon of $\Xi_n$.
	When $n$ is even, there are $6$ zig-zag paths and the Newton polygon is as shown on left hand side of Figure~\ref{fig:nppent}, and when $n$ is odd, there are $4$ zig-zag paths and the Newton polygon as shown on the right hand side of Figure~\ref{fig:nppent}.
	
	\begin{figure}[ht]
		\centering
		\begin{tikzpicture}[yscale=1.5]
			\node[wvert,label=left:$\w_{i}$] (p1) at (0,2) {};
			\node[wvert,label=left:$\w_{i}$]
			(pp1) at (-3,-1) {};
			\node[wvert,label=right:$\w_{i+1}$] (q1) at (8,2) {};
			
			\node[wvert,label=right:$\w_{i+1}$] (qq1) at (5,-1) {};

			\node[bvert] (b1) at (3,2) {};
			\node[bvert] (b2) at (2,1) {};
			\node[bvert] (bb2) at (4,1) {};
			\node[bvert] (b3) at (3,0) {};
			\node[bvert] (bb1) at (0,-1) {};
			
			\node[bvert] (bp) at (0,1) {};
			\node[bvert] (bq) at (6,1) {};
			
			\draw[dashed, gray] (p1) -- (pp1) -- (qq1) -- (q1) -- (p1);	
			\node[wvert,label=above:$\w_{i+1}$] (p2) at (1,1) {};
			\node[wvert,label=left:$\w_{i+1}$] (pp2) at (-1,1) {};
			
			\node[wvert,label=left:$\w_{i+2}$] (p3) at (-2,0) {};
			
			\node[wvert,label=below:$\w_{i+2}$] (q2) at (5,1) {};
			\node[wvert,label=right:$\w_{i+2}$] (qq2) at (7,1) {};
			\node[wvert,label=right:$\w_{i+3}$] (q3) at (6,0) {};
			
			\node[wvert,label=below:$\widetilde \w_{i+1}$] (ww) at (3,1) {};
			\draw[-,text=red,inner sep=1]
			(p1) -- node[above] {$-(-{x_{i+1}}{w})$} (b2) -- (p2)--node[right] {$-1$}(b3)
			--(qq1)
			(p3)--node[right] {$-1$}(bb1)
			(b1)--node[right] {$\frac{y_{i+1}}{w}$}(q2)--(bb2)--node[above] {$-1$}(ww)--(b2)
			(bb2)--(q3)
			(pp2)--(bp)--node[below]{$-1$} (p2)
			(qq2)--(bq)--node[below]{$-1$} (q2)
			;
			\node[blue](no) at (2.5,1.6) {$X_{2i+2}$};
			\node[blue](no) at (4.5,0) {$X_{2i+4}$};
			\node[blue](no) at (.5,0) {$X_{2i+3}$};
			\node[blue](no) at (6.5,0.5) {$X_{2i+5}$};
			\node[blue](no) at (-.5,1.5) {$X_{2i+1}$};
		\end{tikzpicture}
		\caption{The building block graph $H_i$. In the edge label $-(-{x_{i+1}}{w})$, the edge weight is $-{x_{i+1}}$ and the other $-$ is the Kasteleyn sign. We have given white vertices that are identified upon contracting $2$-valent black vertices the same label, since they are mapped to the same point by the twisted TCD map~$P$.}
		\label{fig:pentagram}
	\end{figure}
	\begin{figure}[ht]
		\centering
		\begin{tikzpicture}[scale=0.4,xslant=-2,xscale=4,yscale=4.5]

			\draw[dashed, gray] (0,0)--(8,0)--(11,1)--(3,1)--(0,0);
			
			\foreach[evaluate={\yy=int(\xx+1)}] \xx in {1,2,3,4} {
				\pgfmathtruncatemacro{\label}{\xx}
				\coordinate[wvert,label=above:$p_{\label}$] (p\xx) at (2*\xx+1,1);
				\coordinate[wvert,label=above:$\widetilde p_{\label}$] (q\xx) at (2*\xx+2,1);
				\coordinate[wvert,label=below:$p_{\label}$] (pp\xx) at (2*\xx-2,0);
				\coordinate[wvert,label=below:$\widetilde p_{\label}$] (qq\xx) at (2*\xx-1,0);
				\node[bvert] (b\xx) at (2*\xx+0.3,0.333*2.3) {};
				\node[bvert] (bb\xx) at (2*\xx+1-0.3,0.333*0.7) {};
			}	
			\node[bvert] (bb0) at (1-0.3,0.333*0.7) {};
			\node[bvert] (b5) at (2*5+0.3,0.333*2.3) {};
			\coordinate[wvert,label=above:${p_5=M(p_1)}$] (p5) at (2*5+1,1);
			\coordinate[wvert,label=below:${p_5=M(p_1)}$] (pp5) at (2*5-2,0);
			\foreach[evaluate={\yy=int(\xx+1)}] \xx in {1,2,3,4} {
				\pgfmathtruncatemacro{\label}{2*\xx+1}
				\pgfmathtruncatemacro{\u}{\xx+1}
				\draw[-,text=red,inner sep=1]
				(p\xx)--(b\xx)--node[left] {$y_{\xx}$}(pp\yy)--(bb\xx)--node[right] {$-x_{\u}$}(p\xx);
			}
			\foreach[evaluate={\yy=int(\xx+1)}] \xx in {1,2,3} {
				\pgfmathtruncatemacro{\label}{2*\xx+1}
				\node[blue](no) at (2*\xx+0.5,0.5) {$X_{\label}$};
			}
			\node[blue](no) at (2*4+0.5,0.5) {$X_{1}$};
			\foreach[evaluate={\yy=int(\xx+1)}] \xx in {1,2,3,4} {
				\pgfmathtruncatemacro{\label}{2*\xx}
				\node[blue](no) at (2*\xx-0.5,0.5) {$X_{\label}$};
			}
			
			\foreach[evaluate={\yy=int(\xx-1)}] \xx in {1,2,3,4} {
				\draw[-] (qq\xx)--(bb\yy)
				;
			}
			\foreach[evaluate={\yy=int(\xx+1)}] \xx in {1,2,3,4} {
				\draw[-] (q\xx)--(b\yy)
				;
			}
		\end{tikzpicture}
		\caption{The graph $\Xi_4$ showing the face weights and the pentagram map TCD. Here $\widetilde p_i:=\overline{p_{i-1}p_{i}} \cap \overline{p_{i+1} p_{i+2}}$.}
		\label{fig:pentagramtcd}
	\end{figure}

	\begin{figure}[ht]
		\def\twd{0.5\textwidth}
		\def\scl{0.48}
		\scalebox{0.960}{
			\begin{tabular}{@{}cc@{}}
				\resizebox{\twd}{!}{\begin{tikzpicture}[scale=\scl]
						\draw[rectangle,white] (-4,-2.5)--(9,4);
						\begin{scope}
							[yscale=1.5]
							\node[wvert,label=left:$\w_{i}$] (p1) at (0,2) {};
							\node[wvert,label=left:$\w_{i}$]
							(pp1) at (-3,-1) {};
							\node[wvert,label=right:$\w_{i+1}$] (q1) at (8,2) {};
							
							\node[wvert,label=right:$\w_{i+1}$] (qq1) at (5,-1) {};

							\node[bvert] (b1) at (3,2) {};
							\node[bvert] (b2) at (2,1) {};
							\node[bvert] (bb2) at (4,1) {};
							\node[bvert] (b3) at (3,0) {};
							\node[bvert] (bb1) at (0,-1) {};
							
							\node[bvert] (bp) at (0,1) {};
							\node[bvert] (bq) at (6,1) {};
							
							\draw[dashed, gray] (p1) -- (pp1) -- (qq1) -- (q1) -- (p1);	
							\node[wvert,label=below:$\w_{i+1}$] (p2) at (1,1) {};
							\node[wvert,label=left:$\w_{i+1}$] (pp2) at (-1,1) {};
							
							\node[wvert,label=left:$\w_{i+2}$] (p3) at (-2,0) {};
							
							\node[wvert,label=above:$\w_{i+2}$] (q2) at (5,1) {};
							\node[wvert,label=right:$\w_{i+2}$] (qq2) at (7,1) {};
							\node[wvert,label=right:$\w_{i+3}$] (q3) at (6,0) {};
							
							\node[wvert,label=below:$\widetilde \w_{i+1}$] (ww) at (3,1) {};
							\draw[-,text=red,inner sep=1]
							(p1) -- (b2) -- (p2)--(b3)
							--(qq1)
							(p3)
							(b1)--(q2)--(bb2)--(ww)--(b2)
							(pp2)--(bp)-- (p2)
							(qq2)--(bq)-- (q2)
							;
							
							\draw[-,green,text=red,inner sep=1] (p3)--(bb1)
							(bb2)--(q3)
							;
							\node[red](no) at (4.4,1.9) {$y_{i+1}$};
							\node[red](no) at (1.4,1.9) {$-x_{i+1}$};
							\node[blue](no) at (2.7,1.4) {$X_{2i+2}$};
							\node[blue](no) at (4.5,0) {$X_{2i+4}$};
							\node[blue](no) at (.5,0) {$X_{2i+3}$};
							\node[blue](no) at (6.5,0.5) {$X_{2i+5}$};
							\node[blue](no) at (-.5,1.5) {$X_{2i+1}$};
						\end{scope}
				\end{tikzpicture}}
				&
				\resizebox{\twd}{!}{ \begin{tikzpicture}[scale=\scl]
						\draw[rectangle,white] (-4,-2.5)--(9,4);
						\begin{scope}[yscale=1.5]
							\node[wvert,label=left:$\w_{i}$] (p1) at (0,2) {};
							\node[wvert,label=left:$\w_{i}$]
							(pp1) at (-3,-1) {};
							\node[wvert,label=right:$\w_{i+1}$] (q1) at (8,2) {};
							
							\node[wvert,label=right:$\w_{i+1}$] (qq1) at (5,-1) {};

							\coordinate[] (b1) at (3,2) {};
							\node[bvert] (b2) at (2,1) {};
							\node[bvert] (bb2) at (4,1) {};
							\node[bvert] (b3) at (3,0) {};
							\coordinate[] (bb1) at (0,-1) {};
							
							\node[bvert] (bp) at (0,1) {};
							\node[bvert] (bq) at (6,1) {};
							
							\draw[dashed, gray] (p1) -- (pp1) -- (qq1) -- (q1) -- (p1);	
							\node[wvert,label=below:$\w_{i+1}$] (p2) at (1,1) {};
							\node[wvert,label=left:$\w_{i+1}$] (pp2) at (-1,1) {};
							
							\coordinate[] (p3) at (-2,0) {};
							
							\node[wvert,label=above:$\w_{i+2}$] (q2) at (5,1) {};
							\node[wvert,label=right:$\w_{i+2}$] (qq2) at (7,1) {};
							\coordinate[] (q3) at (6,0) {};
							
							\node[wvert,label=below:$\widetilde \w_{i+1}$] (ww) at (3,1) {};
							\draw[-,text=red,inner sep=1]
							(p1) -- (b2) -- (p2)
							(p3)--(bb1)
							(b1)--(q2)--(bb2)--(ww)--(b2)
							(bb2)--(q3)
							(pp2)--(bp)-- (p2)
							(qq2)--(bq)-- (q2)
							;
							
							\draw[-,green,text=red,inner sep=1] (p2)--(b3) --(qq1)
							;
							\node[red](no) at (4.4,1.9) {$y_{i+1}$};
							\node[red](no) at (1.4,1.9) {$-x_{i+1}$};
							\node[blue](no) at (2.7,1.4) {$X_{2i+2}$};
							\node[blue](no) at (4.5,0) {$X_{2i+4}$};
							\node[blue](no) at (.5,0) {$X_{2i+3}$};
							\node[blue](no) at (6.5,0.5) {$X_{2i+5}$};
							\node[blue](no) at (-.5,1.5) {$X_{2i+1}$};
						\end{scope}
				\end{tikzpicture}}
				\\
				(1) Contract green edges. & (2) Contract green edges.\\
				\resizebox{\twd}{!}{ \begin{tikzpicture}[scale=\scl]
						\draw[rectangle,white] (-4,-2.5)--(9,4);
						\begin{scope}[yscale=1.5]
							\begin{scope}
								\coordinate[] (p1) at (0,2) {};
								\coordinate[]
								(pp1) at (-3,-1) {};
								\coordinate[] (q1) at (8,2) {};
								
								\coordinate[] (qq1) at (5,-1) {};

								\coordinate[] (b1) at (3,2) {};
								\node[bvert] (b2) at (2,1) {};
								\node[bvert] (bb2) at (4,1) {};
								\coordinate[] (b3) at (3,0) {};
								\coordinate[] (bb1) at (0,-1) {};
								
								\node[bvert] (bp) at (0,1) {};
								\node[bvert] (bq) at (6,1) {};
								
								\draw[dashed, gray] (p1) -- (pp1) -- (qq1) -- (q1) -- (p1);	
								\node[wvert,label=below:$\w_{i+1}$] (p2) at (1,1) {};
								\node[wvert,label=left:$\w_{i+1}$] (pp2) at (-1,1) {};
								
								\coordinate[] (p3) at (-2,0) {};
								
								\node[wvert,label=above:$\w_{i+2}$] (q2) at (5,1) {};
								\node[wvert,label=right:$\w_{i+2}$] (qq2) at (7,1) {};
								\coordinate[] (q3) at (6,0) {};
								
								\node[wvert,label=below:$\widetilde \w_{i+1}$] (ww) at (3,1) {};
								\draw[-,text=red,inner sep=1]
								(p1) -- (b2) -- (p2)--(b3)
								--(qq1)
								(p3)--(bb1)
								(b1)--(q2)--(bb2)--(ww)--(b2)
								(bb2)--(q3)
								;
								\draw[-,green,text=red,inner sep=1](pp2)--(bp)-- (p2)
								(qq2)--(bq)-- (q2)
								;
								
								\node[red](no) at (4.4,1.9) {$y_{i+1}$};
								\node[red](no) at (1.4,1.9) {$-x_{i+1}$};
								\node[blue](no) at (2.7,1.4) {$X_{2i+2}$};
								
								\node[blue](no) at (4.5,0) {$X_{2i+4}$};
								\node[blue](no) at (.5,0) {$X_{2i+3}$};
								\node[blue](no) at (6.5,0.5) {$X_{2i+5}$};
								\node[blue](no) at (-.5,1.5) {$X_{2i+1}$};
							\end{scope}
						\end{scope}
				\end{tikzpicture}}
				&
				\resizebox{\twd}{!}{ \begin{tikzpicture}[scale=\scl]
						\draw[rectangle,white] (-1,-1)--(12,5.5);
						\begin{scope}
							\begin{scope}[xslant=-7/4.5]
								
								\begin{scope}[xscale=2,yscale=4.5/5]
									\begin{scope}[shift={(-1,-1)}]
										\draw[dashed, gray] (1,1)--(5,1)--(10,6)--(6,6)--(1,1);	
										\coordinate[wvert,label=left:$\w_{i+1}$] (w1) at (5,5) {};
										\coordinate[wvert,label=right:$\w_{i+2}$] (w2) at (9,5) {};
										\coordinate[wvert,label=below:$\widetilde \w_{i+1}$] (ww) at (7,5) {};
										
										\coordinate[bvert] (b1) at (6,5) {};
										\coordinate[bvert] (b2) at (8,5) {};
										
										\draw[-,text=red,inner sep=1] (w1)--(b1)--(ww)--(b2)--(w2)
										(w2)--(9,6)
										(b2)--(8,4)
										(w1)--(5,1)
										(4,4)--(4,1)
										(b1)--(6,6)
										;
										
										\node[red](no) at (9.5,6) {$y_{i+1}$};
										\node[red](no) at (6.5,6) {$-x_{i+1}$};
										
										\node[blue](no) at (7.5,6) {$X_{2i+2}$};
										\node[blue](no) at (6,3) {$X_{2i+4}$};
										\node[blue](no) at (4.5,3) {$X_{2i+3}$};
										\node[blue](no) at (8.5,4.5) {$X_{2i+5}$};
										\node[blue](no) at (5.5,5.5) {$X_{2i+1}$};
									\end{scope}
								\end{scope}
							\end{scope}
						\end{scope}
				\end{tikzpicture}}
				\\
				(3) Contract green edges. & (4) Translate.\\
				\resizebox{\twd}{!}{\begin{tikzpicture}[scale=\scl]
						\draw[rectangle,white] (-1,-1)--(12,5.5);
						\begin{scope}[xslant=-7/4.5]
							
							\begin{scope}[xscale=2,yscale=4.5/5]
								\begin{scope}
									\draw[dashed, gray] (0,0)--(4,0)--(9,5)--(5,5)--(0,0);	
									\coordinate[wvert,label=left:$\w_{i+1}$] (w1) at (5,5) {};
									\coordinate[wvert,label=right:$\w_{i+2}$] (w2) at (9,5) {};
									\coordinate[wvert,label=above:$\widetilde \w_{i+1}$] (ww) at (7,5) {};
									
									\coordinate[bvert] (b1) at (6,5) {};
									\coordinate[bvert] (b2) at (8,5) {};
									
									\coordinate[wvert,label=left:$\w_{i+1}$] (w11) at (0,0) {};
									\coordinate[wvert,label=right:$\w_{i+2}$] (w21) at (4,0) {};
									\coordinate[wvert,label=below:$\widetilde \w_{i+1}$] (ww1) at (2,0) {};
									
									\coordinate[bvert] (b11) at (1,0) {};
									\coordinate[bvert] (b21) at (3,0) {};
									
									\draw[-,text=red,inner sep=1] (w1)--(b1)--(ww)--(b2)--(w2)
									(w21)--node[right]{$y_{i+1}$}(4,4)
									(b2)--(8,4)
									(w1)--node[right]{$-x_{i+2}$}(5,1)
									(b11)--(1,1)
									;
									
									\draw[-,text=red,inner sep=1] (w11)--(b11)--(ww1)--(b21)--(w21)
									
									;
									\node[blue](no) at (2.5,1.5) {$X_{2i+2}$};
									\node[blue](no) at (6.5,4) {$X_{2i+4}$};
									\node[blue](no) at (4.5,3) {$X_{2i+3}$};
									\node[blue](no) at (8.5,4.5) {$X_{2i+5}$};
									\node[blue](no) at (.5,.5) {$X_{2i+1}$};
								\end{scope}
							\end{scope}
						\end{scope}
				\end{tikzpicture}}
				&
				\resizebox{\twd}{!}{ \begin{tikzpicture}[scale=0.5]
						\draw[rectangle,white] (-1,-1)--(12,5.5);
						\begin{scope}[xslant=3/4.5]
							\begin{scope}[xscale=4,yscale=4.5/4]
								\begin{scope}
									\draw[dashed, gray] (0,0)--(2,0)--(2,4)--(0,4)--(0,0);	
									\coordinate[wvert,label=left:$\w_{i+1}$] (w1) at (0,4) {};
									\coordinate[wvert,label=right:$\w_{i+2}$] (w2) at (2,4) {};
									\coordinate[wvert,label=above:$\widetilde \w_{i+1}$] (ww) at (1,4) {};
									
									\coordinate[bvert] (b2) at (2,3) {};
									\coordinate[bvert] (b22) at (0,3) {};
									
									\coordinate[wvert,label=left:$\w_{i+1}$] (w11) at (0,0) {};
									\coordinate[wvert,label=right:$\w_{i+2}$] (w21) at (2,0) {};
									\coordinate[wvert,label=below:$\widetilde \w_{i+1}$] (ww1) at (1,0) {};
									
									\coordinate[bvert] (b11) at (0,1) {};
									\coordinate[bvert] (b12) at (2,1) {};
									
									\draw[-,text=red,inner sep=1](ww)--(b2)--(w2)
									(w21)--node[right]{$y_{i+1}$}(b22)
									(w1)--node[right]{$-{x_{i+2}}{}$}(b12)
									(b22)--(w1)
									(b12)--(w21)
									;
									\draw[-,text=red,inner sep=1] (w11)--(b11)--(ww1)
									;
									\node[blue](no) at (.5,1.5) {$X_{2i+2}$};
									\node[blue](no) at (1.,3.3) {$X_{2i+4}$};
									\node[blue](no) at (0.4,3) {$X_{2i+3}$};
									
								\end{scope}
							\end{scope}
						\end{scope}
				\end{tikzpicture}}
				\\
				(5) Isotope black vertices. & (6) The graph in Figure~\ref{fig:pentagramtcd}.
			\end{tabular}
		}
		\caption{\label{fig:pentseq} Sequence of steps in the transformation of the building block graph $H_i$ from Figure~\ref{fig:pentagram} into the graph in Figure~\ref{fig:pentagramtcd}.}
	\end{figure}
	
	\begin{figure}[ht]
		\centering
		\begin{tikzpicture}[yscale=0.5]
			\coordinate[lvert,label=below:${(3,0)}$] (no) at (3,0);
			\coordinate[lvert,label=left:${(1,n)}$] (no) at (1,4);
			\coordinate[lvert,label=left:${(2,\frac n 2)}$] (no) at (2,2);
			
			\coordinate[lvert,label=above:${(0,2n)}$] (no) at (0,8);
			\coordinate[lvert,label=right:${(2,n)}$] (no) at (2,4);
			\coordinate[lvert,label=right:${(1,\frac{3n}{2})}$] (no) at (1,6);
			\draw[-] (3,0)--(2,4)--(0,8)--(1,4)--(3,0);
		\end{tikzpicture}\hspace{2cm}
		\begin{tikzpicture}[yscale=0.4]
			\coordinate[lvert,label=below:${(3,0)}$] (no) at (3,0);
			\coordinate[lvert,label=left:${(1,n)}$] (no) at (1,5);
			
			\coordinate[lvert,label=above:${(0,2n)}$] (no) at (0,10);
			\coordinate[lvert,label=right:${(2,n)}$] (no) at (2,5);
			
			\draw[-] (3,0)--(2,5)--(0,10)--(1,5)--(3,0);
			
		\end{tikzpicture}	
		\caption{The Newton polygon $N_{\Xi_n}$ of $\Xi_n$ for even $n$ (left) and odd $n$ (right).}\label{fig:nppent}
	\end{figure}

	\begin{figure}[ht]
		\centering
		\begin{tikzpicture}[scale=0.4]
			\draw[rectangle,white] (-1,-1)--(12,5.5);
			\begin{scope}[xslant=3/4.5]
				\begin{scope}[xscale=4,yscale=4.5/4]
					\begin{scope}
						\draw[dashed, gray] (0,0)--(2,0)--(2,4)--(0,4)--(0,0);	
						\coordinate[wvert,label=left:$\w_{i-1}$] (w1) at (0,4) {};
						\coordinate[wvert,label=right:$\w_{i}$] (w2) at (2,4) {};
						\coordinate[wvert,label=above:$\widetilde \w_{i-1}$] (ww) at (1,4) {};
						
						\coordinate[bvert] (b2) at (2,3) {};
						\coordinate[bvert] (b22) at (0,3) {};
						
						\coordinate[wvert,label=left:$\w_{i-1}$] (w11) at (0,0) {};
						\coordinate[wvert,label=right:$\w_{i}$] (w21) at (2,0) {};
						\coordinate[wvert,label=below:$\widetilde \w_{i-1}$] (ww1) at (1,0) {};
						
						\coordinate[bvert] (b11) at (0,1) {};
						\coordinate[bvert] (b12) at (2,1) {};
						
						\draw[-,text=red,inner sep=1](ww)--(b2)--(w2)
						(w21)--node[below]{$y_{i-1}$}(b22)
						(w1)--node[right]{$-{x_{i}}{}$}(b12)
						(b22)--(w1)
						(b12)--(w21)
						;
						\draw[-,text=red,inner sep=1] (w11)--(b11)--(ww1)
						;
						\def\lw{1}
						
						\draw[-latex, line width=\lw,blue] (w11)--(b11);
						\draw[-latex, line width=\lw,blue] (b11)--(ww1);
						\draw[-latex, line width=\lw,blue] (ww)--(b2);
						\draw[-latex, line width=\lw,blue] (b2)--(w2);
						\draw[-latex, line width=\lw,blue] (w21)--(b12);
						\draw[-latex, line width=\lw,blue] (b12)--(w1);

					\end{scope}
				\end{scope}
			\end{scope}
		\end{tikzpicture} \hspace{10mm}
		\begin{tikzpicture}[scale=0.4]
			\draw[rectangle,white] (-1,-1)--(12,5.5);
			\begin{scope}[xslant=3/4.5]
				\begin{scope}[xscale=4,yscale=4.5/4]
					\begin{scope}
						\draw[dashed, gray] (0,0)--(2,0)--(2,4)--(0,4)--(0,0);	
						\coordinate[wvert,label=left:$\w_{i}$] (w1) at (0,4) {};
						\coordinate[wvert,label=right:$\w_{i+1}$] (w2) at (2,4) {};
						\coordinate[wvert,label=above:$\widetilde \w_{i}$] (ww) at (1,4) {};
						
						\coordinate[bvert] (b2) at (2,3) {};
						\coordinate[bvert] (b22) at (0,3) {};
						
						\coordinate[wvert,label=left:$\w_{i}$] (w11) at (0,0) {};
						\coordinate[wvert,label=right:$\w_{i+1}$] (w21) at (2,0) {};
						\coordinate[wvert,label=below:$\widetilde \w_{i}$] (ww1) at (1,0) {};
						
						\coordinate[bvert] (b11) at (0,1) {};
						\coordinate[bvert] (b12) at (2,1) {};
						
						\draw[-,text=red,inner sep=1](ww)--(b2)--(w2)
						(w21)--node[below]{$y_{i}$}(b22)
						(w1)--node[right]{$-{x_{i+1}}{}$}(b12)
						(b22)--(w1)
						(b12)--(w21)
						;
						\draw[-,text=red,inner sep=1] (w11)--(b11)--(ww1)
						;
						\def\lw{1}
						
						\draw[-latex, line width=\lw,blue] (w1)--(b22);
						\draw[-latex, line width=\lw,blue] (b22)--(w21);
						\draw[-latex, line width=\lw,blue] (w2)--(b2);
						\draw[-latex, line width=\lw,blue] (b2)--(ww);
						\draw[-latex, line width=\lw,blue] (ww1)--(b11);
						\draw[-latex, line width=\lw,blue] (b11)--(w11);

					\end{scope}
				\end{scope}
			\end{scope}
		\end{tikzpicture}
		\caption{The cycles $\sigma_i$ (left) and $\rho_i$ (right) (compare with Figure~\ref{fig:pentagramtcd} and Figure~\ref{fig:pentseq}\,(6)). }
		\label{fig:pentpaths}
	\end{figure}
	
	Consider the cycles $\sigma_i$, $\rho_i$, $i=1,\dots,n$, shown in Figure~\ref{fig:pentpaths}. Let $\xi$ denote the zig-zag path in $\Xi_n$ with homology $(1,-n)$. Let $\lambda \in \C^\times$ and let \smash{$\mathcal L_{\Xi_n}^\lambda$} \big(resp.\ \smash{$\mathcal X_{\Xi_n}^\lambda$}\big)
denote the level set of $\mathcal L_{\Xi_n}$ (resp.\ \smash{$\mathcal X_{\Xi_n}$}) where $\chi_{[\xi]}([\wt]) = \lambda$. Since $\chi_{[\xi]}$ is a Casimir, these level sets are Poisson subvarieties. The choice of $\lambda$ is unimportant: it is an extra Casimir in the GK integrable system compared to the pentagram map and does not affect the twisted TCD map below.

	The homology classes $[\sigma_i]$, $[\rho_i]$, $[\xi]$ freely generate $H_1(\Xi_n,\Z)$, so the coordinate ring
	\[
	\mathcal O_{\mathcal L_{\Xi_n}}^{\lambda}=\C\big[\chi_{[\sigma_i]}^{\pm 1}, \chi_{[\rho_i]}^{\pm 1}\big].
	\]
	Define the birational map \smash{$\pi_n\colon \mathcal X_{N(\Xi_n)}^\lambda \supset \mathcal L_{\Xi_n}^\lambda \ra (\C \setminus \{0,1\})^{2n}$} by
	\[
	\pi_n^*x_i := -\chi_{[\sigma_i]}, \qquad \pi_n^* y_i := \chi_{[\rho_i]},
	\]
	for all $i \in \{1,2,\dots,n\}$.

	\begin{Proposition}
		The map $\pi_n$ is Poisson.
	\end{Proposition}
	\begin{proof}
		The only nonzero intersection numbers are
		\begin{align*}
			\epsilon_{\Xi_n}([\sigma_i],[\sigma_{i + 1}])= 1, \qquad \epsilon_{\Xi_n}([\rho_i],[\rho_{i + 1}])= -1,
		\end{align*}
		where the index $i$ is cyclic, so $\sigma_{n+1}=\sigma_1$.
		Comparing with \eqref{pent:pb}, we see that $\pi_n$ is Poisson (up to an irrelevant global sign which can be made to match by changing the choice of orientation of the conjugated surface).
	\end{proof}

	Let $(p,M) \in \mathcal T_n$ and let \smash{$[\wt] \in \mathcal X_{N(\Xi_n)}^\lambda$} such that $(x,y)^{-1} \circ \pi_n([\wt])=(p,M)$. We choose edge weights and Kasteleyn signs representing $[\wt]$ as in Figure~\ref{fig:pentagram}. Let $\Xi_{n,\A}$ denote the balanced cylinder graph obtained by concatenating the graphs $H_i$ for $i=1,2,\dots,n$ without closing up cyclically.	We have $|W(\Xi_{n,\A})|-|B(\Xi_{n,\A})|=3$. Let \smash{$P\colon W\big(\Xi_{n,\widehat \A}\big) \ra \CP^2$} denote the twisted TCD map associated to $[\wt]$. We label the vertices of $\Xi_{n,\A}$ as in Figure~\ref{fig:pentagram}.
	
	\begin{Proposition} \label{prop:pentagramtcd}
		We have
		\begin{align*}
			P_{\w_i}=p_i, \qquad P_{\widetilde{\w}_i}=\overline{p_{i-1} p_{i}}\cap \overline{p_{i+1} p_{i+2}},
		\end{align*}
		for all $i \in \Z$ modulo the action of $\PGL_3$.
	\end{Proposition}
	\begin{proof}
		Since $P$ is a TCD map, the three white vertices incident to a trivalent black vertex are in a line. Therefore, we have \smash{$P_{\widetilde{\w}_{i+1}}=\overline{P_{\w_i} P_{\w_{i+1}}}\cap \overline{P_{\w_{i+2}} P_{\w_{i+3}}}$}, and so it suffices to show that $P_{\w_i}=p_i$ for all $i \in \Z$. The points $p_i$ are determined modulo $\operatorname{PGL}_3$ by their $x_i$ and $y_i$ coordinates, so it suffices to show that the points $P_{\w_i}$ have the same $x_i$ and $y_i$ coordinates. Since $[\wt]$ is defined so that $(x,y)^{-1} \circ \pi_n([\wt])=(p,M)$, we have $\chi_{[\sigma_i]}([\wt])=-x_i$. On the other hand,
		\[
		\chi_{[\sigma_i]}([\wt])=\mr\big(P_{\w_{i-1}},P_{\w_{i-2}},P_{\widetilde \w_{i-1}},P_{\w_{i+1}},P_{\w_{i}}, P_{\widetilde \w_{i}}\big),
		\]
		so it suffices to show that
		\begin{gather}
			-\mr\big(P_{\w_{i-1}},P_{\w_{i-2}},P_{\widetilde \w_{i-1}},P_{\w_{i+1}},P_{\w_{i}}, P_{\widetilde \w_{i}}\big) \nonumber \\
			\qquad= \cro\big(P_{\w_{i-2}},P_{\w_{i-1}},\overline{P_{\w_{i-2}} P_{\w_{i-1}}}\cap \overline{P_{\w_{i+1}} P_{\w_{i+2}}},P_{\widetilde \w_{i-1}}\big) \nonumber \\
			\qquad=\cro\big(P_{\w_{i-1}},P_{\w_{i-2}},P_{\widetilde \w_{i-1}},\overline{P_{\w_{i-2}} P_{\w_{i-1}}}\cap \overline{P_{\w_{i+1}} P_{\w_{i+2}}}\big),\la{eq:crpen}
		\end{gather}
		where the cross-ratio on the right hand side of the first line is the definition of $x_i$, and the second equality is from reordering the terms in the cross-ratio. Expanding both sides of \eqref{eq:crpen} and canceling common terms, we see that \eqref{eq:crpen} is equivalent to
		\[
		\mr\big(P_{\w_{i-1}},\overline{P_{\w_{i-2}} P_{\w_{i-1}}}\cap \overline{P_{\w_{i+1}} P_{\w_{i+2}}},P_{\widetilde \w_{i-1}},P_{\w_{i+1}},P_{\w_{i}}, P_{\widetilde \w_{i}}\big)=-1,
		\]
		which is Menelaus' theorem applied to the quadrilateral \smash{$P_{\widetilde \w_{i}}P_{\w_{i-1}}P_{\widetilde \w_{i-1}}P_{\w_{i+1}}$}, whose opposite sides intersect at the points $P_{\w_{i}}$ and \smash{$\overline{P_{\w_{i-2}} P_{\w_{i-1}}}\cap \overline{P_{\w_{i+1}} P_{\w_{i+2}}}$}. The proof for $y_i$ is similar.
	\end{proof}

	We call the twisted TCD map $P$ the \textit{pentagram TCD map}. It was studied in \cite[Section 3.1]{AGPR}.
	\begin{Corollary}
		The monodromy matrix of the twisted TCD map $P$ coincides with the monodromy matrix $M$ of the twisted $n$-gon $(p,M)$.
	\end{Corollary}

	Now we compute this monodromy matrix. The Kasteleyn matrix of $H_i$, with rows and columns indexed as in \eqref{k4}, is
	\[
	\aboverulesep=0pt \belowrulesep=0pt
	K_{H_i}(w)=\begin{blockarray}{ccc|cccc}
		\begin{block}{[ccc|ccc]c}
			1&0&0&0&0&0 &{\rm w}_{i+1}\\
			0&1&0&0&0&0 &{\rm w}_{i+2}\\
			0&0&1&0&0&0 &{\rm w}_{i+3}\\
			\cmidrule(lr){1-6}
			0&0&0&0&x_{i+1}w&0 &{\rm w}_{i}\\
			0&0&0&0&0&1 & \w_{i+1}\\
			0&0&0&-1&0&0 & \w_{i+2}\\
			\cmidrule(lr){1-6}
			-1&0&0&0&1&-1&\w_{i+1}\\
			0&-1&1&\frac{y_{i+1}}{w}&0&0&\w_{i+2}\\
			0&0&-1&0&1&0&\widetilde{\w}_{i+1}\\
		\end{block}
	\end{blockarray},
	\]
	using which we get
	\[
	\Pi_{H_i}(w)=\begin{bmatrix}
		0&0&x_{i+1} w\\
		-1&0&1\\
		0&-\frac{w}{y_{i+1}}&\frac{w}{y_{i+1}}
	\end{bmatrix}.
	\]
	We have \smash{$\det \Pi_{H_i}(w)=\frac{x_{i+1}}{y_{i+1}} w^2$}. The monodromy matrix \smash{$M(w)=(-1)^{n}\prod_{i=1}^n \Pi_{H_i}(w)$}, so that
 \be \la{det:pentagram}
	\det M(w)= (-1)^{n}\frac{O_n}{E_n} w^{2n}. \ee
	The following lemma is elementary.
	\begin{Lemma}\la{lem:3x3}
		For a $3 \times 3$ invertible matrix $M$, we have
		\[
		\det(zI-M) = z^3 - \tr Mz^2+\frac{\tr M^{-1}}{\det M^{-1}} z-\det M.
		\]
	\end{Lemma}
	Applying Lemma \ref{lem:3x3} to $M(w)$, multiplying by $E_n$, and using equations \eqref{schinv} and \eqref{det:pentagram}, we~get
	\begin{align*}
		E_nP(z,w)&=E_n z^3-E_n \tr M(w)z^2+E_n\frac{\tr M(w)^{-1}}{\det M(w)^{-1}} z-E_n\det M(w)\\
		&=E_n z^3 \pm \left(1+ \sum_{k=1}^{\floor{\frac n 2}} E_k w^{-k}\right) z^2 w^n \pm
		\left(1+ \sum_{k=1}^{\floor{\frac n 2}} O_k w^k\right)z w^n \pm O_n w^{2n},
	\end{align*}
	where the signs are irrelevant for our purposes. Since the GK Hamiltonians are coefficients of~$P(z,w)$, we have proved the following theorem.
	
	\begin{Theorem}\la{mainthm:pentagram}
		The map $\pi_n$ identifies the GK Poisson structure, Hamiltonians and Casimirs with those of the pentagram map in Theorem~$\ref{thm:ost}$.
	\end{Theorem}
	\begin{Remark}
		Glick \cite{Glick} discovered a cluster algebra structure for the pentagram map, where the cluster variables are given by
		\[
		X_{2i}=-\frac{1}{x_i y_i}, \qquad X_{2i+1}= -x_{i+1} y_{i}.
		\]
		On the other hand, the GK cluster variables are the face weights. Since the cycles $-(\sigma_i+\rho_i)$ and $\sigma_{i+1}+\rho_i$ are the faces of $\Xi_n$, we see that Glick's cluster variables coincide with those of GK.
	\end{Remark}
	
	\section{The cross-ratio dynamics integrable system}
	\label{sec:crdynamics}
	
	In this section, we define cross-ratio dynamics on the space of $\vec{\alpha}$-related pairs of twisted polygons and we recall the integrability results of \cite{AFIT} for that dynamics, generalizing them to the case when $\alpha_i$ may depend on $i$. All the results that we attribute to \cite{AFIT} in this section were stated and proven by them for constant $\alpha$, but the proofs carry over easily to the case when $\alpha_i$ depends~on~$i$.
	
	\subsection[Poisson varieties of twisted polygons in C P\^{}1]{Poisson varieties of twisted polygons in $\boldsymbol{\C \P^1}$}
	
	Let $n\geq2$. A twisted $n$-gon $(p,M)$ in dimension $1$ is called \emph{nondegenerate} if for all $i \in \Z$, we have $p_i \notin \{p_{i+1},p_{i+2}\}$. Let $\widetilde {\mathcal P}_n$ denote the space of nondegenerate twisted $n$-gons. \smash{$\widetilde {\mathcal P}_n$} is an open subvariety of $\big(\CP^1\big)^n \times \PGL_2$, and $\PGL_2$ acts on it by \eqref{eq:pglaction}. The quotient \smash{$\mathcal P_n:=\widetilde{\mathcal P}_n / \PGL_2$} is the moduli space parameterizing projective equivalence classes of nondegenerate twisted $n$-gons. We will sometimes abuse notation and simply write \smash{$p \in \widetilde P_n$} instead of \smash{$(p,M) \in \widetilde P_n$}. Given \smash{$p \in \widetilde{\mathcal P}_n$}, we define the $c$-variables
	\begin{align}\la{def:c}
		c_i := \cro(p_{i-1},p_i,p_{i+2},p_{i+1}) \qquad \text{for all}\ i \in \Z.
	\end{align}
	
	Notice that $c_{i+n}=c_i$ for all $i \in \Z$. Since $p$ is nondegenerate, $c_i(p) \notin \{0,\infty \}$, so the $c$-variables define a morphism
	\begin{align}\la{def:cmor}
		c\colon\ \widetilde{\mathcal P}_n \ra (\C^\times)^n, \qquad
		p \mapsto (c_1,c_2,\dots,c_n).
	\end{align}
	Since each $c_i$ is a cross-ratio, this morphism is $\PGL_2$-invariant, and therefore, \eqref{def:cmor} descends to a morphism $c\colon \mathcal P_n \ra (\C^\times)^n$. Given the $c$-variables and three initial points $p_1$, $p_2$, $p_3$, the~whole polygon is recovered from \eqref{def:c}. Since any three points can be mapped to any other three points by a projective transformation, the $c$-variables characterize a polygon up to projective transformations and so $c$ is an isomorphism. The inverse morphism is given explicitly in~\mbox{\cite[Section~3.2]{AFIT}}.
	
	The following lemma gives an explicit representative for the $\PGL_2$ conjugacy class of the monodromy matrix $M$ in terms of the $c$-variables.
	\begin{Theorem}[\protect{\cite[Lemma 3.2]{AFIT}}] 
		The matrix
		\[
		M=\begin{bmatrix}
			0 & c_1 \\
			-1 & 1
		\end{bmatrix} \begin{bmatrix}
			0 & c_2 \\
			-1 & 1
		\end{bmatrix} \cdots \begin{bmatrix}
			0 & c_n \\
			-1 & 1
		\end{bmatrix}
		\]
		represents the monodromy matrix.
	\end{Theorem}
	
Let $\vec \alpha = (\alpha_i)_{i \in \Z}$ with $\alpha_i \in \C \setminus \{0,1\}$ such that $\alpha_{i+n}=\alpha_i$.
	Two twisted polygons $p, q \in \widetilde P_n$ are said to be \emph{$\vec \alpha$-related}, and denoted \smash{$p \,\raisebox{-0.5pt}{$\arv$}\, q$} if
 \be \la{def:alpha}
	\cro(p_i,q_i,p_{i+1},q_{i+1}) = \alpha_i \qquad \text{for all $i \in \Z$},
	\ee
	and $p$ and $q$ have the same monodromy. Note that condition~\eqref{def:alpha} alone does not imply that $p$ and $q$ have the same monodromy. The relation \smash{\raisebox{-0.4pt}{$\arv$}} is $\PGL_2$-invariant, and therefore, descends to a relation on~$\mathcal P_n$.
	
	Let $\widetilde{\mathcal U}_{n,\vec\alpha}$ denote the space of pairs of $\vec\alpha$-related nondegenerate twisted polygons of length~$n$. It is a~subvariety of \smash{$(\P^1)^{2n} \times \PGL_2$} and $\PGL_2$ acts on it by
	\[
	A \cdot (p_1,\dots,p_n,q_1,\dots,q_n,M) =\big(A(p_1),\dots,A(p_n),A(q_1),\dots,A(q_n),AMA^{-1}\big).
	\]
	
	We define the $u$-variables by
	\begin{align} \la{def:u}
		\qquad u_i := - \cro(p_i,p_{i+1},q_i,p_{i-1}) \qquad\text{for all } i \in \Z.
	\end{align}
	Since $p$ and $q$ are nondegenerate and $\vec\alpha$-related, $u_i \notin \{0,-1,\infty\}$. Therefore, the $u$-variables define a $\PGL_2$-invariant morphism \smash{$u\colon \widetilde{\mathcal U}_{n,\vec\alpha} \ra (\C \setminus \{0,-1\})^n$} which descends to a morphism
	\[
	u\colon \ \mathcal U_{n,\vec\alpha} \ra (\C \setminus \{0,-1\})^n,
	\]
	where \smash{$\mathcal U_{n,\vec\alpha} := \widetilde{\mathcal U}_{n,\vec\alpha}/ \PGL_2$}. When there is no ambiguity on the choice of $\vec\alpha$, we will denote \smash{$\widetilde{\mathcal U}_{n,\vec\alpha}$} and $\mathcal U_{n,\vec\alpha}$ simply by \smash{$\widetilde{\mathcal U}_{n}$} and $\mathcal U_{n}$. Consider the morphism
	\begin{align*}
		\rho_{\vec \alpha} \colon\ \mathcal U_n &\ra \mathcal P_n, \qquad
		(p,q,M)  \mapsto (p,M).
	\end{align*}

	It follows from \cite[Section 4.8]{AFIT} that the diagram
	\[
	\begin{tikzcd}
		\mathcal U_n \arrow[r, "\rho_{\vec \alpha}"] \arrow[d, "u"]
		& \mathcal P_n \arrow[d, "c"] \\
		(\C\setminus\{0,-1\})^n \arrow[r,"\Lambda_{\vec{\alpha}}"]
		& (\C^\times)^n
	\end{tikzcd}
	\]
	commutes, where the map $\Lambda_{\vec{\alpha}}$ is determined by
	\[
		c_i = \frac{\alpha_i}{(1+u_i)\big(1+\frac 1 {u_{i+1}}\big)}.
	\]
	If we are given the $u$-variables, we can recover $p$ up to projective transformations as $c^{-1} \circ \Lambda_{\vec \alpha}(u_1,\dots,u_n)$, and $q$ is then determined by \eqref{def:u}. Therefore, the morphism $u$ induces an isomorphism between $\mathcal U_n$ and $(\C \setminus \{0,-1\})^n$. The set $\mathcal U_n$ is the moduli space parameterizing projective equivalence classes of pairs of $\vec \alpha$-related twisted $n$-gons. The Poisson bracket
	\begin{align}
		\{u_i,u_{i+1}\}_{\mathcal U_n}:=u_i u_{i+1}, \label{eq:upoisson}
	\end{align}
	makes $(\mathcal U_n,\{\cdot, \cdot\}_{\mathcal U_n})$ a Poisson variety while the Poisson bracket
	\[
	\{c_i,c_{i+1}\}_{\vec \alpha} := c_i c_{i+1} \biggl(1-\frac{c_i}{\alpha_i}-\frac{c_{i+1}}{\alpha_{i+1}}\biggr), \qquad
	\{c_i,c_{i+2}\}_{\vec{\alpha}} := -\frac{1}{\alpha_{i+1}}c_i c_{i+1} c_{i+2},
	\]
	makes $(\mathcal P_n,\{\cdot,\cdot\}_{\vec{\alpha}})$ a Poisson variety. For both Poisson brackets, we only give the nonzero values obtained by pairing two coordinate functions. We also define the rescaled coordinates $\overline{c}_i:=\frac{c_i}{\alpha_i}$. In these coordinates, the Poisson bracket $\{\cdot,\cdot\}_{\vec \alpha}$ takes the simpler form
	\[
	\{\overline c_i, \overline c_{i+1}\}_{\vec \alpha} = \overline c_i \overline c_{i+1} \left(1-\overline{c}_i-\overline{c}_{i+1}\right), \qquad
	\{\overline c_i,\overline c_{i+2}\}_{\vec \alpha} = -\overline c_i \overline c_{i+1} \overline c_{i+2}.
	\]

	A computation similar to \cite[Lemma 4.9]{AFIT} shows that $\rho_{\vec \alpha}$ is Poisson. \cite[Corollary 2.7]{AFIT} shows that $\rho_{\vec \alpha}$ is generically finite of degree $2$, that is, for a generic polygon $q \in \mathcal P_n$ there are two polygons $p,r \in \mathcal P_n$ that are ${\vec \alpha}$-related to $q$. Therefore, the maps $C^1_{\vec \alpha}\colon (p,q,M) \mapsto (r,q,M)$ and $C^2_{\vec \alpha}\colon (q,p,M) \mapsto (q,r,M)$ are birational involutions of $\mathcal U_n$, respectively changing the first and the second curve of a pair of $\vec \alpha$-related curves. The space $\mathcal U_n$ also has another involution $j\colon (p,q,M) \mapsto (q,p,M)$ given in coordinates by
\be \la{eq:uv}
	u_i \mapsto v_i:= \frac{1}{u_i} \frac{\alpha_i-1}{\alpha_{i-1}-1}= -\cro(q_i,q_{i+1},p_i,q_{i-1}).
\ee
	
	\subsection{Cross-ratio dynamics and integrability}
	
	Suppose $(q,M) \in \mathcal P_n$ such that $\bigl|\rho_{\vec \alpha}^{-1}(q,M)\bigr|=2$, and suppose $p$, $r$ are the two different polygons $\vec \alpha$-related to $q$. The birational automorphism
	\begin{align*}
		\nu_{\vec \alpha}=j\circ C^1_{\vec \alpha}\colon \ \mathcal U_n \ra \mathcal U_n,\qquad
		(p,q,M) \mapsto (q,r,M)
	\end{align*}
	is called \emph{cross-ratio dynamics}. It can be described in coordinates as follows.
	
	\begin{Proposition}
		\label{prop:uevolution}
		Let $(p,q,M)\in \mathcal U_n$ and denote respectively by $u_i$ and $u'_i$ the coordinates of $(p,q,M)$ and $(q,r,M)=\nu_{\vec \alpha}(p,q,M)$. Then, we have
		\begin{equation}
			\label{eq:ILP}
			u'_i=\frac{\sum\limits_{t=0}^{n-1}\prod\limits_{s=0}^{t-1}v_{i+s}}{\sum\limits_{t=1}^{n}\prod\limits_{s=1}^{t}v_{i+s}},
		\end{equation}
		where the $v_i$ are expressed in terms of the $u_i$ as in formula~\eqref{eq:uv}.
	\end{Proposition}
	
	Proposition~\ref{prop:uevolution} is new compared to \cite{AFIT} and is proved in Section~\ref{subsec:localmovestcd} using geometric $R$-matrices.
	
	For $I \subset [n]$, let $c_I:=\prod_{i \in I} {c_i}$. Similarly, we define $\overline{c}_I$, $\alpha_I$, $u_I$ etc. Let $c_{{\rm even}}:=c_2 c_4 \cdots $
and $c_{{\rm odd}}:=c_1 c_3 \cdots $ denote the product of the even and odd $c$ variables, respectively. In the same way, we define $\alpha_{{\rm even}}$, etc.
	
	A subset $I \subset [i,j]$ is said to be \emph{cyclically sparse} if it contains no pair of consecutive indices where the indices are taken periodic modulo $n$. Define
	\begin{align*}
		&F_k(c):= \sum_{I \text{ cyclically sparse}: \, |I|=k} c_I \qquad\text{for $k=0,\dots, \biggl\lfloor{\frac n 2}\biggr\rfloor$},\\
		&E_{\vec \alpha}:= \frac{1}{\overline c_{[n]}} \left( \sum_{k=0}^{\floor{\frac n 2}} (-1)^k F_k (\overline{c}
		) \right)^2.
	\end{align*}
If $\alpha_i = \alpha$ for all $i$, then we say that $p$ and $q$ are \emph{$\alpha$-related}. In this case, we replace $\vec{\alpha}$ with $\alpha$ and write \smash{$p \stackrel{\alpha}{\sim} q$}, $\{\cdot,\cdot\}_\alpha$, $\rho_\alpha$ etc. This is the setting of \cite{AFIT}, but we defined everything in the more general setting of nonconstant $\vec \alpha$ since it is the natural setting from the point of view of the corresponding GK integrable system.
	
\begin{Theorem}[\protect{\cite[Main Theorem 1]{AFIT}}] 
		Let $\alpha_i=\alpha$ for all $i \in \Z$. We have
		\begin{enumerate}\itemsep=0pt
			\item[$1.$] For $n$ even, $\{\cdot,\cdot \}_{ \alpha}$ has corank $2$ and the subalgebra of Casimirs is generated by $E_{\alpha}$ and~\smash{$\frac{c_{{\rm even}}}{c_{{\rm odd}}}$}.
			\item[$2.$] For $n$ odd, $\{\cdot,\cdot \}_{ \alpha}$ has corank $1$ and the Casimir is $E_{\alpha}$.
			\item[$3.$] For $k=1,2,\dots, \floor{\frac{n+1}{2}} -1$, the functions \smash{$\frac{F_k(c)^2}{c_{[n]}}$} mutually commute and form a maximal set of functionally independent Hamiltonians, making the Poisson variety $\mathcal P_n$ a Liouville integrable system.
		\end{enumerate}
		Moreover, cross-ratio dynamics is discrete integrable in the following sense:
		\begin{enumerate}\itemsep=0pt
			\item[$4.$] $\nu_{\alpha}$ is Poisson.
			\item[$5.$] The pullbacks of the Hamiltonians and the Casimirs to $\mathcal U_n$ by $\rho_{\alpha}$ are invariant under $\nu_{\alpha}$.
		\end{enumerate}
	\end{Theorem}
	
	The following theorem shows that the Hamiltonians can be obtained from the monodromy matrix.
	\begin{Theorem}[\protect{\cite[Theorem 1]{AFIT}}] \la{thm:afit2}
		We have
		\[
		\frac{1}{\det M}\tr^2 M= \frac{1}{c_{[n]}}\left(\sum_{k=0}^{\floor{\frac n 2}} (-1)^k F_k(c)\right)^2.
		\]
	\end{Theorem}
	
	\begin{Remark}
		Note that $\tr M$ is not $\PGL_2$-invariant, but the normalized trace \smash{$\frac{\tr M}{{\sqrt{\det M}}}$} is. However, the normalized trace is not a regular function on $\mathcal P_n$, so we need to square everything to make it so.
	\end{Remark}
	
To get the Hamiltonians from Theorem \ref{thm:afit2}, notice that $F_k(c)$ is the homogeneous degree~$k$ component of \smash{\raisebox{1pt}{$\sqrt{c_{[n]}} \frac{\tr M}{\sqrt{\det M}}$}} up to a sign. This observation will be very useful to prove the coincidence with the AFIT Hamiltonians of the GK Hamiltonians associated to the TCD maps defined in the next section.

\section{TCD maps for cross-ratio dynamics} \label{sec:crtcd12}
In this section, we give two different constructions realizing pairs $(p,q)$ of $\vec\alpha$-related curves as twisted TCD maps on the cylinder graph {$\Gamma_{n,\widehat\A}$}. We provide a detailed account in Section~\ref{sec:hextcd} of a TCD map on a hexagonal lattice. In Section~\ref{subsec:localmovestcd}, we show that cross-ratio dynamics, that is the map $\nu_{\vec \alpha}\colon (p,q,M) \mapsto (q,r,M)$, is identified with a certain semi-local move on the TCD map side. This leads to a proof of Proposition~\ref{prop:uevolution} on the evolution of the $u$ coordinates under cross-ratio dynamics. Finally, in Section~\ref{sec:sqtcd}, we give a second TCD map on a square lattice and describe a semi-local move realizing cross-ratio dynamics.
	
\subsection{Hexagonal TCD map} \label{sec:hextcd}
	\begin{figure}[ht]
		\centering
		\begin{tikzpicture}[scale=1.2,yscale=1.8]
			
			\node[wvert,label=below:${\rm w}_i$] (p1p) at (2,0) {};
			\node[wvert,label=right:${\rm w}_{i+1}$] (p2) at (4,0) {};

			\node[wvert,label=above:${\rm w}_i$] (pp1p) at (4,1.73) {};
			\node[wvert,label=right:${\rm w}_{i+1}$] (pp2) at (6,1.73) {};
			\node[wvert,label=above:${\rm w}_i'$] (q1p) at (3,0.87) {};
			\node[wvert,label=right:${\rm w}_{i+1}'$] (q3) at (5,0.87) {};

			\node[wvert,label=left:${\rm w}_i$] (p1) at (0,0) {};	
			\node[wvert,label=left:${\rm w}_i'$] (q1) at (1,0.87) {};	
			\node[wvert,label=left:${\rm w}_i$] (pp1) at (2,1.73) {};
			
			\node[bvert] (b1p) at (1,0) {};	
			\node[bvert] (b2p) at (2,0.87) {};	
			\node[bvert] (b3p) at (3,1.73) {};
			
			\node[bvert,label=right:${\rm b}_i$] (b1) at (3,0.37) {};
			\node[bvert,label=right:${\rm b}_i'$] (b2) at (4,0.37+0.5*1.73) {};
			
			\draw[dashed,gray] (p1) --(b1p)--(p1p)-- (p2) --(q3)-- (pp2) -- (pp1p) -- (b3p)--(pp1) -- (q1)--(p1);
			
			\draw[] (b1) edge node[above,red]{$1$} (p1p) edge node[left,red]{$1-\alpha_i$} (q1p) edge node[above,red]{$1$} (p2);
			\draw[] (b2) edge node[left,red]{$\frac{1}{u_{i+1}}$} (pp1p) edge node[below,red]{$\frac{1}{u_{i+1}}$} (q1p) edgenode[above,red]{$1$} (q3);		
			
			\node[fvert] (no) at (4,0.5*0.37+.5*0.87){$Y_{i+1}$};
			
			\node[fvert] (no) at (5,0.5*0.37+0.75*1.73){$X_{i+1}$};

			\draw[] (p1)--node[above,red]{$1$}(b1p)--node[above,red]{$-1$}(p1p)
			(q1)--node[above,red]{$1$}(b2p)--node[above,red]{$-1$}(q1p)
			(pp1)--node[above,red]{$1$}(b3p)--node[above,red]{$-1$}(pp1p)
			;	
			
		\end{tikzpicture}
		\caption{The building block graph $K_i$.}\label{fig:hexgraph}
	\end{figure}
	
	\begin{figure}[ht]
		\centering
		\begin{tikzpicture}[scale=0.9,yscale=1.8]
			
			\node[wvert,label=below:$p_2$] (p2) at (2,0) {};
			\node[wvert,label=below:$p_3$] (p4) at (4,0) {};
			\node[wvert,label=below:$p_4$] (px0) at (6,0) {};
			
			\node[wvert,label=above:$p_1$] (pp2) at (2,1.73) {};
			\node[wvert,label=above:$p_2$] (pp4) at (4,1.73) {};
			\node[wvert,label=above:$p_3$] (ppx0) at (6,1.73) {};
			\node[wvert,label=left:$q_1$] (p1) at (1,0.87) {};
			\node[wvert,label=above:$q_2$] (p3) at (3,0.87) {};
			\node[wvert,label=above:$q_3$] (p5) at (5,0.87) {};
			\node[wvert,label=above:$q_4$] (pz1) at (7,0.87) {};
			\node[wvert,label=above:$p_4$] (pz2) at (8,1.73) {};
			
			\node[wvert,label=below:$p_1$] (pz02) at (0,0) {};

			\node[wvert,label=above:${p_5=M(p_1)}$] (qz2) at (10,1.73) {};
			\node[wvert,label=right:${q_5=M(q_1)}$] (qz1) at (9,0.87) {};
			\node[wvert,label=below:${p_5=M(p_1)}$] (qx0) at (8,0) {};

			\draw[dashed,gray] (qx0)-- (qz1) -- (qz2) --(pz2)-- (ppx0)--(pp4) -- (pp2) -- (p1)--(pz02)--(p2)--(p4)--(px0) -- (qx0);	
			
			\node[bvert] (b1) at (1,0.37) {};
			\node[bvert] (b2) at (2,1.24) {};
			\node[bvert] (b3) at (3,0.37) {};
			\node[bvert] (b4) at (4,1.24) {};
			\node[bvert] (b5) at (5,0.37) {};
			\node[bvert] (b7) at (7,0.37) {};
			\node[bvert] (b8) at (8,1.24) {};
			
			\node[bvert] (bx0) at (6,1.24) {};
			\draw[-,text=red,inner sep=1]
			(b1) edge (pz02) edge (p2) edge (p1)
			(b3) edge (p2) edge (p3) edge (p4)
			(b5) edge (p4) edge (p5) edge (px0)
			(bx0) edge (pz1)
			(b2) edge (p1) edge (pp2) edge (p3)
			(b4) edge (p3) edge (pp4) edge (p5)
			(bx0) edge (p5) edge (ppx0)		
			(b8) edge (qz1)	 edge (pz1) edge (pz2)
			(b7) edge (qx0) edge (pz1) edge (px0)
			;
			\node[fvert] (x0) at (8,0.65) {$v_1$};
			\node[fvert] (x0) at (6,0.65) {$v_4$};
			\node[fvert] (x2) at (2,0.65) {$v_2$};
			\node[fvert] (x4) at (4,0.65) {$v_3$};
			\node[fvert] (x1) at (9,1.5) {$u_1$};
			\node[fvert] (x1) at (7,1.5) {$u_4$};
			\node[fvert] (x3) at (3,1.5) {$u_2$};
			\node[fvert] (x5) at (5,1.5) {$u_3$};
		\end{tikzpicture}
		\caption{The graph $\Delta_4$ showing the TCD map and face weights.} \label{fig:hexgraphtcd}
	\end{figure}

	\begin{figure}[ht]
		\centering
		\begin{tikzpicture}[scale=1]
			\draw (0,0) -- (2,0);
			\draw (0,0) -- (0,4);
			
			\draw (0,3) -- (0,4);
			\draw[](2,0) -- (0,4);
			
			\draw[fill=black] (0,0) circle (2pt);
			\draw[fill=black] (1,0) circle (2pt);
			\draw[fill=black] (2,0) circle (2pt);
			\draw[fill=black] (0,4) circle (2pt);

			\draw[fill=black] (0,3) circle (2pt);

			\draw[fill=black] (0,1) circle (2pt);
			\draw[fill=black] (1,1) circle (2pt);
			
			\draw[fill=black] (0,2) circle (2pt);
			\draw[fill=black] (1,2) circle (2pt);


			\node (no) at (-0.5,-0.5) {$(0,0)$};

			\node (no) at (2.5,-0.5) {$(2,0)$};
			\node (no) at (-.5,4.5) {$\left(0,n\right)$};
			\node (no) at (1.5,2.5) {$\left(1,\frac{n}{2}\right)$};
		\end{tikzpicture}
		\hspace{2cm}	\begin{tikzpicture}[scale=1]
			\draw (0,0) -- (2,0);
			\draw (0,0) -- (0,5);

			\draw[](2,0) -- (0,5);
			
			\draw[fill=black] (0,0) circle (2pt);
			\draw[fill=black] (1,0) circle (2pt);
			\draw[fill=black] (2,0) circle (2pt);
			\draw[fill=black] (0,4) circle (2pt);
			\draw[fill=black] (0,5) circle (2pt);

			\draw[fill=black] (0,3) circle (2pt);

			\draw[fill=black] (0,1) circle (2pt);
			\draw[fill=black] (1,1) circle (2pt);
			
			\draw[fill=black] (0,2) circle (2pt);
			\draw[fill=black] (1,2) circle (2pt);


			\node (no) at (-0.5,-0.5) {$(0,0)$};

			\node (no) at (2.5,-0.5) {$(2,0)$};
			\node (no) at (-.5,5.5) {$\left(0,n\right)$};
			
		\end{tikzpicture}
		\caption{The Newton polygon $N_{\Delta_n}$ of $\Delta_n$ for even $n$ (left) and odd $n$ (right).}\label{fig:npdelta}
	\end{figure}
	For $n\geq2$, let $\Delta_n$ denote the torus graph obtained by gluing in cyclic order the graphs $K_i$ in Figure~\ref{fig:hexgraph} for $i =1 ,2,\dots, n$ and contracting two-valent vertices. See Figure~\ref{fig:hexgraphtcd} for a picture of~$\Delta_4$.\looseness=-1
	
	Let $N_{\Delta_n}:=\text{Convex-hull}\{(0,0), (2,0),(0,n)\}$ denote the Newton polygon of $\Delta_n$. For even $n$ there are $n+4$ zig-zag paths and for odd $n$, there are $n+3$ zig-zag paths. The multiset of boundary vectors of $N_{\Delta_n}$ is $\bigl\{(1,0)^2,\bigl(-1,\frac n2\bigr)^2,(0,-1)^n \bigr\}$ in the even case and $\bigl\{(1,0)^2,(-1,n),(0,-1)^n\bigr\}$ in the odd case (see Figure~\ref{fig:npdelta}). We label some of the zig-zag paths as follows:
	\begin{alignat*}{3}
		&\xi_{1}:= \w_1,\bw_1,\w_2,\bw_2,\dots,\w_{n},\bw_{n},\w_{1},  \qquad&& [\xi_1]=(1,0),&\\
		&\xi_{2}:= \w_1',\bw_1',\w_2',\bw_2',\dots,\w_{n}',\bw_{n}',\w_{1}', \qquad&& [\xi_2]=(1,0),&\\
		&\zeta_{i}= \w_{i}, \bw_{i}', \w_{i}', \bw_{i},\w_{i},    \qquad&& [\zeta_{i}]=(0,-1),&
	\end{alignat*}
	where $\zeta_i$ is defined for $i=1,2,\dots,n$. Suppose the faces and zig-zag paths of $\Delta_n$ are labeled as in~Figure~\ref{fig:hexgraph}. The set
\[
\bigl\{X_1^{\pm 1}, \dots,X_n^{\pm 1},Y_1^{\pm 1}, \dots,Y_n^{\pm 1},\chi_{[\zeta_1]}^{\pm 1},\chi_{[\xi_1]}^{\pm 1} \bigr\}
\]
 is a set of generators for~$\mathcal O_{{\mathcal L}_{\Delta_n}}$~and the only relation among them is $\prod_{i=1}^n X_i Y_i=1$. However, it will be more convenient to work with a different set of generators given by \[
 \bigl\{X_1^{\pm 1}, \dots,X_n^{\pm 1},\chi_{[\zeta_1]}^{\pm 1},\dots,\chi_{[\zeta_n]}^{\pm 1},\chi_{[\xi_1]}^{\pm 1}\bigr\}.
 \]
	
We will map a subspace of the dimer parameter space $\mathcal X_{N_{\Delta_n}}$, defined by prescribing the values~of the Casimirs $\chi_{[\zeta_1]},\dots,\chi_{[\zeta_n]}$, to the AFIT parameter space $\mathcal U_n$. That map will transform each $X_i$ into $u_i$. As in Section~\ref{sec:pentagramtcd}, there is an extra Casimir $\chi_{[\xi_1]}$ on the GK side that we also fix to be a constant $\lambda \in \C^\times$.

	Let \smash{$\mathcal X_{N_{\Delta_n},\vec \alpha}^\lambda$} \big(resp.\ \smash{$\mathcal L_{\Delta_n,\vec \alpha}^\lambda$}\big) denote the Poisson subvariety of \smash{$\mathcal X_{N_{\Delta_n}}$} (resp.\ $\mathcal L_{\Delta_n}$) where $\chi_{[\xi]}=\lambda$ and
	\begin{align}
		\chi_{[\zeta_{i}]}= \frac{1}{1-\alpha_{i}}, \label{eq:chidelta}
	\end{align}
	for all $i \in \{1,2,\dots,n\}$. The coordinate ring \smash{$\mathcal O_{\mathcal L_{\Delta_n,\vec \alpha}}^\lambda$} is generated by $\bigl\{X_1^{\pm 1}, \dots,X_n^{\pm 1}\bigr\}$. Recall from Section~\ref{sec:crdynamics} that $\mathcal U_n\cong (\C \setminus \{0,-1\})^n$.
	\begin{Definition}\label{def7.1}
		We define the birational map \smash{$\pi_{\vec \alpha} \colon \mathcal X_{N_{\Delta_n},\vec \alpha}^\lambda \supset \mathcal L_{{\Delta_n},\vec \alpha}^\lambda \!\ra\! (\C \setminus \{0,-1\})^n$} such~that
		\begin{align}\la{def:i}
			\qquad \pi_{\vec \alpha}^*u_i := X_i,
		\end{align}
		for all $i \in \{1,2,\dots,n\}$.
	\end{Definition}

Recall that \smash{$\mathcal L_{{\Delta_n},\vec \alpha}^\lambda$} is a Zariski-dense open subset of \smash{$\mathcal X_{N_{\Delta_n},\vec \alpha}^\lambda$}. Definition~\ref{def7.1} means~that the map $\pi_{\vec \alpha}$ is the unique rational map such that, for \smash{$[\mathrm{wt}] \!\in\! \mathcal L_{{\Delta_n},\vec \alpha}^\lambda$},
	\[
	\pi_{\vec \alpha}([\mathrm{wt}])=(X_1([\mathrm{wt}]),\dots,X_n([\mathrm{wt}])).
	\]
	
	\begin{Lemma}\la{lem:vzzdelta}
		We have $\pi_{\vec \alpha}^*v_i = Y_i$ 	for all $i \in \{1,2,\dots,n\}$.
	\end{Lemma}
	\begin{proof}
		It follows from \smash{$X_i Y_i = \frac{\chi_{[\zeta_{i-1}]}}{\chi_{[\zeta_{i}]}}$} and the equations \eqref{eq:uv} and \eqref{eq:chidelta}.
	\end{proof}

	Next, we check the following.

	\begin{Proposition}\label{p:Poisson}
		The map $\pi_{\vec \alpha}$ is Poisson.
	\end{Proposition}
	\begin{proof}
		The functions $X_1,X_2,\dots,X_{n}$ are local coordinates on \smash{$\mathcal X_{N_{\Delta_n},\vec \alpha}^\lambda$}. The only nonzero Poisson brackets on \smash{$\mathcal X_{N_{\Delta_n},\vec \alpha}^\lambda$} in these coordinates are
		\[
		\{X_i,X_{i+1}\}=X_i X_{i+1},\qquad i=1,2,\dots,n.
		\]
		We compute that
		\[
		{\pi_{\vec \alpha}^*\{u_i,u_{i+1}\}_{\mathcal U_n}=\pi_{\vec \alpha}^*(u_iu_{i+1})=X_i X_{i+1}=\{X_i,X_{i+1}\}=\bigl\{\pi_{\vec \alpha}^*u_i,\pi_{\vec \alpha}^*u_{i+1}\bigr\}.}
\tag*{\qed}
  \]
\renewcommand{\qed}{}
\end{proof}

	Let $(p,q,M)\in\mathcal U_n$ and let \smash{$[\wt] \in \mathcal X_{N_{\Delta_n},\vec \alpha}^\lambda$} be such that $u^{-1} \circ \pi_{\vec \alpha}([\wt])=(p,q,M)$. We~choose edge weights representing $[\wt]$ and Kasteleyn signs as in Figure~\ref{fig:hexgraph}.
	Note that we have
\[
|W(\Delta_{n,\A})|-|B(\Delta_{n,\A})|=2.
\]
 Let \smash{$P\colon W(\Delta_{n,\widehat \A}) \ra \CP^1$} denote the twisted TCD map associated to $[\wt]$. The graph $\Delta_{n,\widehat \A}$ is a~union of infinitely many copies of the building block graph~$K_i$, $i \in \Z$. We label the vertices of~$\Delta_{n,\widehat \A}$ as in Figure~\ref{fig:hexgraph}.
	
	\begin{Lemma}
		We have $P_{\w_i}=p_i$ and $P_{\w'_i}=q_i$ for all $i \in \Z$, up to a {common} projective transformation.
	\end{Lemma}
	\begin{proof}
		The polygons $p$ and $q$ are determined up to projective transformations by the cross-ratios \eqref{def:alpha} and \eqref{def:u}. Using Lemma \ref{lem:mrcyl}, we see from \eqref{def:i} and Lemma \ref{lem:vzzdelta} that the $P_\w$ have the same cross-ratios.
	\end{proof}

	The TCD map for $n=4$ is shown in Figure~\ref{fig:hexgraphtcd}.
	
	\begin{Corollary}\la{cor:mon}
		The monodromy matrix of the twisted TCD map $P$ coincides with the monodromy matrix $M$ of the pair of polygons $(p,q)$.
	\end{Corollary}
	
	We now compute this monodromy matrix. The Kasteleyn matrix of $K_i$ is
	\[
	\aboverulesep=0pt \belowrulesep=0pt
	K_{K_i}(w)=\begin{blockarray}{cccc}
		\begin{block}{[cc|cc]}
			1 & 0 & 0&0 \\
			0 & 1 & 0&0\\
			\cmidrule(lr){1-4}
			0&0&1&0\\
			0&0&0&1\\
			\cmidrule(lr){1-4}
			1&\frac{1}{u_{i+1}}&-1&0\\
			(1-\alpha_i)w&\frac{1}{u_{i+1}}&0&1\\	
		\end{block}
	\end{blockarray}
	\]
	so that
	\begin{align*}
		\Pi_{K_i}(w)=
		\begin{bmatrix}
			1&\frac{1}{u_{i+1}}\\
			(1-\alpha_{i})w &\frac{1}{u_{i+1}}
		\end{bmatrix} = \frac{1}{u_{i+1}}\begin{bmatrix}
			u_{i+1}&1\\
			u_{i+1}(1-\alpha_{i})w &1
		\end{bmatrix}.
	\end{align*}
	
	By Proposition \ref{prop:prod}, the monodromy matrix is $\Pi(1)$, where
	\[
	\Pi(w)= \Pi_{H_{1}}(w) \Pi_{H_{2}}(w) \cdots \Pi_{H_{n}}(w).
	\]
	We have \smash{$\det \Pi_{K_i}(w)=\frac{1-w(1-\alpha_i)}{u_{i+1}}$}, so we get
	\be \la{eq:det}
	\det \Pi(1)=\frac{\alpha_{[n]}}{u_{[n]}}.
	\ee
	
	The following lemma is elementary.
	\begin{Lemma}\la{lem:22}
		Suppose $M$ is a $2 \times 2$ matrix and $P(z)=\det (z I + M)$ is its characteristic polynomial. Then $P(z)=z^2 + \tr M z+ \det M$.
	\end{Lemma}
	
	Let $P(z,w)$ denote the characteristic polynomial of $\Delta_n$, normalized so that it has its Newton polygon as {on the left picture of} Figure~\ref{fig:npdelta} with the vertex in the bottom left corner corresponding to the monomial $1$, and such that the coefficient of $z^2$ is $u_{[n]}$. To find the normalization explicitly, we know from Theorem \ref{p:sc} and Lemma \ref{lem:22} that $P(z,w)$ is up to normalization equal to $\det (zI+\Pi(w)) = z^2+\tr \Pi(w) z + \det \Pi(w)$, which is equal to
	\begin{align*}
		z^2 &+z \frac{1}{u_{[n]}} \tr \prod_{i=1}^{ n }\begin{bmatrix}
			u_{i+1}&1\\
			u_{i+1}(1-\alpha_{i})w &1
		\end{bmatrix} + \frac{1}{u_{[n]}} \prod_{i=1}^n (1-w(1-\alpha_i)),
	\end{align*}
	so
	\begin{align*}
		P(z,w)=u_{[n]}z^2 &+z \tr \prod_{i=1}^{ n }\begin{bmatrix}
			u_{i+1}&1\\
			u_{i+1}(1-\alpha_{i})w &1
		\end{bmatrix} + \prod_{i=1}^n (1-w(1-\alpha_i)).
	\end{align*}
	
	Let $H_{(1,k)}$ denote the coefficient of $z w^k$ in $P(z,w)$ for $k=0,\dots,\floor{\frac{n+1}{2}}$, so that when ${k \in \bigl\{1,2,\dots,\floor{\frac{n+1}{2}} -1\bigr\}}$, they are the Hamiltonians of the cluster integrable system. Then, we have
	\begin{align*}
		\sum_{k=0}^{\floor{\frac n 2}} H_{(1,k)} w^k &= \tr \prod_{i=1}^{ n }\begin{bmatrix}
			u_{i+1}&1\\
			u_{i+1}(1-\alpha_{i})w &1
		\end{bmatrix}.
	\end{align*}

	Make the substitution $\beta_i=1-\alpha_i$ and consider the above expression as a polynomial in $\beta_1,\beta_2,\dots,\beta_{n}$. Since each $(1-\alpha_i)$ inside the matrices appears with a $w$, the homogeneous component of degree $k$ in $\beta_1,\beta_2,\dots,\beta_{n}$ is $H_{(1,k)} w^{k}$. The main result of this section is the following simple procedure for converting between the AFIT Hamiltonians \cite{AFIT} and the GK dimer Hamiltonians for $\Delta_n$.
	\begin{Theorem}
		Let \smash{$k \in \bigl\{1,2,\dots,\floor{\frac{n+1}{2}} -1\bigr\}$} and let $\alpha_i=\alpha$ for all $i$. We have
		\begin{enumerate}\itemsep=0pt
			\item[$1.$] The homogeneous degree $k$ component of \smash{$\sum_{d=0}^{\floor{\frac{n+1}{2}} -1}H_{(1,d)}$} as a polynomial in the variables $\alpha_1,\dots,\alpha_n$ is, up to a sign, equal to
\[
\sqrt{ \frac{\alpha_{[n]}}{X_{[n]}}} \pi_{\vec \alpha}^* \circ \Lambda_{\vec \alpha}^*\biggl(\frac{F_k(c)}{\sqrt{c_{[n]}}}\biggr),
\]
which is the product of the Casimir \smash{$\sqrt{ \frac{\alpha_{[n]}}{X_{[n]}}}$} with the pullback of an AFIT Hamiltonian.
			\item[$2.$] The homogeneous degree $k$ component of \smash{$\sum_{d=0}^{\floor{\frac{n+1}{2}} -1}H_{(1,d)}$} as a polynomial in $1-\alpha_1,\dots,\allowbreak{1-\alpha_n}$ is the GK dimer Hamiltonian $H_{(1,k)}$.
		\end{enumerate}
	\end{Theorem}
	
	\begin{proof}
		
		Only the first item remains to be proved. By Corollary \ref{cor:mon}, $\Pi(1)$ is conjugated to \smash{$\pi_{\vec \alpha}^* \circ \Lambda_{\vec \alpha}^* M$} in $\PGL_2$. Since $\frac{1}{\det M}\tr^2 M$ is a $\PGL_2$-conjugacy invariant, we have
		\[
		\pi_{\vec \alpha}^* \circ \Lambda_{\vec \alpha}^* \left(\frac{1}{\sqrt{\det M}}\tr M \right)=\pm \frac{1}{\sqrt{\det (\Pi(1)})}\tr( \Pi(1)) .
		\]
		Using Theorem \ref{thm:afit2} and \eqref{eq:det}, we get
		\[
		\sqrt{ \frac{\alpha_{[n]}}{X_{[n]}}} \sum_{k=0}^{\floor{\frac n 2}} (-1)^k \pi_{\vec \alpha}^* \circ \Lambda_{\vec \alpha}^*\biggl(\frac{F_k(c)}{\sqrt{c_{[n]}}}\biggr)=\pm \sum_{k=0}^{\floor{\frac n 2}}H_{(1,k)}.
		\]
		Note that the left-hand side is a polynomial in $\alpha_i$ since $\pi_{\vec \alpha}^* \circ \Lambda_{\vec \alpha}^*\sqrt{c_{[n]}}$ has a factor $\sqrt{\alpha_{[n]}}$ which cancels the same factor in the numerator. Then, the homogeneous component of degree $k$ in the variables $\alpha_1, \alpha_2, \dots,\alpha_{n}$ on the left is
\[
(-1)^k \sqrt{ \frac{\alpha_{[n]}}{X_{[n]}}} \pi_{\vec \alpha}^* \circ \Lambda_{\vec \alpha}^*\biggl(\frac{F_k(c)}{\sqrt{c_{[n]}}}\biggr).
\]
		Finally, note that \smash{$X_{[n]}=\frac {\chi_{[\xi_2]}} {\chi_{[\xi_1]}}$} is a Casimir, hence so is \smash{$\sqrt{ \frac{\alpha_{[n]}}{X_{[n]}}}$}.
	\end{proof}

	\subsection{Cross-ratio dynamics via a semi-local move}
	\label{subsec:localmovestcd}
	
	\begin{figure}[ht]
		\centering
		\begin{tikzpicture}[scale=0.8,baseline={(current bounding box.center)}]
			\def\scl{1}
			\def\shif{5}
			\begin{scope}[scale=\scl,yscale=1.8]
				\node[wvert,label=above:$p_0$] (p0) at (1,0.87) {};
				\node[wvert,label=above:$p_1$, label=left:$\times$] (p1) at (3,0.87) {};

				\node[wvert,label=below:$q_1$] (q1) at (2,0) {};
				\node[wvert,label=below:$q_2$] (q2) at (4,0) {};

				\node[wvert,label=above:$q_0$] (qq0) at (2,1.74) {};

				\node[bvert] (b1) at (2,1.24) {};
				\node[bvert] (b2) at (4,1.24) {};

				\node[bvert] (b4) at (1,0.37) {};
				\node[bvert] (b5) at (3,0.37) {};

				\draw[] (b1)edge(qq0)edge(p0)edge(p1)
				(b2) edge (p1)
				(b4) edge (p0) edge (q1)
				(b5) edge (q1) edge (q2) edge (p1)
				;
				\node[](no) at (5,0.62) {$\rightarrow$};
				
				
			\end{scope}
			\begin{scope}[shift={(\shif,0)},scale=\scl,yscale=1.8]
				\node[] (no) at (2.2,0.52) {$\times$};
				\node[wvert,label=above:$p_0$] (p0) at (1,0.87) {};
				\node[wvert,label=above:$p_1$] (p1) at (3,0.87) {};

				\node[wvert,label=below:$q_1$] (q1) at (2,0) {};
				\node[wvert,label=below:$q_2$] (q2) at (4,0) {};

				\node[wvert,label=above:$q_0$] (qq0) at (2,1.74) {};

				\node[bvert] (b1) at (2,1.24) {};
				\node[bvert] (b2) at (4,1.24) {};

				\node[bvert] (b4) at (1,0.37) {};
				\node[bvert] (b5) at (3,0.37) {};

				\draw[] (b1)edge(qq0)edge(p0)
				(b2) edge (p1)
				(b4) edge (p0) edge (q1)
				(b5) edge (q1) edge (q2) edge (p1)
				;
				\node[bvert] (bn2) at (2.4,.87) {};
				
				\node[wvert,label=below:$p_1$] (pn2) at (1.8,0.87) {};
				\draw[] (b1) -- (pn2) -- (bn2) -- (p1);
				\node[](no) at (5,0.62) {$\rightarrow$};
				
				
			\end{scope}
			\begin{scope}[shift={(2*\shif,0)},scale=\scl,yscale=1.8]
				\node[wvert,label=above:$p_0$] (p0) at (1,0.87) {};
				\node[wvert,label=above:$p_1$] (p1) at (3,0.87) {};

				\node[wvert,label=below:$q_1$] (q1) at (2,0) {};
				\node[wvert,label=below:$q_2$] (q2) at (4,0) {};

				\node[wvert,label=above:$q_0$] (qq0) at (2,1.74) {};

				\node[bvert] (b1) at (2,1.24) {};
				\node[bvert] (b2) at (4,1.24) {};

				\node[bvert] (b4) at (1,0.37) {};
				\node[bvert] (b5) at (3,0.37) {};

				\draw[] (b1)edge(qq0)edge(p0)
				(b2) edge (p1)
				(b4) edge (p0) edge (q1)
				(b5) edge (q1) edge (q2) edge (p1)
				;
				\node[bvert,label=above:$\times$] (bn2) at (2.4,.87) {};
				
				\node[wvert,label=below:$p_1$] (pn2) at (1.8,0.87) {};
				\draw[] (b1) -- (pn2) -- (bn2) -- (p1);
				
				\draw[] (bn2) edge [bend right=30] (q1) edge [bend left=30] (q1) ;
				\node[](no) at (5,0.62) {$\rightarrow$};
				
				
			\end{scope}
			\begin{scope}[shift={(3*\shif,0)},scale=\scl,yscale=1.8]
				\node[wvert,label=above:$p_0$] (p0) at (1,0.87) {};
				\node[wvert,label=above:$p_1$] (p1) at (3,0.87) {};

				\node[wvert,label=below:$q_1$] (q1) at (2,0) {};
				\node[wvert,label=below:$q_2$] (q2) at (4,0) {};

				\node[wvert,label=above:$q_0$] (qq0) at (2,1.74) {};

				\node[bvert] (b1) at (2,1.24) {};
				\node[bvert] (b2) at (4,1.24) {};

				\node[bvert] (b4) at (1,0.37) {};
				\node[bvert] (b5) at (3,0.37) {};

				\node[bvert] (bn1) at (2.2,0.87) {};
				
				\node[bvert] (bn2) at (2.6,.87) {};
				
				\node[wvert,label=below:$p_1$] (pn2) at (1.8,0.87) {};
				\node[wvert,label=above:$x_1$] (x1) at (2.4,1) {};
				
				\draw[](b1)--(pn2)--(bn1)--(x1)--(bn2)--(p1)
				(q1) edge (bn2) edge (bn1)
				(qq0) -- (b1) -- (p0) -- (b4) -- (q1) -- (b5)--(q2)
				(b5)--(p1)--(b2)
				;

				
			\end{scope}

		\end{tikzpicture}	
		\caption{When following the arrows from left to right, this depicts Step 1 of the sequence, namely the insertion of $x_1$. The vertices/faces where moves are applied are marked with $\times$'s.}
		\label{fig:insertr}
	\end{figure}
	\begin{figure}[ht]
		\centering
		\begin{tikzpicture}[scale=0.8,baseline={(current bounding box.center)}]
			
			\def\scl{1}
			\def\shif{5}
			
			\begin{scope}[shift={(0,0)},scale=\scl,yscale=1.8]

				\node[wvert,label=below:$q_i$] (q1) at (2,0) {};
				\node[wvert,label=below:$q_{i+1}$] (q2) at (4,0) {};
				
				\node[wvert,label=above:$p_i$] (p1) at (3,0.87) {};
				\node[wvert,label=left:$q_i$] (qq1) at (4,1.74) {};
				\node[wvert,label=right:$p_{i+1}$] (p2) at (5,0.87) {};
				\node[bvert] (b6) at (5,0.37) {};
				\node[bvert] (b2) at (4,1.24) {};
				
				\node[bvert] (b5) at (3,0.37) {};

				\node[bvert] (bn2) at (2.6,.87) {};
				
				\node[wvert,label=above:$x_i$] (x1) at (2.4,1) {};
				
				\draw[](x1)--(bn2)--(p1)--(b2)
				(q1)--(bn2)
				(q1)--(b5)--(q2)
				(b5)--(p1)
				(qq1)--(b2)--(p2)--(b6)--(q2)
				;
				
				\node[](no) at (6,0.62) {$\rightarrow$};
				\node[] (no) at (2.6,0.5*0.87+0.1) {$\times$};	
			\end{scope}

			\begin{scope}[shift={(\shif,0)},scale=\scl,yscale=1.8]

				\node[wvert,label=below:$q_i$] (q1) at (2,0) {};
				\node[wvert,label=below:$q_{i+1}$] (q2) at (4,0) {};
				
				\node[wvert,label=above:$p_i$,label=right:$\times$] (p1) at (3,0.87) {};
				\node[wvert,label=left:$q_i$] (qq1) at (4,1.74) {};
				\node[wvert,label=right:$p_{i+1}$] (p2) at (5,0.87) {};
				\node[bvert] (b6) at (5,0.37) {};
				\node[bvert] (b2) at (4,1.24) {};
				
				\node[bvert] (b5) at (2.3,0.2) {};

				\node[bvert] (bn2) at (3.1,.67) {};
				
				\node[wvert,label=above:$x_i$] (x1) at (2.4,1) {};
				
				\draw[](x1)--(bn2)--(p1)--(b2)--(p2)
				(qq1)--(b2)
				(q1)--(b5)--(x1)
				(b5)--(q2)--(bn2)	
				(p2)--(b6)--(q2)
				;
				
				\node[](no) at (6,0.62) {$\rightarrow$};
				
			\end{scope}

			\begin{scope}[shift={(2*\shif,0)},scale=\scl,yscale=1.8]

				\node[wvert,label=below:$q_i$] (q1) at (2,0) {};
				\node[wvert,label=below:$q_{i+1}$] (q2) at (4,0) {};
				
				\node[wvert,label=left:$x_{i+1}$] (x2) at (4.4,1) {};
				\node[wvert,label=left:$q_i$] (qq1) at (4,1.74) {};
				\node[wvert,label=right:$p_{i+1}$] (p2) at (5,0.87) {};
				\node[bvert] (b6) at (5,0.37) {};
				\node[bvert] (b2) at (4.6,0.87) {};
				
				\node[bvert] (b5) at (2.3,0.2) {};

				\node[bvert] (bn2) at (4.2,1.13) {};
				
				\node[wvert,label=above:$x_i$] (x1) at (2.4,1) {};
				
				\draw[]
				(b2)--(p2)
				(q1)--(b5)
				(b5)--(x1) -- (bn2)--(qq1)
				(bn2)--(x2)--(b2)
				(b5)--(q2)
				(p2)--(b6)--(q2)--(b2)
				
				;
				
				
			\end{scope}

		\end{tikzpicture}			
		\caption{Step 2 in the sequence replacing $p_i$ with $x_{i+1}$.}
		\label{fig:crsequence}	
	\end{figure}
	
	We prepare with an observation on multi-ratios.
	\begin{Lemma}\label{lem:dskpincrdynamics}
		Assume $p,q$ are nondegenerate twisted $n$-gons such that $p$ is $\vec \alpha$-related to $q$. Suppose $x_i,x_{i+1} \in \CP^1$ are two points such that $-\cro(q_i,x_i,q_{i+1},x_{i+1})=\alpha_i$. Then,
		\begin{align*}
			\mr(p_i,q_i,x_i,x_{i+1},q_{i+1},p_{i+1}) = -1
		\end{align*}
		holds for all $i\in \Z$.
	\end{Lemma}
	\begin{proof}
		The lemma follows from the following two equations
		\begin{align*}
			&\cro(p_i,q_i,p_{i+1},q_{i+1})= \cro(q_i,x_i,q_{i+1},x_{i+1}),\\
			&\mr(p_i,q_i,x_i,x_{i+1},q_{i+1},p_{i+1})= -\frac{\cro(p_i,q_i,p_{i+1},q_{i+1})}{\cro(q_i,x_i,q_{i+1},x_{i+1})}.
\tag*{\qed}
\end{align*}
\renewcommand{\qed}{}
\end{proof}
	
	Now we construct the sequence of moves.
	
	\begin{figure}[ht]
		\centering
		\begin{tikzpicture}[scale=0.8,baseline={(current bounding box.center)}]
			\def\scl{1}
			\def\shif{5}
			\begin{scope}[shift={(3*\shif,0)},scale=2*\scl,yscale=1.8]
				\node[wvert,label=left:$M^{-1}(x_n)$] (p0) at (1,0.87) {};
				\node[wvert,label=above:$x_1$] (p1) at (3,0.87) {};

				\node[wvert,label=below:$q_1$] (q1) at (2,0) {};
				\node[wvert,label=below:$q_2$] (q2) at (4,0) {};

				\node[wvert,label=above:$q_0$] (qq0) at (2,1.74) {};

				\node[bvert] (b1) at (2,1.24) {};
				\node[bvert] (b2) at (4,1.24) {};

				\node[bvert] (b4) at (1,0.37) {};
				\node[bvert] (b5) at (3,0.37) {};

				\node[bvert] (bn1) at (2.2,0.87) {};
				
				\node[bvert] (bn2) at (2.6,.87) {};
				
				\node[wvert] (pn2) at (1.8,0.87) {};
				\node[wvert,label=above:$p_1$] (x1) at (2.4,1) {};
				
				\draw[](b1)--(pn2)--(bn1)--(x1)--(bn2)--(p1)
				(q1) edge (bn2) edge (bn1)
				(qq0) -- (b1) -- (p0) -- (b4) -- (q1) -- (b5)--(q2)
				(b5)--(p1)--(b2)
				;
				
				\node[](no) at (1.6,0.77){$M^{-1}(x_{n+1})$};
				
			\end{scope}

		\end{tikzpicture}	
		\caption{Local configuration near $p_1$ in Step 3.}
		\label{fig:step3}
	\end{figure}
	
	\begin{Theorem}
		\label{thm:crdynamicslocalmoves}
		Suppose \smash{$[\wt] \in \mathcal X_{N_{\Delta_n},\vec \alpha}^\lambda$} is such that $u^{-1} \circ \pi_{\vec \alpha} ([\wt])=(p,q,M)$. Consider the sequence of moves shown in Figures $\ref{fig:insertr}$, $\ref{fig:crsequence}$, and let $\mu$ denote the induced birational map of weights. Then the pair of curves $(q,r,M)$ is $u^{-1} \circ \pi_{\vec \alpha} \circ \mu ([\wt])$.	In other words, the following diagram commutes:
		
		\[
		{\begin{tikzcd}[every label/.append style = {font = \normalsize}, row sep=large, column sep = huge]
				\mathcal X_{N_{\Delta_n},\vec \alpha}^\lambda \arrow[r, "u^{-1} \circ \pi_{\vec \alpha}"] \arrow[d, "\mu"']
				& \mathcal U_n \arrow[d, "\nu_{\vec \alpha}"] \\
				\mathcal X_{N_{\Delta_n},\vec \alpha}^\lambda \arrow[r,"u^{-1} \circ \pi_{\vec \alpha}"]
				& \mathcal U_n
		\end{tikzcd}}.
		\]
\end{Theorem}
	
\begin{proof}
By Lemma \ref{lem:tcddef}, it suffices to trace what happens to the twisted TCD map associated to $[\wt]$ through the sequence of moves. We identify $\A$ with the fundamental domain in $\widehat \A$ containing $p_1,\dots,p_n$, and the other fundamental domains are denoted $\A+ m \gamma_z$, $m \in \Z$. Recall that $P_{\w+\gamma_z}=M(P_\w)$ for a twisted TCD map. The moves are applied $\gamma_z$-periodically in $\widehat \A$, so we only describe what happens in $\A$. We proceed in four steps:
\begin{enumerate}\itemsep=0pt
			\item 
First, we insert a new point $x_1$ into the twisted TCD map, which, based on a choice made in Step 3, will turn out to be $r_1$; see Figure~\ref{fig:insertr} for an illustration (in $\A + m \gamma_z$, this means that we insert $M^m(x_1)$). To do this, we begin by using the resplit move to split $p_1$ into two copies of $p_1$ and a new black vertex, which we denote $\bw$. Then, we add a bigon with vertices $\bw$ and $q_1$, such that the bigon is inside the face also bounded by $p_0$ and $p_1$. In~order for this bigon insertion not to change the rest of the TCD map, we require that the weights of the two edges of this bigon sum to $0$. Finally, we split $\bw$ into two new black vertices of degree three while also generating the new white vertex corresponding to $x_1$. The result is a version of the graph in the middle of Figure~\ref{fig:bigon} in which all black vertices are trivalent so that we are in the realm of TCD maps.
			\item 
Let $x_i \in \CP^1$ for $i \in \{2,\dots,n+1\}$ be defined by $-\cro(q_i,x_i,q_{i+1},x_{i+1})=\alpha_i$.
			We apply a~spider move followed by a resplit that replaces $p_i$ with $x_{i+1}$; see Figure~\ref{fig:crsequence}. When we apply the resplit, \eqref{eq:menelaus} and Lemma~\ref{lem:dskpincrdynamics} imply that the newly created point is $x_{i+1}$. We~apply these moves until we have replaced $p_n$ with a new copy of $x_{n+1}$.
			\item 
After the moves of Step 2,
there is still one copy of $p_1$ left in the graph. Locally near $p_1$, the TCD map looks like Figure~\ref{fig:step3}. This is the same as the rightmost graph in Figure~\ref{fig:insertr}, except that the roles of $p$ and $x$ are interchanged. There are two choices of $x_1$ that make $M^{-1} (x_{n+1})=x_1$, namely $p_1$ and $r_1$. However, $x_1=p_1$ makes the weight of the face to the right of the bigon equal to $0$, and is therefore disallowed. We then apply the sequence of moves in Step 1 in reverse order to recover the hexagonal graph with $p$ replaced by $r$. 		
			
			\item Finally, we translate $\Delta_n$ by \smash{$\frac 1 {2} \gamma_w$} to interchange $q$ and $r$.
\hfill\qed
\end{enumerate}
\renewcommand{\qed}{}
\end{proof}
	
Since the weights of the two edges of a bigon sum to zero, the weight of any dimer cover that uses one of the edges of the bigon is canceled by the weight of the dimer cover that uses the other edge. Therefore, the spectral curve, the Casimirs and the Hamiltonians of the dimer model are unchanged upon inserting a bigon. Since the dimer Casimirs and Hamiltonians are also preserved by the elementary transformations \cite[Theorem~4.7]{GK13}, we obtain as a corollary of Theorem~\ref{thm:main1}.

	\begin{Corollary}\label{cor:alternative}
		The AFIT Casimirs and Hamiltonians are invariant under cross-ratio dy\-namics.
	\end{Corollary}
	
	We can now prove Proposition~\ref{prop:uevolution} on the evolution of the $u_i$ coordinates under cross-ratio dynamics.
	
	\begin{proof}[Proof of Proposition~\ref{prop:uevolution}]
		
		The sequence of moves in Theorem~\ref{thm:crdynamicslocalmoves} is a sequence of local moves on TCD maps starting and ending with a TCD map on $\Delta_n$. It can be seen as a sequence of dimer local moves starting and ending with $\Delta_n$. Up to contractions and expansions of degree $2$ vertices, the sequence of dimer local moves is exactly the one described above for the geometric $R$-matrix transformation. More precisely, the bigon is initially added between the white vertex carrying $q_1$ and the black vertex resulting from the contraction of the white vertex carrying $p_1$. Then, the sequence of $n$ spider moves crosses a string of hexagons and finally the bigon gets deleted at the end of this sequence. Since contractions and expansions of degree $2$ vertices do not change the face weights, we deduce that the evolution of the face weights for the sequence in Theorem~\ref{thm:crdynamicslocalmoves} is given by the geometric $R$-matrix transformation.
		
		Let $X'_i$ and $Y'_i$ be the face weights after applying the birational map $\mu$ of Theorem~\ref{thm:crdynamicslocalmoves}, which includes in Step~4 a translation by $\tfrac{1}{2}\gamma_w$. In our case, the face weights of the strip of hexagonal faces crossed by the spider moves are the $X_i$ and they become $Y'_i$ after the transformation, i.e.,
		\[
		(Y'_1,\dots,Y'_n)=\Phi(X_1,\dots,X_n).
		\]
		By Theorem~\ref{thm:ILP}, we have
		\[
			Y'_i=\frac{\sum\limits_{t=0}^{n-1}\prod\limits_{s=0}^{t-1}X_{i+s}}{\sum\limits_{t=1}^{n}\prod\limits_{s=1}^{t}X_{i+s}}.
		\]
		Formula~\eqref{eq:ILP} follows from the fact that \smash{$\pi_{\vec \alpha}^*v_i = X_i$} and \smash{$\pi_{\vec \alpha}^*u'_i = Y'_i$}.
	\end{proof}
	
	We can also use this correspondence with the geometric $R$-matrix transformation to provide an alternative proof that the cross-ratio dynamics map $\nu_{\vec\alpha}$ is Poisson.
	
	\begin{Corollary}
		The map $\nu_{\vec\alpha}$ from $\mathcal U_n$ to itself is Poisson for the bracket $\{\cdot,\cdot\}_{\mathcal U_n}$.
	\end{Corollary}
	
\begin{proof}In \cite{ILP2}, it is shown that the geometric $R$-matrix transformation can rewritten as a~composition of cluster mutations for a quiver obtained from the original dimer quiver by adding several edges and vertices. In particular, the Poisson bracket on the space of cluster variables~\cite{GSV1} induces on the space of $a_i$ variables the bracket given by $\{a_i,a_{i+1}\}=a_ia_{i+1}$ for every $i$. Since cluster mutations induce Poisson maps \cite{GSV1}, we deduce that the map $\Phi$ sending $(a_1,\dots,a_n)$ to $(a'_1,\dots,a'_n)$ is a Poisson map. Alternatively, this result follows from a direct computation on Poisson brackets using formula~\eqref{eq:Rmatrix}.
		
		The Poisson bracket $\{\cdot,\cdot\}_{\mathcal U_n}$ on $\mathcal U_n$ is given by the same formula~\eqref{eq:upoisson} as the bracket for the $a_i$. Furthermore, $\nu_{\vec\alpha}$ is given by the composition of the Poisson map $\Phi$ with the map transforming each component into its inverse (which is also Poisson), thus $\nu_{\vec\alpha}$ is Poisson.
	\end{proof}
	
\subsection{Other dynamics}\label{subsec:otherdynamics}
	In this subsection, we describe all the other integrable dynamics that are defined on the phase space of the cross-ratio dynamics integrable system, called \textit{generalized cluster modular transformations} in \cite{GeRa}.
	
By definition, each side $e$ of the Newton polygon (which we think of as oriented counterclockwise around the boundary of $N$) corresponds to a subset $\mathcal Z_e : = \{ \beta \in \mathcal Z\mid e \in \Z_{>0}[\beta]\}$ of zig-zag paths. Let $|e|_\Z$ denote the \textit{integral length} of $e$, i.e., the number of lattice points in $e$ minus one or equivalently, the number of primitive line segments in $e$. Let $p\colon \R^2 \ra \T$ denote the universal covering map of the torus, and let $\widetilde \Gamma$ denote the biperiodic graph $p^{-1}(\Gamma)$ in $\R^2$. Let $\beta_1,\dots,\beta_{|e|_{\Z}}$ denote the zig-zag paths in $\mathcal Z_e$ in cyclic order around the torus from right to left. Their lifts to $\R^2$ form a collection of bi-infinite parallel zig-zag paths $\widetilde \beta_i$, $i\in \Z,$ in $\widetilde \Gamma$ labeled in order from right to left, such that $p\big(\widetilde \beta_i\big)=\beta_{j}$, where $1\leq j\leq |e|_\Z$ and $j \equiv i $ mod $|e|_\Z$.
	
	An \textit{extended affine permutation of period $k$} is a bijection $w\colon \Z \ra \Z$ such that ${w(i+k)}={w(i)+k}$. Let $\hat S_k$ denote the group of extended affine permutations with period $k$. We will write extended affine permutations $w$ in \emph{window notation} $[w(1), \dots, w(k)]$. Define $\tau\!:=\![2,3,\dots ,k,\allowbreak {k+1}]$, $s_i:= [1,2,\dots ,i-1,i+1,i,i+2,\dots ,k]$ for $1\leq i\leq k-1$ and $s_0=s_k:= [0,2,3,\dots ,{k-2},\allowbreak{k-1},{k+1}]$. Then, $\widehat S_k$ is the group generated by $\tau,s_0,\dots,s_{k-1}$ modulo the relations
	\[
	s_i^2=1, \qquad s_i s_{i+1} s_i = s_{i+1} s_i s_{i+1}, \qquad s_i s_j=s_j s_i \quad \text{if $|i-j|>1$} ,\qquad \tau s_{i}\tau ^{-1}=s_{i+1}.
	\]
	There is a group homomorphism $\disp\colon \widehat S_k \ra \Z$ given by $\disp(w):=\frac 1 k \sum_{i=1}^k (f(i)-i)$. Given a~Newton polygon $N$, let $E(N)$ denote the set of edges of $N$. Let $L_N$ denote the kernel of the group homomorphism
	\[
	\prod_{e \in E(N)} \widehat S_{|e|_\Z} \xrightarrow[]{\sum_{e \in E(N)} \disp} \Z,
	\]
	i.e., $L_N$ consists of an extended affine permutation $w^e$ for each edge $e$ of $N$ such that the total displacement $\sum_{e \in E(N)} \disp(w^e)$ is $0$.
	
One of the main results of \cite{GeRa} is that each $w=(w^e)_{e \in E(N)} \in L_N$ determines an automorphism of the cluster Poisson variety $\mathcal X_N$ given by a sequence of isotopies, elementary transformations and geometric $R$-matrix transformations that take a graph $\Gamma$ to itself, called a {generalized cluster modular transformation}. Generalized cluster modular transformations are determined by what they do to zig-zag paths and the correspondence is as follows. Each generalized cluster modular transformation $\phi$ lifts to an $H_1(\T,\Z)$-periodic sequence of isotopies, elementary transformations and geometric $R$-matrix transformations in $\widetilde \Gamma$ that take $\widetilde \Gamma$ to itself. Therefore, each zig-zag path~$\widetilde \beta_i$ of $\widetilde \Gamma$ ends up at the initial location of a parallel zig-zag path $\widetilde \beta_j$. Then, $w^e(i):=j$ where~$e$ is the edge of $N$ in the direction $[\beta]$. We have an injective group homomorphism (coming from the translation action of $H_1(\T,\Z)$)
	\begin{align*}
		j\colon\ H_1(\T,\Z) \hookrightarrow L_N,\qquad
		m \mapsto \big(\tau_\rho^{\langle e,m \rangle} \big)_{e \in E(N)},
	\end{align*}
	where $\langle \cdot,\cdot \rangle$ denotes the intersection form in $\T$. Explicitly, identifying $H_1(\T,\Z)$ with $\Z^2$ using the basis $(\gamma_z,\gamma_w)$, if $(a,b), (c,d) \in \Z^2$, then $\langle (a,b),(c,d) \rangle := ad-bc$. The \textit{generalized cluster modular group}, the group of all generalized cluster transformations is isomorphic to the quotient of $L_N$ by the subgroup $j(H_1(\T,\Z))$.
	
	Let $e_\rightarrow$, \smash{$e_{\myarrow[135]}$} and $e_\downarrow$ denote the edges of $N_{\Delta_n}$ given by the vectors $(2,0),(-2,n)$ and $(0,-n)$, respectively.
	For $a \in \{ \rightarrow, \myarrow[135],\downarrow\}$, we denote the generators of \smash{$\widehat S_{|e_a|_\Z}$} by $\tau_a$ and $s_{i,a}$, ${0 \leq i \leq |e_a|_\Z-1}$. The subgroup \smash{$j(H_1(\T,\Z)) = \bigl\langle \tau_{\myarrow[135]}^n\tau_\downarrow^{-n}, \tau_{\rightarrow}^{-2}\tau_{\myarrow[135]}^2 \bigr\rangle$}. The generators of the generalized cluster modular group are
	\begin{enumerate}\itemsep=0pt
		\item $s_{i,\rightarrow}$, $0 \leq i \leq 1$: Let $(p,q)$ be a pair of $\vec \alpha$-related twisted $n$-gons, and let $o$ (resp.~$r$) denote the other twisted $n$-gon $\vec \alpha$-related to $p$ (resp.~$q$). Then, $s_{0,\ra}$ (resp.\ $s_{1,\ra}$) is given by $(p,q) \mapsto (r,q)$ (resp.\ $(p,o)$). Here, we are indexing the two horizontal zig-zag paths so that the zig-zag path containing the points of $p$ gets label $1$ and the one containing the points of $q$ gets label $2$ (see the left hand side of Figure~\ref{fig:crtcdintro}).
		
		\item $s_{i,\downarrow}$ for $i = 0,1,\dots,n-1$: Given a pair of $\vec{\alpha}$-related twisted $n$-gons $(p,q)$, let $s_i \cdot \vec \alpha$ be defined by $(s_i \cdot \vec \alpha)_j := \alpha_{s_i(j)}$. It is not difficult to see that there is a unique $(s_i \cdot \vec \alpha)$-related pair of twisted $n$-gons $(p',q')$ such that $p'_k=p_k$ and $q'_k=q_k$ if $k \neq i+1$. The transformation is~${(p,q) \mapsto (p',q')}$.
		\item Only when $n$ is even, $s_{i,\myarrow[135]}$, $0 \leq i \leq 1$: There is a graph automorphism exchanging the zig-zag paths in directions $(1,0)$ and $\big(-1,\frac n 2\big)$ which sends $(p,q)$ to $(p',q')$, where
		\[
		p'_k = \begin{cases}
			p_k&\text{if $k$ is odd,}\\
			q_k&\text{if $k$ is even,}
		\end{cases}\qquad q'_k = \begin{cases}
			q_k&\text{if $k$ is odd,}\\
			p_k&\text{if $k$ is even}.
		\end{cases}
		\]
		Note that $(p',q')$ is a pair of $\vec \alpha'$-related twisted $n$-gons where $\alpha'_i := \frac{1}{\alpha_i}$ and the monodromy is the same as $(p,q)$. Under this automorphism, \smash{$s_{i,\myarrow[135]}$} becomes $s_{i,\rightarrow}$. 	
		\item \smash{$\tau_{\ra} \tau_{\myarrow[135]}^{-1}$}: This is the transformation $(p,q) \mapsto (q,p)$.
		\item \smash{$ \tau_{\myarrow[135]} \tau_{\downarrow}^{-1}$}: Given a twisted $n$-gon $p$, let $\mathrm{shift}(p)$ be the twisted $n$-gon defined by $\mathrm{shift}(p)_i:=p_{i+1}$. The transformation is $(p,q) \mapsto (\mathrm{shift}(p),\mathrm{shift}(q))$.
	\end{enumerate}
	In terms of the generators above, Theorem~\ref{thm:crdynamicslocalmoves} says that cross-ratio dynamics is the generalized cluster modular transformation \smash{$\tau_{\ra} \tau_{\myarrow[135]}^{-1} s_{1,\ra}$}. The above discussion shows that there are other integrable discrete dynamics on the same space whose interactions with each other can be written down explicitly as the relations in the generalized cluster modular group. The most interesting one occurs when $n$ is even where, if we conjugate cross-ratio dynamics by the automorphism in item 3, we get a generalized cluster modular transformation that we call \textit{switch dynamics}.
	
Let us mention the interpretation of the three geometrically non-trivial operations above in terms of {discrete integrable systems}. Recall from the beginning of Section~\ref{subsec:crdynintro} that a solution of the cross-ratio dynamics system may be seen as a map \smash{$f\colon \Z^2 \rightarrow \CP^1$}, such that ${f_{i,0}=p_i}$ and ${f_{i,1}=q_i}$. We view each row as a twisted $n$-gon and the four vertices on the boundary of any \textit{quad}, i.e., $(1 \times 1)$-square, are required to satisfy the cross-ratio condition $\cro({f_{i,j}, f_{i,j+1},f_{i+1,j},f_{i+1,j+1}})=\alpha_i$. Then, cross-ratio dynamics corresponds to a vertical translation by one unit in $\Z^2$. A priori, there is a one-parameter family of choices for $r$ (resp.~$o$) such that the cross-ratios formed by the quads between $q$ and $r$ (resp.~$o$ and~$p$) are~$\alpha_i$, but after imposing the monodromy condition in the twisted case, there is only one choice.
	
	Similarly, switch dynamics is related to the corresponding \emph{discrete equation of Toda type} \cite{bsintegrablequads} of the cross-ratio dynamics system. The Toda type equation is a lattice equation satisfied by the even (resp.\ odd) parity vertices of a solution of the cross-ratio dynamics system. Thus, the even (resp.\ odd) vertices of the cross-ratio dynamics system satisfy an equation independently of the odd (resp.\ even) vertices. Moreover, a solution of the Toda type equation on the even (resp.\ odd) vertices can be extended to a solution of the cross-ratio dynamics system, but there is a one-parameter family of such extensions on $\Z^2$. Again, upon imposing the monodromy condition in the twisted case, the one-parameter family reduces to only two extensions. Given one extension, there is thus exactly one other such extension, and this defines switch dynamics by alternately choosing the other odd (resp.\ even) extension.

	\begin{figure}[ht]
		\centering
		\begin{tikzpicture}[scale=1.4, baseline={(current bounding box.center)}]
			
			\node[wvert,label=below:$p_0$] (p0) at (0,0) {};
			\node[wvert,label=below:$p_1$] (p1) at (1,0) {};
			\node[wvert,label=below:$p_2$] (p2) at (2,-.4) {};
			\node[wvert,label=below:$p_3$] (p3) at (3,0) {};
			\node[wvert,label=below:$p_4$] (p4) at (4,0) {};
			\node[wvert,label=above:$q_0$] (q0) at (0,1) {};
			\node[wvert,label=above:$q_1$] (q1) at (1,1) {};
			\node[wvert,label=above:$q_2$] (q2) at (2,.6) {};
			\node[wvert,label=above:$q'_2$] (qq2) at (2,1.4) {};
			\node[wvert,label=above:$q_3$] (q3) at (3,1) {};
			\node[wvert,label=above:$q_4$] (q4) at (4,1) {};
			
			\draw[-]
			(p0) -- (p1) -- (p2) -- (p3) -- (p4)
			(q0) -- (q1) -- (q2) -- (q3) -- (q4)
			(p0) -- (q0) (p1) -- (q1) (p2) -- (q2) (p3) -- (q3) (p4) -- (q4)
			(q1) -- (qq2) -- (q3)
			;
			
			\node[blue] at (0.5, -0.8) {$\alpha_0$};
			\node[blue] at (1.5, -0.8) {$\alpha_1$};
			\node[blue] at (2.5, -0.8) {$\alpha_2$};
			\node[blue] at (3.5, -0.8) {$\alpha_3$};
			\node[blue] at (-0.7, 0.5) {$1$};
			
			\draw[-, blue]
			(0.5, -0.5) -- (0.5, 1.5)
			(3.5, -0.5) -- (3.5, 1.5)
			(1.5, -0.5) -- (1.5, 0.4) to[out=90,in=210] (2,1) to[out=30,in=270] (2.5, 1.5)
			(2.5, -0.5) -- (2.5, 0.4) to[out=90,in=330] (2,1) to[out=150,in=270] (1.5, 1.5)
			(-0.5, 0.5) -- (0.5,0.5) to[out=0,in=180] (2,0.1) to[out=0,in=180] (3.5,0.5) -- (4.5, 0.5)
			;
			
		\end{tikzpicture}
		\begin{tikzpicture}[scale=1.4, baseline={(current bounding box.center)}]
			
			\node[wvert,label=below:$p_0$] (p0) at (0,0) {};
			\node[wvert,label=below:$p_1$] (p1) at (1,0) {};
			\node[wvert,label=below:$p_2$] (p2) at (2,-.4) {};
			\node[wvert,label=below:$p_3$] (p3) at (3,0) {};
			\node[wvert,label=below:$p_4$] (p4) at (4,0) {};
			\node[wvert,label=above:$q_0$] (q0) at (0,1) {};
			\node[wvert,label=above:$q_1$] (q1) at (1,1) {};
			\node[wvert,label=above:$q'_2$] (qq2) at (2,1.4) {};
			\node[wvert,label=below:$p'_2$] (pp2) at (2,.4) {};
			\node[wvert,label=above:$q_3$] (q3) at (3,1) {};
			\node[wvert,label=above:$q_4$] (q4) at (4,1) {};
			
			\draw[-]
			(p0) -- (p1) -- (p2) -- (p3) -- (p4)
			(q0) -- (q1) -- (qq2) -- (q3) -- (q4)
			(p0) -- (q0) (p1) -- (q1) (pp2) -- (qq2) (p3) -- (q3) (p4) -- (q4)
			(p1) -- (pp2) -- (p3)
			;
			
			\node[blue] at (0.5, -0.8) {$\alpha_0$};
			\node[blue] at (1.5, -0.8) {$\alpha_1$};
			\node[blue] at (2.5, -0.8) {$\alpha_2$};
			\node[blue] at (3.5, -0.8) {$\alpha_3$};
			\node[blue] at (-0.7, 0.5) {$1$};
			
			\draw[-, blue]
			(0.5, -0.5) -- (0.5, 1.5)
			(3.5, -0.5) -- (3.5, 1.5)
			(2.5, 1.5) -- (2.5, 0.6) to[out=270,in=30] (2,0) to[out=210,in=90] (1.5, -0.5)
			(1.5, 1.5) -- (1.5, 0.6) to[out=270,in=150] (2,0) to[out=330,in=90] (2.5, -0.5)
			(-0.5, 0.5) -- (0.5,0.5) to[out=0,in=180] (2,0.9) to[out=0,in=180] (3.5,0.5) -- (4.5, 0.5)
			;
			
		\end{tikzpicture}			
		\caption{Illustration of $s_{1,\downarrow}$ around the cube formed by $p_1$, $p_2$, $p_3$, $p'_2$, $q_1$, $q_2$, $q_3$ and $q'_2$. On these pictures, we are requiring that the cross-ratio of the four points around any quad be equal to the ratio of the parameters attached to the two blue lines crossing the quad. The three blue lines with parameters~$\alpha_1$,~$\alpha_2$ and $1$ cycle around the cube.}
		\label{fig:consistency}
	\end{figure}	
	
	Finally, $s_{i,\downarrow}$ corresponds to introducing a ``fault'' in the lattice; see Figure~\ref{fig:consistency}. On that figure we start with the points $p_1$, $p_2$, $p_3$, $q_1$, $q_2$, $q_3$ satisfying the two cross-ratio equations corresponding to the two front faces of the cube. We then compute $q'_2$ using the equation ${\cro\big(q_2,q_1,q_3,q'_2\big)=\alpha_2/\alpha_1}$ corresponding to the top face. There are three possible ways to compute~$p'_2$, using any equation corresponding to one of the three hidden faces of the cube. They actually give the same result and this fact is called 3D-consistency of the cross-ratio system \cite{bsintegrablequads}.
	
	\subsection{Square TCD map}\label{sec:sqtcd}

	\begin{figure}[ht]
		\centering
		\begin{tikzpicture}[scale=1.7] 		\def\nnp{3};
			\def\nn{2};
			\def\nnm{1};
			\draw[dashed,gray] (1,0) rectangle (\nnp,4);
			\foreach[evaluate={\yy=int(\xx+1)}] \xx in {1} {
				\coordinate[wvert,label=left:$\w_{2i-1}$] (p\xx) at (\xx,3);
				\coordinate[wvert,label=left:$\w_{2i-1}'$] (q\xx) at (\xx,1);		
				\coordinate[bvert,label=-30:$\bw_{2i-1}$] (b\xx) at (\xx+0.6,3);		
				\coordinate[bvert,label=-30:$\bw_{2i-1}'$] (c\xx) at (\xx+0.6,1);	
				
			}
			\foreach[evaluate={\yy=int(\xx+1)}] \xx in {2} {
				\coordinate[wvert,label=above:$\w_{2i}'$] (p\xx) at (\xx,4);
				\coordinate[wvert,label=below:$\w_{2i}'$] (pp\xx) at (\xx,0);
				\coordinate[wvert,label=left:$\w_{2i}$] (q\xx) at (\xx,2);
			}
			\foreach[evaluate={\yy=int(\xx+1)}] \xx in {1} {
				\draw[white,text=red,every node/.style={fill=white}]
				(b\xx) edge node[below] {\ifthenelse{\xx>1}{$1$}{$1$}} (p\xx) edge node[left] {$- u_{2i-1}$} (p\yy) edge node[left] {$1$} (q\yy)
				(c\xx) edge node[below] {\ifthenelse{\xx>1}{$1$}{$1$}} (q\xx) edge node[left] {$-\frac{1-\alpha_{2i-1}}{u_{2i-1}} w$} (q\yy) edge node[left] {$1$} (pp\yy)
				;
				\draw[-]
				(b\xx) edge (p\xx) edge (p\yy) edge (q\yy)
				(c\xx) edge (q\xx) edge (q\yy) edge (pp\yy)
				;
				
			}
			\coordinate[wvert,label=right:$\w_{2i+1}$] (p11) at (\nn+1,3);
			\coordinate[wvert,label=right:$\w_{2i+1}'$] (q11) at (\nn+1,1);
			\foreach[evaluate={\yy=int(\xx-1)}] \xx in {3} {
				\coordinate[bvert,label=5:$\bw_{2i}$] (b\xx) at (\xx-0.6,3);		
				\coordinate[bvert,label=5:$\bw_{2i}'$] (c\xx) at (\xx-0.6,1);		
				\draw[white,text=red,every node/.style={fill=white}]
				(b\xx) edge node[below] {$1$} (p11) edge node[right] {$ { \alpha_{2i-1} u_{2i-1} u_{2i} } $} (p\yy) edge node[right] {$\alpha_{2i-1}$} (q\yy)
				(c\xx) edge node[below] {$1$} (q11) edge node[right] {$\alpha_{2i-1}u_{2i-1} u_{2i}$} (pp\yy) edge node[right] {$\alpha_{2i-1} (1-\alpha_{2i})w$} (q\yy)
				;
				\draw[-]
				(b\xx) edge (p11) edge (p\yy) edge (q\yy)
				(c\xx) edge (q11) edge (pp\yy) edge (q\yy)			
				;
				\node[fill=white,text=blue] (no) at (1,2) {$Y_{2i-1}$};
				\node[fill=white,text=blue] (no) at (3,2) {$Y_{2i+1}$};
				\node[text=blue] (no) at (2,1) {$Y_{2i}$};
				\node[fill=white,text=blue] (no) at (1,4) {$X_{2i-1}$};
				\node[fill=white,text=blue] (no) at (3,4) {$X_{2i+1}$};
				\node[fill=white,text=blue] (no) at (2,3) {$X_{2i}$};
				\node[fill=white,text=blue] (no) at (1,0) {$X_{2i-1}$};
				\node[fill=white,text=blue] (no) at (3,0) {$X_{2i+1}$};
				
			}						
		\end{tikzpicture}\hspace{11mm}
		\begin{tikzpicture}[scale=1.7]		\def\nnp{3};
			\def\nn{2};
			\def\nnm{1};
			\draw[dashed, gray] (1,0) rectangle (\nnp,4);
			\foreach[evaluate={\yy=int(\xx+1)}] \xx in {1} {
				\coordinate[wvert,label=left:$\w_{n}$] (p\xx) at (\xx,3);
				\coordinate[wvert,label=left:$\w_{n}'$] (q\xx) at (\xx,1);

			}
			\foreach[evaluate={\yy=int(\xx+1)}] \xx in {2} {
				\coordinate[bvert,label=above:$\bw_{n}$] (p\xx) at (\xx,4);
				\coordinate[bvert,label=below:$\bw_{n}$] (pp\xx) at (\xx,0);
				\coordinate[bvert,label=left:$\bw_{n}'$] (q\xx) at (\xx,2);
			}
			
			\coordinate[wvert,label=right:$\w_{1}$] (p11) at (\nn+1,3);
			\coordinate[wvert,label=right:$\w_{1}'$] (q11) at (\nn+1,1);
			
			\draw[-,text=red]
			(p2) edge node[above left] {$1$} (p1) (p11) edge node[above right] {$\frac 1 {\alpha_n}$} (p2)
			(q2) edge node[above left] {$1$} (q1)
			(q11) edge node[above right] {$\frac{1}{\alpha_n u_n}$} (q2)
			(p11) edge node[above left] {$-\frac{1-\alpha_n}{\alpha_n u_n}w$} (q2)
			(q11) edge node[above left] {$-\frac 1 {\alpha_n}$} (pp2)
			;

			\node[fill=white,text=blue] (no) at (3,2) {$Y_{1}$};
			\node[fill=white,text=blue] (no) at (2,1) {$X_{n}$};
			\node[fill=white,text=blue] (no) at (1,4) {$X_{n}$};
			\node[fill=white,text=blue] (no) at (3,4) {$X_{1}$};
			\node[fill=white,text=blue] (no) at (2,3) {$Y_{n}$};
			
			\node[fill=white,text=blue] (no) at (3,0) {$X_{1}$};			
		\end{tikzpicture}
		
		\caption{The building block graphs $G_i$ and $G_n^{{\rm odd}}$.} \label{fig:gi}
	\end{figure}
	
	\begin{figure}[ht]
		\centering
		\begin{tikzpicture}[scale=1]
			\def\nnp{7};
			\def\nn{6};
			\def\nnm{5};
			\foreach[evaluate={\yy=int(\xx+1)}] \xx in {1, 3, ..., \nnm} {
				\pgfmathtruncatemacro{\label}{\xx}
				\coordinate[wvert,label=above:$p_{\label}$] (p\xx) at (\xx,3);
				\coordinate[wvert,label=above:$q_{\label}$] (q\xx) at (\xx,1);		
				\coordinate[bvert] (b\xx) at (\xx+0.4,3);		
				\coordinate[bvert] (c\xx) at (\xx+0.4,1);	
				\node[fvert] (fw) at (\xx+1,3) {$u_\yy$};		
				\node[fvert] (fw) at (\xx+1,1) {$v_\yy$};		
			}
			\foreach[evaluate={\yy=int(\xx+1)}] \xx in {2, 4, ..., \nn} {
				\pgfmathtruncatemacro{\label}{\xx}
				\coordinate[wvert,label=above:$q_{\label}$] (p\xx) at (\xx,4);
				\coordinate[wvert,label=below:$q_{\label}$] (pp\xx) at (\xx,0);
				\coordinate[wvert,label=left:$p_{\label}$] (q\xx) at (\xx,2);
			}
			\foreach[evaluate={\yy=int(\xx+1)}] \xx in {1, 3, ..., \nnm} {
				\draw[-]
				(b\xx) edge (p\xx) edge (p\yy) edge (q\yy)
				(c\xx) edge (q\xx) edge (q\yy) edge (pp\yy)
				;
			}
			\foreach[evaluate={\yy=int(\xx+1)}] \xx in {0, 2, 4, ..., \nnm} {
				\node[fvert] (fw) at (\xx+1,0) {$u_\yy$};		
				\node[fvert] (fw) at (\xx+1,2) {$v_\yy$};		
				\node[fvert] (fw) at (\xx+1,4) {$u_\yy$};		
			}
			\coordinate[wvert,label=above:$p_{7}$] (p7) at (\nn+1,3);
			\coordinate[wvert,label=above:$q_{7}$] (q7) at (\nn+1,1);
			\foreach[evaluate={\yy=int(\xx-1)}] \xx in {3, 5, 7} {
				\coordinate[bvert] (b\xx) at (\xx-0.4,3);		
				\coordinate[bvert] (c\xx) at (\xx-0.4,1);		
				\draw[-]
				(b\xx) edge (p\xx) edge (p\yy) edge (q\yy)
				(c\xx) edge (q\xx) edge (pp\yy) edge (q\yy)
				;
			}		
			\node[fvert] (fw) at (7,0) {$u_1$};
			\node[fvert] (fw) at (7,2) {$v_1$};
			\node[fvert] (fw) at (7,4) {$u_1$};
			
		\end{tikzpicture}\hspace{10mm}
		\begin{tikzpicture}[scale=1]
			\def\nnp{6};
			\def\nn{5};
			\def\nnm{4};
			\foreach \xx in {1, 3, 5} {
				\pgfmathtruncatemacro{\label}{\xx}
				\coordinate[wvert,label=above:$p_{\label}$] (p\xx) at (\xx,3);
				\coordinate[wvert,label=above:$q_{\label}$] (q\xx) at (\xx,1);		
			}
			\foreach[evaluate={\yy=int(\xx-1)}] \xx in {2, 4} {
				\pgfmathtruncatemacro{\label}{\xx}
				\coordinate[wvert,label=above:$q_{\label}$] (p\xx) at (\xx,4);
				\coordinate[wvert,label=below:$q_{\label}$] (pp\xx) at (\xx,0);
				\coordinate[wvert,label=left:$p_{\label}$] (q\xx) at (\xx,2);
			}
			\foreach[evaluate={\yy=int(\xx+1)}] \xx in {1, 3} {
				\coordinate[bvert] (b\xx) at (\xx+0.4,3);		
				\coordinate[bvert] (c\xx) at (\xx+0.4,1);		
				\draw[-]
				(b\xx) edge (p\xx) edge (p\yy) edge (q\yy)
				(c\xx) edge (q\xx) edge (q\yy) edge (pp\yy)
				;
				\node[fvert] (fw) at (\xx+1,3) {$u_\yy$};		
				\node[fvert] (fw) at (\xx+1,1) {$v_\yy$};		
			}
			\foreach[evaluate={\yy=int(\xx-1)}] \xx in {3, 5} {
				\coordinate[bvert] (b\xx) at (\xx-0.4,3);		
				\coordinate[bvert] (c\xx) at (\xx-0.4,1);		
				\draw[-]
				(b\xx) edge (p\xx) edge (p\yy) edge (q\yy)
				(c\xx) edge (q\xx) edge (pp\yy) edge (q\yy)
				;
			}		
			\foreach[evaluate={\yy=int(\xx+1)}] \xx in {0, 2, 4, ..., \nnm} {
				\node[fvert] (fw) at (\xx+1,0) {$u_\yy$};		
				\node[fvert] (fw) at (\xx+1,2) {$v_\yy$};		
				\node[fvert] (fw) at (\xx+1,4) {$u_\yy$};		
			}
			\coordinate[wvert,label=above:$p_{6}$] (p1) at (7,3);
			\coordinate[wvert,label=above:$q_{6}$] (q1) at (7,1);
			\coordinate[bvert] (bn) at (6,4);
			\coordinate[bvert] (cn) at (6,2);
			\coordinate[bvert] (dn) at (6,0);
			\draw[-]
			(bn) edge (p5) edge (p1)
			(cn) edge (q5) edge (q1) edge (p1)
			(dn) edge (q1)
			;
			\node[fvert] (fw) at (7,0) {$u_1$};
			\node[fvert] (fw) at (7,2) {$v_1$};
			\node[fvert] (fw) at (7,4) {$u_1$};
			
		\end{tikzpicture}	
		\caption{The relevant bipartite graphs to describe a pair of curves of length $n=6$ (left) and $n=5$ (right). Here the top and the bottom sides of each graph are identified, yielding the cylinder graphs $\Gamma_{6,\A}$ and $\Gamma_{5,\A}$. If we additionally identify the left and right sides of each graph, we obtain the torus graphs $\Gamma_6$ and $\Gamma_5$. In blue we indicate how the face weights are related to the coordinates on the space of pairs of $\protect\vec\alpha$-related twisted polygons.}
		\label{fig:pairofcurvestcd}
	\end{figure}
	
	\begin{figure}[ht]
		\centering
		\begin{tikzpicture}[scale=1]
			\draw (0,0) -- (2,0);
			\draw (0,0) -- (0,2);
			\draw [dashed] (0,2) -- (0,3);
			\draw (0,3) -- (0,4);
			\draw (2,3) -- (2,4);
			\draw (2,0) -- (2,2);
			\draw [dashed] (2,2) -- (2,3);
			\draw (0,4) -- (2,4);
			\draw[fill=black] (0,0) circle (2pt);
			\draw[fill=black] (1,0) circle (2pt);
			\draw[fill=black] (2,0) circle (2pt);
			\draw[fill=black] (0,4) circle (2pt);
			\draw[fill=black] (1,4) circle (2pt);
			\draw[fill=black] (2,4) circle (2pt);
			\draw[fill=black] (0,3) circle (2pt);
			\draw[fill=black] (1,3) circle (2pt);
			\draw[fill=black] (2,3) circle (2pt);
			\draw[fill=black] (0,1) circle (2pt);
			\draw[fill=black] (1,1) circle (2pt);
			\draw[fill=black] (2,1) circle (2pt);
			\draw[fill=black] (0,2) circle (2pt);
			\draw[fill=black] (1,2) circle (2pt);
			\draw[fill=black] (2,2) circle (2pt);
			

			\node (no) at (-0.5,-0.5) {$(0,0)$};

			\node (no) at (2.5,-0.5) {$(2,0)$};
			\node (no) at (-.5,4.5) {$\left(0,\frac{n}{2}\right)$};

			\node (no) at (2.5,4.5) {$\left(2,\frac{n}{2}\right)$};
		\end{tikzpicture}\hspace{2cm}
		\begin{tikzpicture}[scale=1]
			\draw (0,0) -- (0,1);		
			\draw [dashed] (0,1) -- (0,2);
			\draw (0,2) -- (0,3);
			\draw (0,0) -- (2,0);
			\draw (2,0) -- (2,1);
			\draw [dashed] (2,1) -- (2,2);
			\draw (0,3) -- (2,4);
			\draw (2,2) -- (2,4);
			
			\draw[fill=black] (2,0) circle (2pt);
			
			\draw[fill=black] (2,4) circle (2pt);
			\draw[fill=black] (1,0) circle (2pt);
			\draw[fill=black] (0,0) circle (2pt);
			\draw[fill=black] (0,3) circle (2pt);
			\draw[fill=black] (1,3) circle (2pt);
			\draw[fill=black] (2,3) circle (2pt);
			\draw[fill=black] (0,1) circle (2pt);
			\draw[fill=black] (1,1) circle (2pt);
			\draw[fill=black] (2,1) circle (2pt);
			\draw[fill=black] (0,2) circle (2pt);
			\draw[fill=black] (1,2) circle (2pt);
			\draw[fill=black] (2,2) circle (2pt);
			
			\node (no) at (-0.5,-0.5) {$(0,0)$};
			\node (no) at (2.5,-0.5) {$(2,0)$};

		\end{tikzpicture}	
		\caption{The Newton polygon $N_{\Gamma_n}$ of $\Gamma_n$ for even $n$ (left) and odd $n$ (right).}
		\label{fig:npg} \label{fig:npgodd}
	\end{figure}
	
	For $n$ even, consider the graph $\Gamma_n$ in $\T$ for which a fundamental domain is obtained by gluing the cylinder graphs $G_i$ shown on the left side of Figure~\ref{fig:gi} for $i \in \bigl\{1,2,\dots,\frac{n}{2}\bigr\}$ in the order \smash{$G_{1} G_{2} \cdots G_{\frac n 2}$}, so that $\w_{2n+1}$ is identified with $\w_1$ and $\w_{2n+2}$ with $\w_2$ (see the left side of Figure~\ref{fig:pairofcurvestcd} for $\Gamma_6$). Similarly, for odd $n$, let $\Gamma_n$ be obtained by gluing the graphs \smash{$G_{1} G_{2} \cdots G_{\frac {n-3} 2} G_{\frac {n-1} 2}G_n^{\rm odd} $} from left to right and identifying $\w_{2n+1}$ with $\w_1$ and $\w_{2n+2}$ with $\w_2$ (see the right side of Figure~\ref{fig:pairofcurvestcd} for $\Gamma_5$). Here \smash{$G_n^{\rm odd}$} is the graph represented on the right picture of Figure~\ref{fig:gi}.

	We label some of the zig-zag paths of $\Gamma_n$ as follows:
	\begin{alignat*}{3}
		&\xi_{1}:= \w_1,\bw_1,\w_2,\bw_2,\dots,\w_{n},\bw_{n},\w_{1}, \qquad&& [\xi_1]=(1,0),&\\
		&\xi_{2}:= \w_1',\bw_1',\w_2',\bw_2',\dots,\w_{n}',\bw_{n}',\w_{1}', \qquad&& [\xi_2]=(1,0),&\\
		&\zeta_{2i-1}:= \w_{2i}, \bw_{2i-1}, \w_{2i}', \bw_{2i-1}',\w_{2i}, \qquad&& [\zeta_{2i-1}]=(0,1),&\\
		&\zeta_{2i}:= \w_{2i}, \bw_{2i-1}', \w_{2i}', \bw_{2i-1},\w_{2i}, \qquad&& [\zeta_{2i}]=(0,-1),&
	\end{alignat*}
	where $i \in \bigl\{1,2,\dots,\floor{\frac n 2}\bigr\}$, and for $n$ odd, we also define $\zeta_n := \w_1 \bw_n \w_1' \bw_n' \w_1$ where $[\zeta_n]= (0,1)$. Using the basis $(\gamma_z,\gamma_w)$, we identify $H_1(\T,\Z)$ with $\Z^2$. Then, the Newton polygon of $\Gamma_n$ is
	\begin{gather*}
	N_{\Gamma_n}= \begin{cases}
		\operatorname{Convex-hull}\bigl\{(0,0), (2,0),\bigl(0,\frac n 2 \bigr), \bigl(2,\frac n 2 \bigr)\bigr\} &\text{if $n$ is even, (Figure~\ref{fig:npg}, left),}\\
		\text{Convex-hull}\bigl\{(0,0), (2,0),\bigl(0,\frac{n-1}{2}\bigr), \bigl(2,\frac{n+1}{ 2}\bigr)\bigr\} &\text{if $n$ is odd (Figure~\ref{fig:npg}, right).}
	\end{cases}
	\end{gather*}
	
	For any $n$, the faces of $\Gamma_n$ are labeled by their face weights $X_i$, $Y_i$ for $i \in \{1,2,\dots,n\}$ as shown in Figure~\ref{fig:gi}. A set of generators for the coordinate ring \smash{$\mathcal O_{\mathcal L_{\Gamma_n}}$} is given by	
	\[
	\bigl\{X_1^{\pm 1}, \dots,X_n^{\pm 1},\chi_{[\zeta_1]}^{\pm 1},\dots,\chi_{[\zeta_n]}^{\pm 1},\chi_{[\xi_1]}^{\pm 1}\bigr\}.
	\]
Let $\lambda \in \C^\times$ and let \smash{$\mathcal X_{N_{\Gamma_n},\vec \alpha}^\lambda$} \big(resp.\ \smash{$\mathcal L_{\Gamma_n,\vec \alpha}^\lambda$}\big) denote the Poisson subvariety of \smash{$\mathcal X_{N_{\Gamma_n}}$} \big(resp.\ \smash{$\mathcal L_{\Gamma_n}$}\big), where $\chi_{[\xi_1]}=\lambda$ and
	\[
	\chi_{[\zeta_{k}]} := \begin{cases}
		1-\alpha_{k}&\text{if $k$ is odd},\\
		\frac{1}{1-\alpha_{k}}&\text{if $k$ is even},
	\end{cases} \]
	where $k \in \{1,2,\dots,n\}$. The coordinate ring \smash{$\mathcal O_{\mathcal L_{\Gamma_n,\vec \alpha}}^\lambda$} is generated by $\bigl\{X_1^{\pm 1}, \dots,X_n^{\pm 1}\bigr\}$. We define the birational map
\[
\pi_{\vec \alpha} \colon \ \mathcal X_{N_{\Gamma_n},\vec \alpha}^\lambda \supset \mathcal L_{\Gamma_n,\vec \alpha}^\lambda \ra (\C \setminus \{0,-1\})^n
\]
 by $\pi_{\vec \alpha}^*u_i := X_i$ for all $i \in \{1,2,\dots,n\}$. Similarly to Lemma~\ref{lem:vzzdelta} and Proposition~\ref{p:Poisson}, \smash{$\pi_{\vec \alpha}^*v_i = Y_i$} for all $i \in \{1,2,\dots,n\}$ and \smash{$\pi_{\vec \alpha}$} is Poisson.

	Let $(p,q,M)\in\mathcal U_n$ and let \smash{$[\wt] \in \mathcal X_{N_{\Gamma_n},\vec \alpha}^\lambda$} be such that
\[
u^{-1} \circ \pi_{\vec \alpha}([\wt])=(p,q,M).
\]
The left picture of Figure~\ref{fig:gi} shows edge weights and Kasteleyn signs. Let \smash{$P\colon W\bigl(\Gamma_{n,\widehat \A}\bigr) \ra \CP^1$} denote the twisted TCD map associated to $[\wt]$. We label the vertices of \smash{$\Gamma_{n,\widehat \A}$} as in Figure~\ref{fig:gi}.
	
	\begin{Lemma}\la{lem:tcddef}
		We have $($see Figure $\ref{fig:pairofcurvestcd})$
		$P_{\w_i}=p_i$ and $P_{\w_i'}=q_i$ for all $i \in \Z$, up to a {common} projective transformation. The monodromy matrix of the twisted TCD map $P$ coincides with the monodromy matrix $M$ of the pair of polygons $(p,q)$.
	\end{Lemma}
	
	The Kasteleyn matrix of $G_i$ is
	\[
	\aboverulesep=0pt \belowrulesep=0pt
	K_{G_i}(w)=\begin{blockarray}{cccc}
		\begin{block}{[cc|cc]}
			1 & 0 & 0&0 \\
			0 & 1 & 0&0\\
			\cmidrule(lr){1-4}
			0&0&1&0\\
			0&0&0&1\\
			\cmidrule(lr){1-4}
			{\alpha_{2i-1} u_{2i-1} u_{2i} } & {\alpha_{2i-1} u_{2i-1} u_{2i} }& 1 & -u_{2i-1} \\
			\alpha_{2i-1} {(1-\alpha_{2i})}{w} &\alpha_{2i-1} & -\frac{1-\alpha_{2i-1}}{u_{2i-1}} w & 1\\
		\end{block}
	\end{blockarray}
	\]
	so that
	\begin{align*}
		&\Pi_{G_i}(w)=
		\frac{\alpha_{2i-1} }{ ( {1-\alpha_{2i-1})w-1}}
		\begin{bmatrix}
			u_{2i-1}(u_{2i}+{(1-\alpha_{2i})w)} &
			u_{2i-1}(1+u_{2i}) \\
			{{({1-\alpha_{2i}})w}+u_{2i}{{(1-\alpha_{2i-1})}{w}}} &
			1+u_{2i}({{1-\alpha_{2i-1}}})w
		\end{bmatrix},\\
		&\det \Pi_{G_i}(w)=\biggl(\frac{\alpha_{2i-1} }{{(1-\alpha_{2i-1})}{w}-1}\biggr) ({{(1-\alpha_{2i})}{w}-1}) {\alpha_{2i-1} u_{2i-1} u_{2i}},\\
		&\det \Pi_{G_i}(w)= {\alpha_{2i-1} \alpha_{2i}u_{2i-1} u_{2i}}.
	\end{align*}
	Similarly, we compute
	\begin{align*}
		&\Pi_{G_n^{{\rm odd}}}(w)=
		\frac{\alpha_{n} }{({1-\alpha_{n}}){w}-1}
		\begin{bmatrix}
			1 & (1-\alpha_n)w\\
			u_n & u_n
		\end{bmatrix},\la{eq:matodd} \\
		&\det \Pi_{G_n^{{\rm odd}}}(w)=\frac{\alpha_{n} }{
			({1-\alpha_{n}}){w}-1} (-{\alpha_{n} u_{n} }),\nonumber\\
		&\det \Pi_{G_n^{{\rm odd}}}(1)=\alpha_n u_n. \nonumber
	\end{align*}
	Therefore, for all $n$, we get
	\[
	\det \Pi(1)={\alpha_{[n]} u_{[n]}}.
	\]

	Let $P(z,w)$ denote the characteristic polynomial of $\Gamma_n$, normalized so that
	\[
	\det (zI + \Pi(w))=\frac{ \alpha_{{\rm odd}}}{\prod_{i \text{ odd} }^{} ((1-\alpha_{i})w-1)} P(z,w).
	\]
	Let $H_{(1,k)}$ denote the coefficient of $z w^k$ in $P(z,w)$ for $k=0,\dots,\frac n 2 $, so that when $k \in \bigl\{1,2,\dots,\allowbreak\frac n 2 -1\bigr\}$, they are the Hamiltonians of the cluster integrable system. Then, we have
	\begin{align*}
		\sum_{k=0}^{\frac n 2} H_{(1,k)} w^k = \frac{\prod_{i=1}^{\frac n 2} ((1-\alpha_{2i-1})w-1)} {\alpha_{{\rm odd}}} \tr \Pi(w).
	\end{align*}
	Since each $(1-\alpha_i)$ inside each of the matrices whose product is $\Pi(w)$ appears with a $w$, the homogeneous component of degree $k$ in $1-\alpha_1,1-\alpha_2,\dots,\alpha_{n}$ is $H_{(1,k)} w^{k}$.
	\begin{Theorem}
		Let $k \in \bigl\{1,2,\dots,\floor{\frac{n+1}{2} }-1\bigr\}$ and let $\alpha_i=\alpha$ for all $i$. We have:
		\begin{enumerate}\itemsep=0pt
			\item[$1.$] The homogeneous degree $k$ component of \smash{$\sum_{d=0}^{\floor{\frac n 2}}H_{(1,d)}$} as a polynomial in the variables $\alpha_1,\dots,\alpha_n$ is, up to a sign, equal to
\[
\sqrt{X_{[n]} \alpha_{[n]}} \pi_{\vec \alpha}^* \circ \Lambda_{\vec \alpha}^*\biggl(\frac{F_k(c)}{\sqrt{c_{[n]}}}\biggr),
\]
which is the product of the Casimir \smash{$\sqrt{X_{[n]} \alpha_{[n]}}$} with the pullback of an AFIT Hamiltonian.
			\item[$2.$] The homogeneous degree $k$ component of \smash{$\sum_{d=0}^{\floor{\frac n 2}}H_{(1,d)}$} as a polynomial in $1-\alpha_1,\dots,1-\alpha_n$ is the dimer Hamiltonian $H_{(1,k)}$.
		\end{enumerate}
	\end{Theorem}

	\begin{figure}[ht]
		\centering
		\def\scl{0.8}
		\begin{tikzpicture}[scale=\scl,baseline={([yshift=-1mm]current bounding box.center)}]
			\node[wvert,label=right:$q_2$] (q2) at (0,-4) {};
			\node[wvert,label=right:$q_2$] (qq2) at (0,0) {};
			\node[wvert,label=right:$p_2$] (p2) at (0,-2) {};
			\node[wvert,label=below:$q_1$] (q1) at (-1,-3) {};
			\node[wvert,label=above:$p_1$,label=below:$\times$] (p1) at (-1,-1) {};
			\node[wvert,label=left:$q_0$] (qn) at (-2,-4) {};
			\node[wvert,label=left:$q_0$] (qqn) at (-2,0) {};
			\node[wvert,label=left:$p_0$] (pn) at (-2,-2) {};
			\node[bvert] (cn) at (-1.6,-3) {};
			\node[bvert] (bn) at (-1.6,-1) {};
			\node[bvert] (c1) at (-0.4,-3) {};
			\coordinate[bvert] (b1) at (-0.4,-1) {};
			\draw[-]
			(b1) edge (p1) edge (p2) edge (qq2)
			(bn) edge (p1) edge (pn) edge (qqn)
			(c1) edge (q1) edge (p2) edge (q2)
			(cn) edge (q1) edge (pn) edge (qn)
			;
		\end{tikzpicture}\hspace{2mm}$\rightarrow$
		\begin{tikzpicture}[scale=\scl,baseline={([yshift=-1mm]current bounding box.center)}]
			\node[wvert,label=right:$q_2$] (q2) at (0,-4) {};
			\node[wvert,label=right:$q_2$] (qq2) at (0,0) {};
			\node[wvert,label=right:$p_2$] (p2) at (0,-2) {};
			\node[wvert,label=below:$q_1$] (q1) at (-1,-3) {};
			\node[wvert,label=above:$p_1$] (p1) at (-0.5,-1) {};
			\node[wvert,label=above:$p_1$] (pp1) at (-1.5,-1) {};
			\node[wvert,label=left:$q_0$] (qn) at (-2,-4) {};
			\node[wvert,label=left:$q_0$] (qqn) at (-2,0) {};
			\node[wvert,label=left:$p_0$] (pn) at (-2,-2) {};
			\node[bvert] (cn) at (-1.6,-3) {};
			\coordinate[bvert] (bn) at (-1.9,-1) {};
			\node[bvert] (c1) at (-0.4,-3) {};
			\node[bvert] (b1) at (-0.1,-1) {};
			\coordinate[bvert] (bb) at (-1,-1) {};
			\draw[-]
			(b1) edge (p1) edge (p2) edge (qq2)
			(bn) edge (pp1) edge (pn) edge (qqn)
			(c1) edge (q1) edge (p2) edge (q2)
			(cn) edge (q1) edge (pn) edge (qn)
			(bb) edge (p1) edge (pp1)
			;
			\node[] (no) at (-1,-2) {$\times$};
		\end{tikzpicture}\hspace{2mm}$\rightarrow$
		\begin{tikzpicture}[scale=\scl,baseline={([yshift=-1mm]current bounding box.center)}]
			\node[wvert,label=right:$q_2$] (q2) at (0,-4) {};
			\node[wvert,label=right:$q_2$] (qq2) at (0,0) {};
			\node[wvert,label=right:$p_2$] (p2) at (0,-2) {};
			\node[wvert,label=below:$q_1$] (q1) at (-1,-3) {};
			\node[wvert,label=above:$p_1$] (p1) at (-0.5,-1) {};
			\node[wvert,label=above:$p_1$] (pp1) at (-1.5,-1) {};
			\node[wvert,label=left:$q_0$] (qn) at (-2,-4) {};
			\node[wvert,label=left:$q_0$] (qqn) at (-2,0) {};
			\node[wvert,label=left:$p_0$] (pn) at (-2,-2) {};
			\node[bvert] (cn) at (-1.6,-3) {};
			\node[bvert] (bn) at (-1.9,-1) {};
			\node[bvert] (c1) at (-0.4,-3) {};
			\node[bvert] (b1) at (-0.1,-1) {};
			\coordinate[bvert,label=above:$\times$] (bb) at (-1,-1) {};
			\draw[-]
			(b1) edge (p1) edge (p2) edge (qq2)
			(bn) edge (pp1) edge (pn) edge (qqn)
			(c1) edge (q1) edge (p2) edge (q2)
			(cn) edge (q1) edge (pn) edge (qn)
			(bb) edge (p1) edge (pp1) edge[bend right=20] (q1) edge[bend left=20] (q1)
			;
		\end{tikzpicture}\hspace{2mm}$\rightarrow$
		\begin{tikzpicture}[scale=\scl,baseline={([yshift=-1mm]current bounding box.center)}]
			\node[wvert,label=right:$q_2$] (q2) at (0,-4) {};
			\node[wvert,label=right:$q_2$] (qq2) at (0,0) {};
			\node[wvert,label=right:$p_2$] (p2) at (0,-2) {};
			\node[wvert,label=below:$q_1$] (q1) at (-1,-3) {};
			\node[wvert,label=above:$p_1$] (p1) at (-0.5,-1) {};
			\node[wvert,label=above:$p_1$] (pp1) at (-1.5,-1) {};
			\node[wvert,label=above:$x_1$] (r1) at (-1,-1) {};
			\node[wvert,label=left:$q_0$] (qn) at (-2,-4) {};
			\node[wvert,label=left:$q_0$] (qqn) at (-2,0) {};
			\node[wvert,label=left:$p_0$] (pn) at (-2,-2) {};
			\node[bvert] (cn) at (-1.6,-3) {};
			\node[bvert] (bn) at (-1.9,-1) {};
			\node[bvert] (c1) at (-0.4,-3) {};
			\node[bvert] (b1) at (-0.1,-1) {};
			\node[bvert] (bb) at (-0.7,-2) {};
			\node[bvert] (bbb) at (-1.3,-2) {};
			\draw[-]
			(b1) edge (p1) edge (p2) edge (qq2)
			(bn) edge (pp1) edge (pn) edge (qqn)
			(c1) edge (q1) edge (p2) edge (q2)
			(cn) edge (q1) edge (pn) edge (qn)
			(bb) edge (p1) edge (r1) edge (q1)
			(bbb) edge (r1) edge (pp1) edge (q1)
			;
		\end{tikzpicture}		
		\caption{Step 1: Insertion of $x_1$.}
		\label{fig:insertr1}
	\end{figure}
	
	\begin{figure}[ht]
		\def\scl{0.8}
		\begin{tikzpicture}[scale=\scl,baseline={(current bounding box.center)}]
			\node[wvert,label=below:$q_{i+2}$] (q3) at (0,-3) {};
			\node[wvert,label=above:$p_{i+2}$] (p3) at (0,-1) {};
			\node[wvert,label=right:$q_{i+1}$] (q2) at (-1,-4) {};
			\node[wvert,label=right:$q_{i+1}$] (qq2) at (-1,0) {};
			\node[wvert,label=right:$p_{i+1}$] (p2) at (-1,-2) {};
			\node[wvert,label=below:$q_i$] (q1) at (-2,-3) {};
			\node[wvert,label=above:$p_i$,label=below:$\times$] (p1) at (-2,-1) {};
			\node[bvert] (c3) at (-0.6,-3) {};
			\node[bvert] (b3) at (-0.6,-1) {};
			\node[bvert] (c1) at (-1.4,-3) {};
			\node[bvert] (b1) at (-1.4,-1) {};
			\node[wvert,label=above:$x_i$] (r1) at (-2.5,-1) {};
			\node[bvert] (bb1) at (-2.25,-2) {};
			\draw[-]
			(b1) edge (p1) edge (p2) edge (qq2)
			(b3) edge (p3) edge (p2) edge (qq2)
			(c1) edge (q1) edge (p2) edge (q2)
			(c3) edge (q3) edge (p2) edge (q2)
			(bb1) edge (p1) edge (q1) edge (r1)
			;
		\end{tikzpicture}$\rightarrow$\hspace{-1mm}
		\begin{tikzpicture}[scale=\scl,baseline={(current bounding box.center)}]
			\node[wvert,label=below:$q_{i+2}$] (q3) at (0,-3) {};
			\node[wvert,label=above:$p_{i+2}$] (p3) at (0,-1) {};
			\node[wvert,label=right:$q_{i+1}$] (q2) at (-1,-4) {};
			\node[wvert,label=right:$q_{i+1}$] (qq2) at (-1,0) {};
			\node[wvert,label=right:$p_{i+1}$] (p2) at (-1,-2) {};
			\node[wvert,label=below:$q_i$] (q1) at (-2,-3) {};
			\node[wvert,label=left:$x_{i+1}$] (r2) at (-1.5,-2) {};
			\node[bvert] (c3) at (-0.6,-3) {};
			\node[bvert] (b3) at (-0.6,-1) {};
			\node[bvert] (c1) at (-1.4,-3) {};
			\node[bvert] (b1) at (-1.5,-1) {};
			\node[wvert,label=above:$x_i$] (r1) at (-2,-1) {};
			\node[bvert] (bb1) at (-1.5,-2.5) {};
			\draw[-]
			(b1) edge (r1) edge (r2) edge (qq2)
			(b3) edge (p3) edge (p2) edge (qq2)
			(c1) edge (q1) edge (p2) edge (q2)
			(c3) edge (q3) edge (p2) edge (q2)
			(bb1) edge (r2) edge (q1) edge (p2)
			;
			\node[] (no) at (-1.5,-2.8) {$\times$};
		\end{tikzpicture}$\rightarrow$\hspace{-1mm}
		\begin{tikzpicture}[scale=\scl,baseline={(current bounding box.center)}]
			\node[wvert,label=below:$q_{i+2}$] (q3) at (0,-3) {};
			\node[wvert,label=above:$p_{i+2}$] (p3) at (0,-1) {};
			\node[wvert,label=right:$q_{i+1}$] (q2) at (-1,-4) {};
			\node[wvert,label=right:$q_{i+1}$] (qq2) at (-1,0) {};
			\node[wvert,label=right:$p_{i+1}$] (p2) at (-1,-2) {};
			\node[wvert,label=below:$q_i$] (q1) at (-2,-3) {};
			\node[wvert,label=left:$x_{i+1}$] (r2) at (-1.5,-2) {};
			\node[bvert] (c3) at (-0.6,-3) {};
			\node[bvert] (b3) at (-0.6,-1) {};
			\node[bvert] (c1) at (-1.4,-3) {};
			\node[bvert] (b1) at (-1.5,-1) {};
			\node[wvert,label=above:$x_i$] (r1) at (-2,-1) {};
			\node[bvert] (bb1) at (-1,-3) {};
			\draw[-]
			(b1) edge (r1) edge (r2) edge (qq2)
			(b3) edge (p3) edge (p2) edge (qq2)
			(c1) edge (q1) edge (r2) edge (q2)
			(c3) edge (q3) edge (p2) edge (q2)
			(bb1) edge (r2) edge (q2) edge (p2)
			;
			\node[] (no) at (-0.8,-3) {$\times$};
		\end{tikzpicture}$\rightarrow$\hspace{-1mm}
		\begin{tikzpicture}[scale=\scl,baseline={(current bounding box.center)}]
			\node[wvert,label=below:$q_{i+2}$] (q3) at (0,-3) {};
			\node[wvert,label=above:$p_{i+2}$] (p3) at (0,-1) {};
			\node[wvert,label=right:$q_{i+1}$] (q2) at (-1,-4) {};
			\node[wvert,label=right:$q_{i+1}$] (qq2) at (-1,0) {};
			\node[wvert,label=right:$p_{i+1}$,label=above:$\times$] (p2) at (-1,-2) {};
			\node[wvert,label=below:$q_i$] (q1) at (-2,-3) {};
			\node[wvert,label=left:$x_{i+1}$] (r2) at (-1.5,-2) {};
			\node[bvert] (c3) at (-0.75,-2.5) {};
			\node[bvert] (b3) at (-0.6,-1) {};
			\node[bvert] (c1) at (-1.4,-3) {};
			\node[bvert] (b1) at (-1.5,-1) {};
			\node[wvert,label=above:$x_i$] (r1) at (-2,-1) {};
			\node[bvert] (bb1) at (-0.6,-3) {};
			\draw[-]
			(b1) edge (r1) edge (r2) edge (qq2)
			(b3) edge (p3) edge (p2) edge (qq2)
			(c1) edge (q1) edge (r2) edge (q2)
			(c3) edge (q3) edge (p2) edge (r2)
			(bb1) edge (r2) edge (q2) edge (q3)
			;
		\end{tikzpicture}$\rightarrow$\hspace{-1mm}
		\begin{tikzpicture}[scale=\scl,baseline={(current bounding box.center)}]
			\node[wvert,label=below:$q_{i+2}$] (q3) at (0,-3) {};
			\node[wvert] (p3) at (0,-1) {}; 
			\node[wvert,label=right:$q_{i+1}$] (q2) at (-1,-4) {};
			\node[wvert,label=right:$q_{i+1}$] (qq2) at (-1,0) {};
			\node[wvert,label=above:$x_{i+2}$] (r3) at (-0.5,-1) {};
			\node[wvert,label=below:$q_i$] (q1) at (-2,-3) {};
			\node[wvert,label=left:$x_{i+1}$] (r2) at (-1,-2) {};
			\node[bvert] (c3) at (-1,-1) {};
			\node[bvert] (b3) at (-0.25,-2) {};
			\node[bvert] (c1) at (-1.4,-3) {};
			\node[bvert] (b1) at (-1.5,-1) {};
			\node[wvert,label=above:$x_i$] (r1) at (-2,-1) {};
			\node[bvert] (bb1) at (-0.6,-3) {};
			\draw[-]
			(b1) edge (r1) edge (r2) edge (qq2)
			(b3) edge (p3) edge (r3) edge (q3)
			(c1) edge (q1) edge (r2) edge (q2)
			(c3) edge (r3) edge (qq2) edge (r2)
			(bb1) edge (r2) edge (q2) edge (q3)
			;
		\end{tikzpicture}			
		\caption{Step 2: Replacing $(p_i,p_{i+1})$ with $(x_{i+1},x_{i+2})$.}
		\label{fig:crsequence1}	
	\end{figure}
	
	\begin{figure}[ht]
		\def\scl{1.5}
		\centering
		\begin{tikzpicture}[scale=\scl,baseline={([yshift=-1mm]current bounding box.center)}]
			\node[wvert,label=right:$q_2$] (q2) at (0,-4) {};
			\node[wvert,label=right:$q_2$] (qq2) at (0,0) {};
			\node[wvert,label=right:$x_2$] (p2) at (0,-2) {};
			\node[wvert,label=below:$q_1$] (q1) at (-1,-3) {};
			\node[wvert,label=above:$x_1$] (p1) at (-0.5,-1) {};
			\node[wvert,label=above:$M^{-1}(x_{n+1})$] (pp1) at (-1.5,-1) {};
			\node[wvert,label=right:$p_1$] (r1) at (-1,-1) {};
			\node[wvert,label=left:$q_0$] (qn) at (-2,-4) {};
			\node[wvert,label=left:$q_0$] (qqn) at (-2,0) {};
			\node[wvert,label=left:$M^{-1}(x_n)$] (pn) at (-2,-2) {};
			\node[bvert] (cn) at (-1.6,-3) {};
			\node[bvert] (bn) at (-1.9,-1) {};
			\node[bvert] (c1) at (-0.4,-3) {};
			\node[bvert] (b1) at (-0.1,-1) {};
			\node[bvert] (bb) at (-0.7,-2) {};
			\node[bvert] (bbb) at (-1.3,-2) {};
			\draw[-]
			(b1) edge (p1) edge (p2) edge (qq2)
			(bn) edge (pp1) edge (pn) edge (qqn)
			(c1) edge (q1) edge (p2) edge (q2)
			(cn) edge (q1) edge (pn) edge (qn)
			(bb) edge (p1) edge (r1) edge (q1)
			(bbb) edge (r1) edge (pp1) edge (q1)
			;
		\end{tikzpicture}
		\caption{Step 3: Local configuration near $p_1$ after Step 2. Apply Step 1 backwards to obtain $\Gamma_n$.}\label{fig:squarestep3}
	\end{figure}
	
	As in Theorem~\ref{thm:crdynamicslocalmoves}, we have the following.
	\begin{Theorem} \label{thm:sqseq}
		Suppose \smash{$[\wt] \in \mathcal X_{N_{\Gamma_n},\vec \alpha}^\lambda$} is such that $u^{-1} \circ \pi_{\vec \alpha} ([\wt])=(p,q,M)$. Consider the~sequence of moves shown in Figures $\ref{fig:insertr1}$, $\ref{fig:crsequence1}$ and $\ref{fig:squarestep3}$, and let $\mu$ denote the induced birational map of weights. Then, the following diagram commutes:
		\[
		{\begin{tikzcd}[every label/.append style = {font = \normalsize}, row sep=large, column sep = huge]
				\mathcal X_{N_{\Gamma_n},\vec \alpha}^\lambda \arrow[r, "u^{-1} \circ \pi_{\vec \alpha}"] \arrow[d, "\mu"']
				& \mathcal U_n \arrow[d, "\nu_{\vec \alpha}"] \\
				\mathcal X_{N_{\Gamma_n},\vec \alpha}^\lambda \arrow[r,"u^{-1} \circ \pi_{\vec \alpha}"]
				& \mathcal U_n
		\end{tikzcd}}.
		\]
		
	\end{Theorem}

	We now explain how the sequence in Theorem~\ref{thm:sqseq} is related to geometric $R$-matrices. Let~${n\geq2}$. We first transform the graphs $\Gamma_n$ into graphs $\widetilde{\Gamma}_n$ as follows. Recall that $\Gamma_n$ possesses two zig-zag paths with homology $(1,0)$, the path $\xi_1$ which goes through all the white vertices carrying the $q_i$'s and the path $\xi_2$ which goes through all the white vertices carrying the~$p_i$'s. We modify the path $\xi_2$ by contracting all the 2-valent white vertices carrying points of the form $p_{2i-1}$ with $1\leq 2i-1\leq n$ (this concerns every white vertex carrying a point of the form~$p_{2i-1}$ except in the odd $n$ case the trivalent white vertex which carries $p_1$) and we horizontally expand all the white vertices carrying points of the form $p_{2i}$ with $1\leq 2i\leq n$, in such a~way that each of the two new white vertices created by such an expansion is connected to one black neighbor in $\xi_1$ and to another black neighbor in $\xi_2$. We call $\widetilde{\Gamma}_n$ the graph obtained from this procedure. Using the terminology of \cite{Chepuri}, this corresponds to putting the 2-loop graph~$\Gamma_n$ in its 1-expanded form. We have depicted $\widetilde{\Gamma}_6$ on Figure~\ref{fig:hexstrip}.
	
	\begin{figure}[ht]
		\centering
		\begin{tikzpicture}
			\node[wvert,label=left:$q_2$] (q2) at (2,0) {};
			\node[wvert,label=left:$q_4$] (q4) at (5,0) {};
			\node[wvert,label=left:$q_6$] (q6) at (8,0) {};
			\node[wvert,label=left:$q_2$] (qq2) at (2,4) {};
			\node[wvert,label=left:$q_4$] (qq4) at (5,4) {};
			\node[wvert,label=left:$q_6$] (qq6) at (8,4) {};
			\node[wvert,label=below:$q_1$] (q1) at (0.5,1) {};
			\node[wvert,label=below:$q_3$] (q3) at (3.5,1) {};
			\node[wvert,label=below:$q_5$] (q5) at (6.5,1) {};
			\node[wvert,label=below:$q_7$] (qq1) at (9.5,1) {};
			\node[wvert,label=above right:$p_2$] (p2) at (1,2) {};
			\node[wvert,label=above left:$p_2$] (pp2) at (3,2) {};
			\node[wvert,label=above right:$p_4$] (p4) at (4,2) {};
			\node[wvert,label=above left:$p_4$] (pp4) at (6,2) {};
			\node[wvert,label=above right:$p_6$] (p6) at (7,2) {};
			\node[wvert,label=above left:$p_6$] (pp6) at (9,2) {};
			\node[bvert] (c2) at (2,2) {};
			\node[bvert] (c4) at (5,2) {};
			\node[bvert] (c6) at (8,2) {};
			\node[bvert] (bb1) at (1,1) {};
			\node[bvert] (b3) at (3,1) {};
			\node[bvert] (bb3) at (4,1) {};
			\node[bvert] (b5) at (6,1) {};
			\node[bvert] (bb5) at (7,1) {};
			\node[bvert] (b1) at (9,1) {};
			\node[bvert] (d1) at (0.5,3) {};
			\node[bvert] (d3) at (3.5,3) {};
			\node[bvert] (d5) at (6.5,3) {};
			\node[bvert] (dd1) at (9.5,3) {};
			
			\draw[-]
			(bb1) edge (p2) edge (q1) edge (q2)
			(bb3) edge (p4) edge (q3) edge (q4)
			(bb5) edge (p6) edge (q5) edge (q6)
			(b1) edge (qq1) edge (pp6) edge (q6)
			(b3) edge (q3) edge (pp2) edge (q2)
			(b5) edge (q5) edge (pp4) edge (q4)
			(c2) edge (p2) edge (pp2)
			(c4) edge (p4) edge (pp4)
			(c6) edge (p6) edge (pp6)
			(d1) edge (qq2) edge (p2)
			(d3) edge (qq2) edge (qq4) edge (pp2) edge (p4)
			(d5) edge (qq4) edge (qq6) edge (pp4) edge (p6)
			(dd1) edge (qq6) edge (pp6)
			;
			
			\node[label=left:$\rightarrow$] (n) at (q1) {};
			\node[label=left:$\rightarrow$] (n) at (d1) {};
			
			\node[] (x) at (0.5,2) {$\times$};
			\node[] (x) at (3.5,2) {$\times$};
			\node[] (x) at (6.5,2) {$\times$};
			\node[] (x) at (2,1) {$\times$};
			\node[] (x) at (5,1) {$\times$};
			\node[] (x) at (8,1) {$\times$};
		\end{tikzpicture}
		\caption{The graph $\widetilde{\Gamma}_6$. The two arrows on the left indicate the two endpoints to which a bigon is attached and the stars indicate the string of hexagonal faces traversed by spider moves before coming back to the starting point and deleting the bigon.}
		\label{fig:hexstrip}
	\end{figure}
	
	The sequence in Theorem~\ref{thm:sqseq} is a sequence of local moves on TCD maps starting and ending with a TCD map on $\Gamma_n$. It can be seen as a sequence of dimer local moves starting and ending with $\Gamma_n$. Since the transformation of $\Gamma_n$ into $\widetilde\Gamma_n$ only involves contractions and expansions of degree $2$ vertices, the sequence can be turned into a sequence of dimer local moves starting and ending with $\widetilde\Gamma_n$. Up to contractions and expansions of degree $2$ vertices, the latter sequence of dimer local moves is exactly the one described above for the geometric $R$-matrix transformation.
	
\appendix

		\section{Schur complement}
		\label{sec:schur}
		Suppose $M=\bigl[\begin{smallmatrix}
			A &B\\C&D
		\end{smallmatrix}\bigr]$
		is a $(p+q) \times (r+q)$ block matrix with $D$ an invertible $q \times q$ matrix. Let $V:=\coker M$. Let $e_i$ denote the $i$th basis column vector, and let $v_i \in V$ denote the image of $e_i$ under the cokernel map. Let $M/D:=A-B D^{-1} C$ denote the Schur complement.
		
		\begin{Theorem}[Schur determinant formula \cite{Schur}] \la{thm::schur1}
			If $M$ is a square matrix $(p=r)$, then we have%
			\[
			\det (M) = \det (D) \det (M/D).
			\]
		\end{Theorem}
		
		\begin{Theorem}\la{thm::schur}
			$\coker M/D \cong \coker M$, and under this identification the cokernel map of $M/D$ is $e_i \mapsto v_i$ for $i=1,\dots,p$.
		\end{Theorem}
		\begin{proof}
			After a change of basis, $M$ takes the block diagonal form $\bigl[\begin{smallmatrix} M/D &0\\ 0&D \end{smallmatrix}\bigr]$,
			\[
			M=\begin{bmatrix}
				I & B D^{-1}\\
				0 & I
			\end{bmatrix}
			\begin{bmatrix} M/D &0\\ 0&D \end{bmatrix}
			\begin{bmatrix}
				I & 0\\
				D^{-1}C & I
			\end{bmatrix}.
			\]
			
			Since $D$ is invertible, we have $\coker D=0$. Therefore, we have
			\[
			\coker M \cong \coker M/D \oplus \coker D \cong \coker M/D.
			\]
			Under the change of basis $\big[\begin{smallmatrix}
				I & B D^{-1}\\
				0 & I
			\end{smallmatrix}\bigr]$ of $\C^{p+q}$, we have
			\[
			e_i \mapsto \begin{bmatrix}
				I & B D^{-1}\\
				0 & I
			\end{bmatrix}^{-1} e_i = e_i\qquad \text{for }i =1,2,\dots,p,
			\]
			from which we see that the cokernel map of $M/D$, when we identify $\coker M/D$ with ${\coker M\!=\!V}$, is given by $e_i \mapsto v_i$.
		\end{proof}

\subsection*{Acknowledgements}

TG thanks Nick Ovenhouse for discussions about networks in a cylinder. SR thanks Ivan Izmestiev for discussions on cross-ratio dynamics during a visit at TU Wien, Anton Izosimov for comments on Newton polygons in the first version of this paper and Rei Inoue for exchanges on the geometric $R$-matrix. NA was supported by the Deutsche Forschungsgemeinschaft (DFG) Collaborative Research Center TRR 109 ``Discretization in Geometry and Dynamics'' and by the ENS-MHI chair funded by MHI. NA and SR were partially supported by the Agence Nationale de la Recherche, Grant Number ANR-18-CE40-0033 (ANR DIMERS). SR was also partially supported by the CNRS grant Tremplin@INP, which funded a visit of NA to Paris-Saclay.


\pdfbookmark[1]{References}{ref}
\LastPageEnding

\end{document}